\documentclass[numberedappendix]{emulateapj}

\usepackage{natbib}

\usepackage{ulem}

\usepackage{subfigure}

\usepackage{amsmath}
\begin{document}

\newcommand{\unit}[1]{\ensuremath{\, \mathrm{#1}}}

\renewcommand*{\thefootnote}{\fnsymbol{footnote}}

\title{One Plane for All: Massive Star-Forming and Quiescent Galaxies Lie on the Same Mass Fundamental Plane at $z\sim0$ and $z\sim0.7$}

\author{Rachel Bezanson\altaffilmark{1}\footnotemark[$\dagger$], Marijn Franx\altaffilmark{2}, and Pieter G. van Dokkum\altaffilmark{3}}
\slugcomment{Submitted to the Astrophysical Journal}

\altaffiltext{1}{Steward Observatory, Department of Astronomy, University of Arizona, AZ 85721, USA}
\altaffiltext{2}{Sterrewacht Leiden, Leiden University, NL-2300 RA Leiden, Netherlands}
\altaffiltext{3}{Department of Astronomy, Yale University, New Haven, CT 06520-8101}
\footnotetext[$\dagger$]{Hubble Fellow}

\shortauthors{Bezanson, Franx, \& van Dokkum}
\shorttitle{One Plane for All}

\begin{abstract}
Scaling relations between galaxy structures and dynamics have been studied extensively for early and late-type galaxies, both in the local universe and at high redshifts. The abundant differences between the properties of disky and elliptical, or star-forming and quiescent, galaxies seem to be characteristic of the local Universe; such clear distinctions begin to disintegrate as observations of massive galaxies probe higher redshifts. In this Paper, we investigate the existence the mass fundamental plane of all massive galaxies ($\sigma\gtrsim100\unit{km\,s^{-1}}$). This work includes local galaxies ($0.05<z<0.07$) from the SDSS, in addition to 31 star-forming and 72 quiescent massive galaxies at intermediate redshift ($z\sim0.7$) with absorption line kinematics from deep Keck-DEIMOS spectra and structural parameters from HST imaging. In two parameter scaling relations, star-forming and quiescent galaxies differ structurally and dynamically. However, we show that massive star-forming and quiescent galaxies lie on nearly the same mass fundamental plane, or the relationship between stellar mass surface density, stellar velocity dispersion, and effective radius. The scatter in this relation (measured about $\log\sigma$) is low: $0.072\unit{dex}$  ($0.055\unit{dex}$ intrinsic) at $z\sim0$ and $0.10\unit{dex}$ ($0.08\unit{dex}$ intrinsic) at $z\sim0.7$. This three dimensional surface is not unique: virial relations, with or without a dependence on luminosity profile shapes, can connect galaxy structures and stellar dynamics with similar scatter. This result builds on the recent finding that mass fundamental plane has been stable for early-type galaxies since $z\sim2$ \citep{bezanson:13b}. As we now find this also holds for star-forming galaxies to $z\sim0.7$, this implies that these scaling relations of galaxies will be minimally susceptible to progenitor biases due to the evolving stellar populations, structures, and dynamics of galaxies through cosmic time.
\end{abstract}


\section{Introduction}

Galaxy bimodality seems to be a fundamental property of the Universe, especially at the present day. In the local Universe, galaxies are either forming stars or not and as a result of their differing stellar populations their colors are generally blue or red \citep[e.g.][]{blanton:03}. In fact, the existence of massive quenched galaxies has been demonstrated as early as $z\sim4$, only a couple billion years after the Big Bang \citep[e.g.][]{straatman:14}. Traditionally, scaling relations between global properties (sizes, luminosities or masses, and kinematics) of disk and elliptical galaxies have been studied separately to constrain their formation and evolutionary models. This approach is intuitive as late and early-type galaxies differ in most ways in the local Universe. Structurally, the stars in star-forming galaxies are generally flattened into disklike formations, following exponential light profiles. Galaxies with quiescent stellar populations are rounder spheroids which follow de Vaucouler light profiles in projection. Furthermore, at fixed mass, quiescent galaxies are more compact than their star-forming counterparts \citep[e.g.][]{shen:03}. These populations also appear to differ dynamically: star-forming galaxies are primarily supported by rotation and quiescent galaxies are dominated by dispersion support. 

The relationship between rotational velocity and galaxy luminosity (or stellar mass), called the Tully-Fischer relation, describes the fundamental scaling of disk (star-forming) galaxies \citep[e.g.][]{tullyfisher,bell:01}. Early-type (quiescent) galaxies lie on a similar scaling relation between the luminosity and velocity dispersion, called the Faber-Jackson relation \citep{faberjackson}. The scatter around this relation tightens when galaxy sizes are included, indicating that quiescent galaxies are better described by the three parameter Fundamental Plane \citep[e.g.][]{djorgovski:87,dressler:87}. These relations have been used to constrain aspects of galaxy formation such as the growth of disks within dark matter halos \citep[e.g.][]{fall:80,blumenthal:86,mmw:98} and variations in the global mass-to-light ratios of elliptical galaxies \citep[e.g.][]{faber:87}. Observationally, measurements of these scaling relations have been extended to high redshift, adding the dimension of time to further constrain formation of star-forming disks \citep[e.g.][]{vogt:96,weiner:06,kassin:07,miller:12} or the aging of quiescent spheroids \citep[e.g.][]{dokkum:96,wel:04,dokkummarel:07,holden:10,toft:12}. 

However, as observations of distant galaxies push to earlier epochs in the high redshift universe, building evidence suggests that the clear distinctions between galaxy populations begin to break down. Populations of massive star-forming galaxies, which must be the progenitors of many of today's massive galaxies, increase in number density at higher redshift \citep[e.g.][]{bell:04,brammer:11,whitaker:12a,muzzin:13,tomczak:13}. Dividing lines between structural properties and stellar populations become blurred as quiescent galaxies may appear more disklike at higher redshifts ($z\gtrsim1$) \citep[e.g.][]{wel:11,weinzirl:11,bruce:12,chevance:12,chang:13}. As a result of the decreasing number density of quiescent galaxies with redshift, selection criteria based on either structural morphology or stellar populations designed to identify galaxies through cosmic time will be biased against a subset of the progenitors of those early galaxies. This ``progenitor bias'' \citep[e.g.][]{dokkum:96} will become increasingly important as we connect the evolution of galaxies through earlier times. 

As galaxy populations become less clearly bimodal at earlier epochs and individual galaxies likely transition between star-forming and quiescent periods, it would be preferable to define a flexible framework that allows for star-forming and quiescent galaxies to be studied together to allow for this ambiguity. In this paper, we examine the scaling relations between galaxy structures (light profile shapes and sizes), stellar masses (from stellar population synthesis modeling), and dynamics (velocity dispersions) for star-forming and quiescent galaxies alike. We examine these relations in the local Universe ($0.05<z<0.07$) utilizing data from the Sloan Digital Sky Survey (SDSS) and at higher redshift ($z\sim0.7$) using a deep spectroscopic survey collected with the DEIMOS spectrograph on Keck II in the COSMOS and UKIDSS UDS fields.

The Paper is organized as follows: \S \ref{sec:data} provides an overview of the two datasets included in this analysis and describes the measured and derived properties of galaxies in those samples. \S \ref{sec:2d} examines the two dimensional scaling relations between structural and dynamical properties of galaxies, highlighting the differences between star-forming and quiescent massive galaxies. \S \ref{sec:massfp} demonstrates the existence of a unified mass fundamental plane for both star-forming and quenched galaxies. In \S \ref{sec:sigma} we assess the ability of a variety of scaling relations to include both star-forming and quiescent galaxy populations, comparing measured scatter about each relation. Finally, in \S \ref{sec:discuss} we discuss the results of this work and highlight the implications for future studies of galaxy evolution. Throughout this paper we assume standard $\mathrm{\Lambda CDM}$ concordance cosmology with $H_0 = 70 \unit{km\,s^{-1} Mpc^{-1}}$, $\Omega_M = 0.3$, and $\Omega_{\Lambda} = 0.7$. All magnitudes are quoted in the AB system. 

\section{Data}
\label{sec:data}

\subsection{SDSS Sample at $z\sim0$}

For our study of local galaxies, we use a sample of galaxies at $0.05<z<0.07$ from DR7 of the Sloan Digital Sky Survey \citep{dr7}, selected as described in \citet{bezanson:13b}. Galaxies are selected to have reliable spectroscopic measurements with: \textit{keep\_flag}=1, \textit{z\_warning}=0, \textit{sciencePrimary}=1, $S/N>20$. Stellar mass-to-light ($M_{\star}/L$) ratios are acquired from the MPA-JHU galaxy catalog \citep{brinchmann:04}. Best-fit Sersic profiles in the $r'$-band are included from \citet{simard:11} and total stellar masses are calculated by scaling $M_{\star}/L$ to the total luminosity of the best-fit S\'ersic profile (see equation \ref{eq:mass}). Effective radii are circularized $r_e = R_{hl}\sqrt{b/a}$, where $R_{hl}$ is the semi-major half light radius and $b/a$ is the axis ratio. Velocity dispersions are taken from David Schlegel's \textit{spZbest} catalog ($v\_disp$) and are aperture corrected from the $3''$ SDSS fiber to an effective radius using:

\begin{equation}
\sigma_{re}=\sigma_{ap}(r_{ap}/r_{\mathrm{e}})^{0.066}
\label{eq:apcor}
\end{equation}

\noindent from \citep{cappellari:06}. Only galaxies with $<10\%$ errors in velocity dispersion are included in the final sample.

Both star-forming and quiescent galaxies are included in this sample; the two populations are distinguished based on their colors. We calculate K-corrections to observed colors to $z=0$ using KCORRECT \citep{blanton:03, blanton:07}. Finally, we adopt the $u-r$ and $r-z$ rest-frame color-cuts from \citep{holden:12} to identify quiescent galaxies as:

\begin{align}
&(u-r) > 2.26, \\
&(r-z) < 0.75, \\
&\mathrm{and\,} (u-r) > 0.76 + 2.5(r-z).
\end{align}

\noindent These criteria have been demonstrated to separate star-forming and quiescent galaxies with only $\sim18\%$ contamination.

\subsection{Keck-DEIMOS Sample at $z\sim0.7$} \label{sect:deimos}

\begin{figure*}[!t]
\centering
\includegraphics[width=0.45\textwidth]{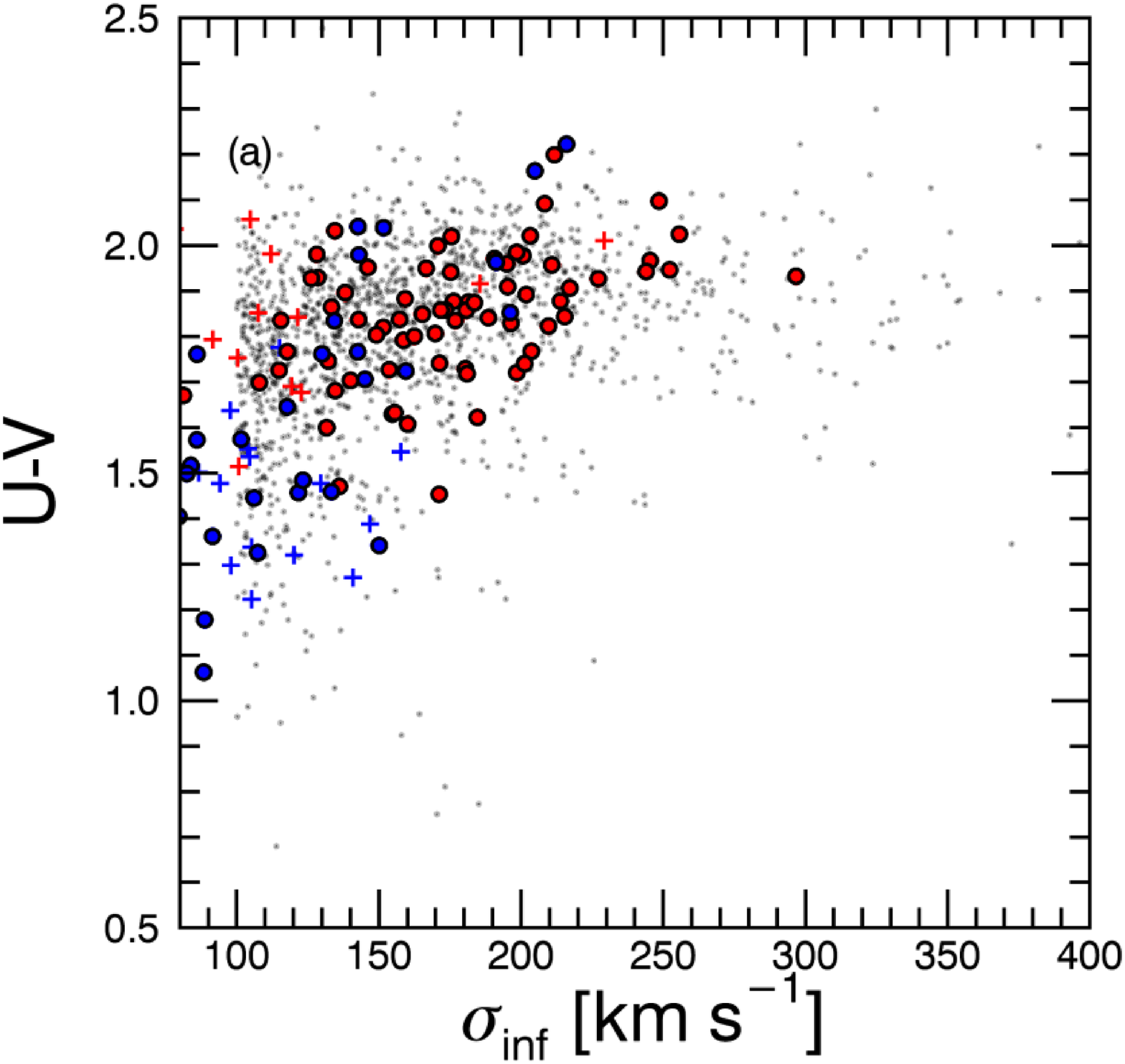}
\includegraphics[width=0.45\textwidth]{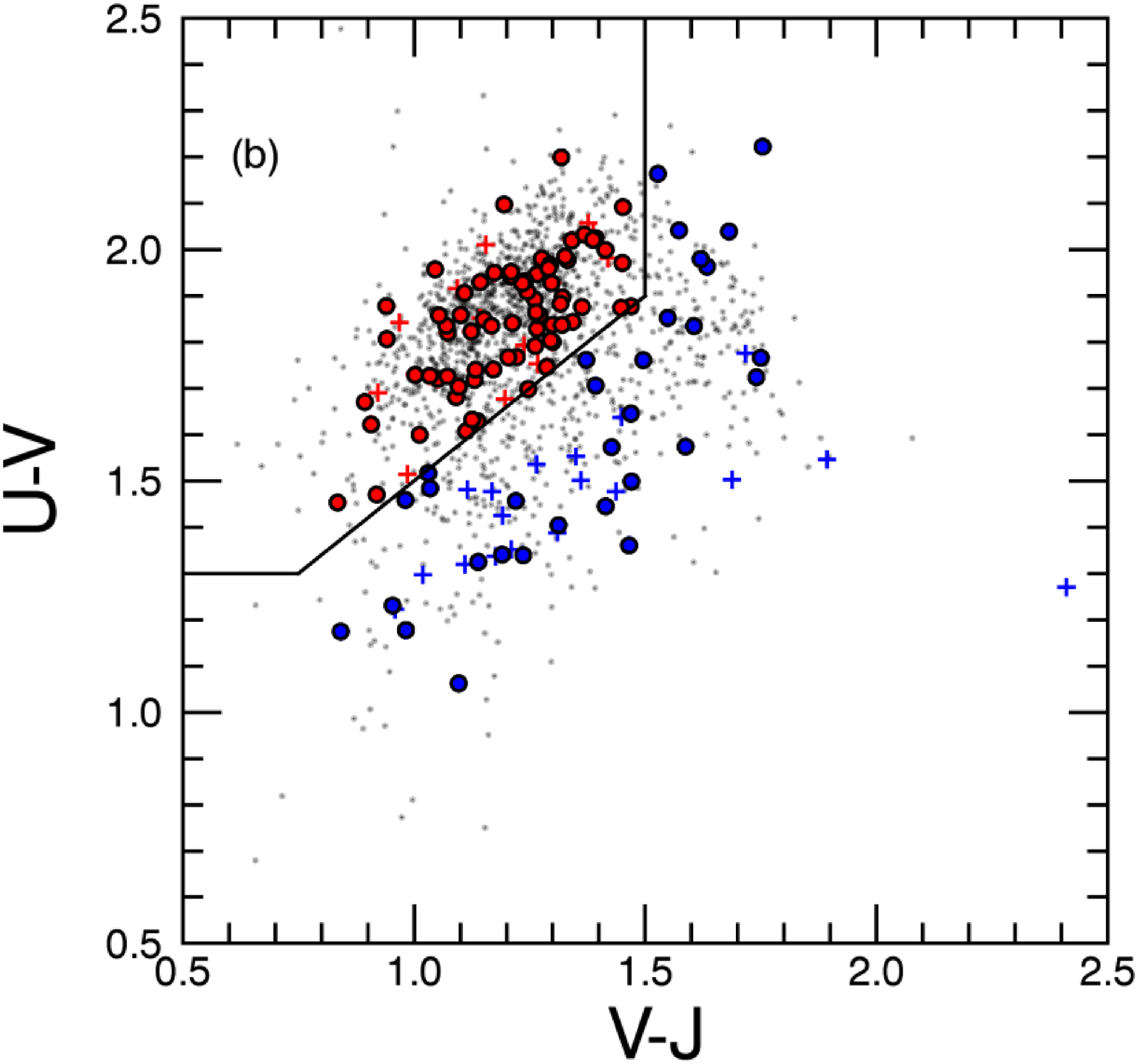}
\caption{Selection criteria for the $z\sim0.7$ spectroscopic sample (quiescent galaxies in red, star-forming galaxies in blue) relative to $0.4<z_{phot}<0.95$ galaxies in the NMBS-Cosmos field. Filled circles represent successful measurements of velocity dispersions (with $<15\%$ statistical error), crosses represent galaxies with spectra that have been excluded from this sample, either due to failed extraction or insufficient S/N. (a) U-V rest-frame color versus inferred velocity dispersion: galaxies are selected to span a range in both quantities. (b) U-V versus V-J rest-frame colors, used to distinguish between star-forming (lower right of solid black dividing lines, indicating \citet{whitaker:12a} empirical distinctions) and quiescent galaxies (upper left).}
\label{fig:selection}
\end{figure*}
Spectra for a total of 162 targeted galaxies were collected using the $1200\unit{mm^{-1}}$ grating, centered at $7800\unit{\AA}$.  Spectra were reduced and extracted using the Spec2d pipeline \citep{cooper:12, newmandeep:12}. Telluric corrections are applied by fitting models for atmospheric absorption, scaled to fit spectra in each mask. The resulting spectra have an average spectral range of $\sim$6500--$9200\mathrm{\AA}$.  The instrumental resolution, as measured from sky lines was $\sim1.6\unit{\AA}$ at $\sim$7800$\unit{\AA}$, which corresponds to $R\sim$5000 or $\Delta\,v\sim60\unit{km\,s^{-1}}$.

We observed a sample of galaxies at $0.4<z<0.9$ from the Newfirm Medium Band Survey (NMBS)-Cosmos \citep{whitaker:11} and UKIDSS UDS fields \citep{williams:09}, focusing on overlap with the CANDELS \citep{candels,candelsb}/3DHST \citep{3dhst,skelton3dhst} fields, using DEIMOS \citep{faber:deimos} on Keck II from January 19-21, 2012. Three masks were observed: two in NMBS-Cosmos with total exposures of 13.67 and 5.67 hours and one in UDS for 7.67 hours. Weather and seeing conditions throughout the run were very good, with average seeing ranging from $\sim 0.5-0.7\arcsec$.

Galaxies were selected to span a range in \emph{inferred velocity dispersion} \citep[see e.g.][]{bezanson:11} prioritizing galaxies with $\sigma_{\mathrm{inf}}\gtrsim100\unit{km\,s^{-1}}$. We adopt the following definition of inferred velocity dispersion in this Section and revisit possible definitions in \S \ref{sec:sigma}:

\begin{equation}
\sigma_{\mathrm{inf,V(n)}} = \sqrt{\frac{G M_{\star}}{0.557 k_{V}R_e}},
\end{equation}
in which the Virial constant depends on S\'ersic index as:

 \begin{equation}
 k_{\mathrm{V}}(n)\approx 8.87-0.831n+0.0241n^2
 \label{eq:kv_n}
 \end{equation} \citep{cappellari:06}. This includes the average ratio $\langle\frac{M_{\star}}{M_{dyn}}\rangle \approx0.557$ measured from a similar sample of galaxies in the SDSS in \citet{bezanson:11}. 

\noindent This corresponds to a selection in size and mass ($\log M_{\star}>10$), with no preselection on morphology. Therefore the dataset includes early and late-type, or alternatively quiescent and star-forming, galaxies. Selection of this targets relative to the NMBS-Cosmos photometric catalog is presented in Figure\,\ref{fig:selection}. Additional properties of the sample (large symbols) are presented in Figure\,\ref{fig:completeness} relative to galaxies in the NMBS-Cosmos field within the same redshift range (gray points). In addition to the $\sigma$ selection, targets were selected within $0.4\lesssim z_{phot}\lesssim1$ and brighter than $I=23.5$. Priority was given to spanning the observed range of $U-V$ color, inferred velocity dispersion, S\'ersic index, and SFR. Finally, additional low mass ($\sigma\lesssim100\unit{km s^{-1}}$) galaxies within the same redshift range were added to fill spectroscopic masks. These filler galaxies are apparent at the low-$\sigma$ end of Figure\,\ref{fig:selection}(a) (and later in Figure\,\ref{fig:siginflssfr}) and were biased towards brighter (and bluer) galaxies.

In some cases (13 galaxies), we failed to successfully extract a sufficient one-dimensional spectrum for one or both of the red and blue chips, these spectra are excluded from the final sample.  The remaining 148 galaxies are included in this work. 

\begin{deluxetable*}{ccccccccccc}[!t]
\tablecaption{DEIMOS $z\sim0.7$ Sample - Galaxy Properties}
\tablefontsize{\footnotesize}

\tablehead{
\colhead{id} & \colhead{RA} & \colhead{Dec} & \colhead{z} & \colhead{$R_e$} & \colhead{n} & \colhead{Filter} & \colhead{log Stellar Mass} & \colhead{$\sigma_{\mathrm{aperture}}$} & \colhead{$\sigma_{Re}$} & \colhead{Exposure Time} \\
\colhead{} & \colhead{[$^{o}$]} & \colhead{[$^{o}$]} & \colhead{} & \colhead{[kpc]} & \colhead{} & \colhead{} & \colhead{[$M_{\odot}$]} & \colhead{[$km\,s^{-1}$]} & \colhead{[$km\,s^{-1}$]} & \colhead{[s]} \\
}
\startdata
C1971 &    150.104 &      2.198 &      0.682 &   4.4 &   4.7 &  WFC3-F160W  & 10.96 &   214 $\pm$ 8 &   211 $\pm$ 8 & 49200 \\
C2335 &    150.097 &      2.205 &      0.424 &   0.8 &   2.7 &  WFC3-F160W  & 10.13 &   131 $\pm$ 5 &   143 $\pm$ 6 & 49200 \\
C3382 &    150.084 &      2.222 &      0.560 &   5.9 &   1.5 &  WFC3-F160W  & 10.62 &   158 $\pm$ 7 &   152 $\pm$ 7 & 49200 \\
C3420 &    150.119 &      2.223 &      0.839 &   2.1 &   3.3 &  WFC3-F160W  & 10.83 &   245 $\pm$ 15 &   255 $\pm$ 16 & 49200 \\
C3751 &    150.121 &      2.227 &      0.733 &   3.6 &   2.9 &  WFC3-F160W  & 11.07 &   171 $\pm$ 7 &   172 $\pm$ 7 & 49200 \\
C3769 &    150.121 &      2.230 &      0.732 &   2.0 &   0.8 &  WFC3-F160W  & 10.10 &   153 $\pm$ 14 &   159 $\pm$ 15 & 49200 \\
C4987 &    150.116 &      2.250 &      0.747 &   1.4 &   4.2 &  WFC3-F160W  & 10.47 &   236 $\pm$ 13 &   252 $\pm$ 14 & 49200 \\
C5585 &    150.104 &      2.261 &      0.642 &   1.9 &   1.7 &  WFC3-F160W  & 10.15 &   133 $\pm$ 9 &   138 $\pm$ 9 & 49200 \\
C6205 &    150.086 &      2.272 &      0.728 &   1.5 &   4.7 &  WFC3-F160W  & 10.65 &   225 $\pm$ 6 &   238 $\pm$ 7 & 49200 \\
C6574 &    150.083 &      2.278 &      0.835 &   1.6 &   1.8 &  WFC3-F160W  & 10.49 &   241 $\pm$ 34 &   255 $\pm$ 36 & 49200 \\

\enddata

\tablecomments{Table \ref{tab:data} is published in its entirety in the electronic edition of this article, a portion is shown here for guidance regarding its form and content.}
\label{tab:data}

\end{deluxetable*}

\subsection{Imaging and Photometric Catalogs}

Stellar population analysis and rest-frame color estimates for galaxies in this sample are based on multi-wavelength broad and medium-band photometric data from two fields: the UKIRT Infrared Deep Sky Survey (UKIDSS) - Ultra Deep Survey (UDS) and Newfirm Medium Band Survey (NMBS) - COSMOS. Additionally, high resolution, space-based imaging taken using the Advanced Camera for Surveys (ACS) and the Wide Field Camera 3 (WFC3) on the Hubble Space Telescope (HST) exists in both fields, also described in \S \ref{subsec:derived}.

In the COSMOS field, we utilize v5.1 NMBS catalogs, which we briefly summarize below; see \citet{whitaker:11} for a full description. This dataset is designed around deep medium-band Near infrared (NIR) imaging in $J1,J2,J3,H1,H2$ bands from the Mayall $4.0\unit{m}$ telescope in addition to a multitude of ancillary data. Optical imaging is included from the Deep Canada-France-Hawaii Telescope Legacy survey (CFHTLS) \citep{erben:09, hildebrandt:09} ($u',g',r',i',z'$) and deep Subaru imaging in $B_J,V_J,r^+,i^+,z^+$ and 12 medium band optical filters \citep{taniguchi:07}. Ultraviolet data is included in the NUV and FUV from the Galaxy Evolution Explorer \citep[GALEX][]{martin:05}. Additional IR measurements are included in the NIR from the WIRcam Deep Survey \citep[WIRDS][]{bielby:12} ($J,H,K$) and in the mid-IR four channels of Spitzer-IRAC data and Spitzer-MIPS $24\mu m$ fluxes.

We utilize a $K$-selected v.4.1 catalog in the UKIDSS UDS field \citep{williams:09}. In addition to the deep NIR Imaging ($J,H,K$) from the UKIDSS-UDS survey \citep{warren:07}, this catalog includes optical imaging in the field is included from the Subaru/XMM-Newton Deep Survey (SXDS) \citep{furusawa:08} ($U,B,V,R,i,z$) and four channels of Spitzer-IRAC data.

\subsection{Derived Properties: Sersic Profiles, Stellar Populations, and Rest-Frame Colors} \label{subsec:derived}

\begin{figure*}[!t]
\centering
\includegraphics[width=0.92\textwidth]{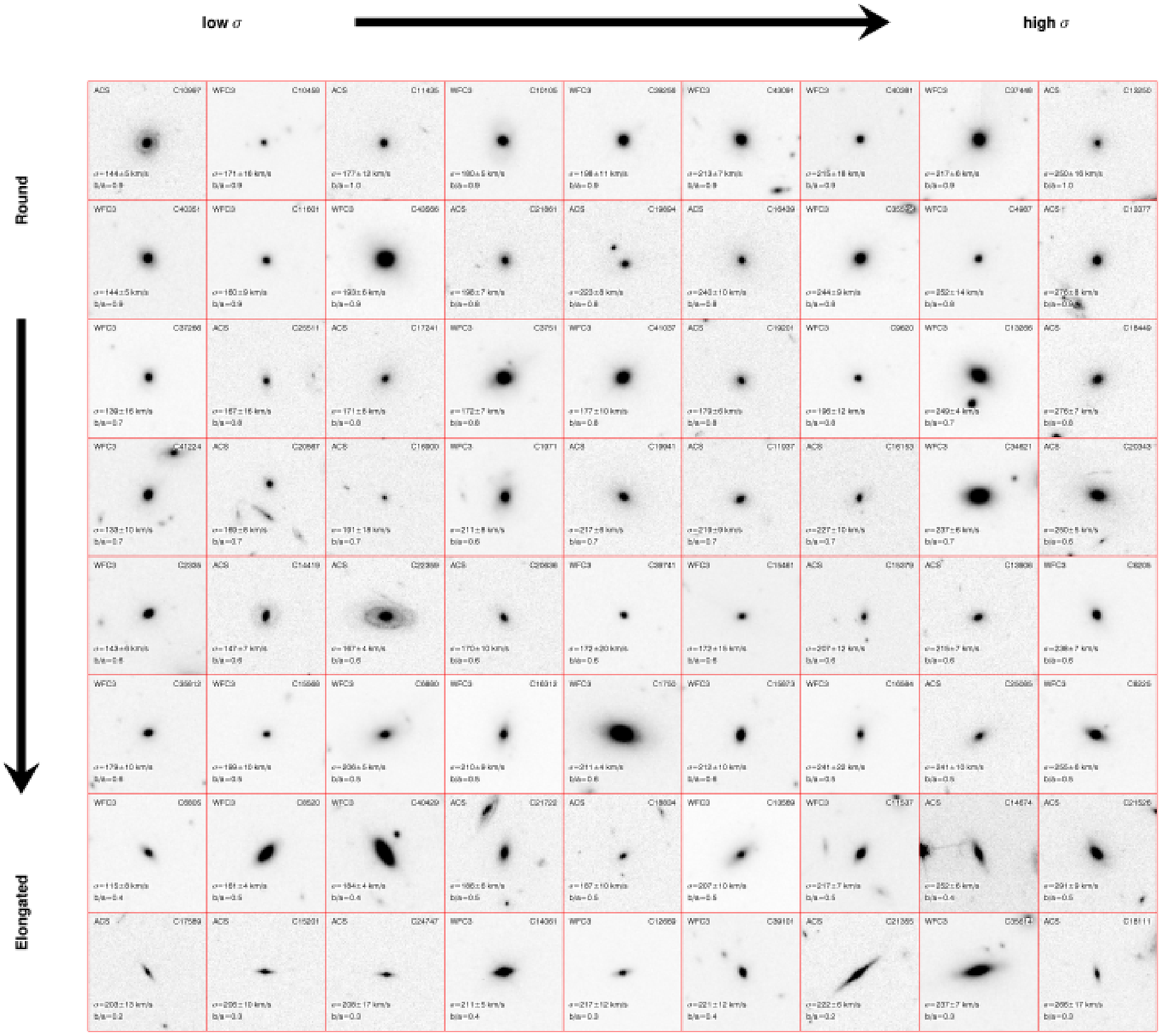}
\includegraphics[width=0.92\textwidth]{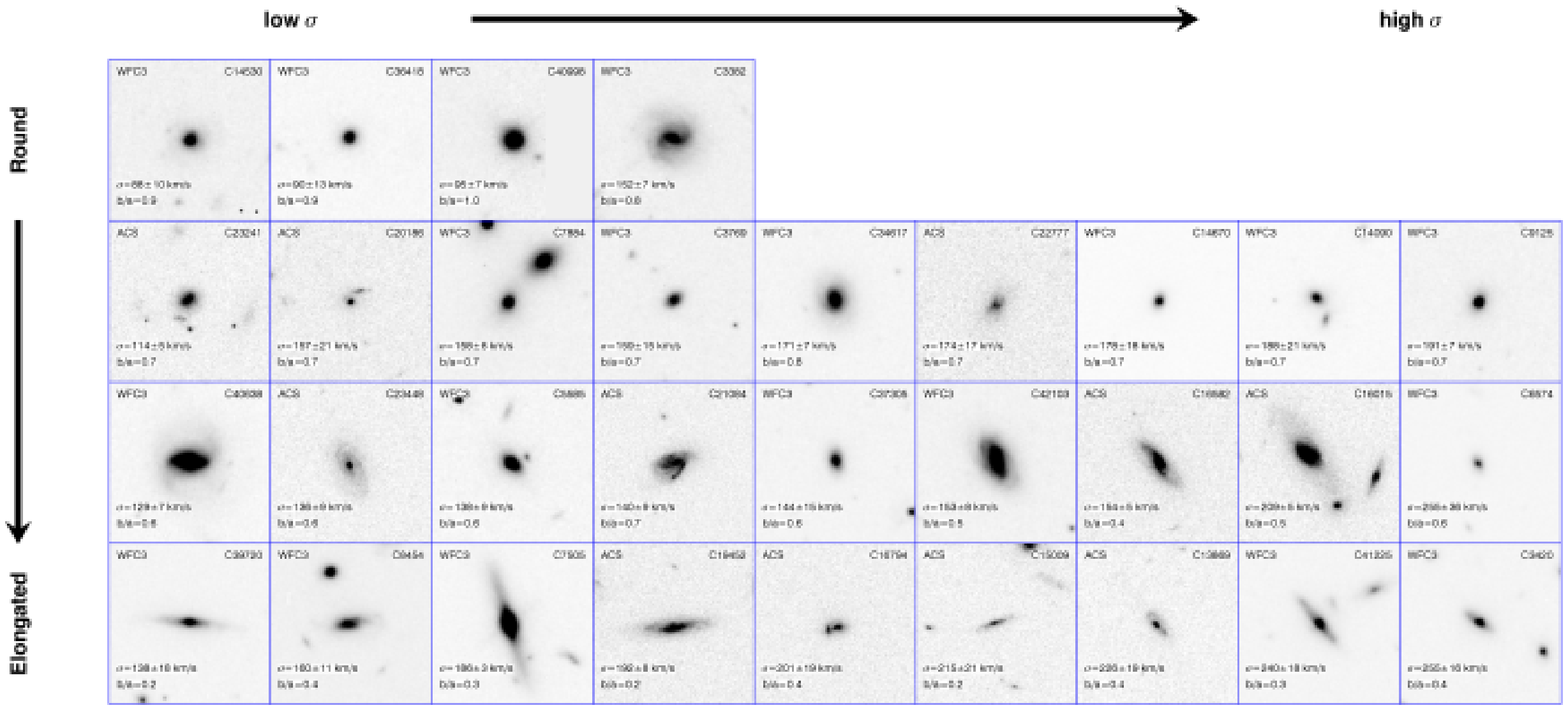}

\caption{Images of quiescent (top panels with red outlines) and star-forming (bottom panels with blue outlines) galaxies, ordered by axis ratio (vertical) and velocity dispersion (horizontal). Images are $9\arcsec \times 9\arcsec$ and are taken from mosaic images of the COSMOS and UDS fields. Images labeled (left upper corner) WFC3 are extracted from CANDELS v1.0 F160W mosaics \citep{candels,candelsb} and ACS are from COSMOS v2.0 ACS mosaics \citep{cosmosacs,massey:10}. Galaxy IDs are indicated in upper right corner of each panel.}
\label{fig:images}
\end{figure*}

Galaxy morphologies are measured by fitting 2D S\'ersic models using GALFIT \citep{galfit} to HST imaging, either from CANDELS F160W WFC3 imaging \citep{wel:12} when available (74 galaxies) or from ACS F814W imaging \citep{bezanson:11} in NMBS-Cosmos (59 galaxies). Figure\,\ref{fig:images} includes galleries of $9\arcsec\times9\arcsec$ cutouts of quiescent and star-forming galaxies; either from CANDELS v1.0 mosaics \citep{candels,candelsb} or COSMOS ACS v2.0 mosaics \citep{cosmosacs,massey:10}. Quoted sizes are circularized such that $r_e=\sqrt{ab}$, where $a$ and $b$ are the semi-major and semi-minor axes of the half-light ellipse. Errors in size estimates are assumed to be $\sim10\%$ when not quoted. We exclude 15 galaxies with only ground-based sizes for this work. 

Utilizing the excellent spectral coverage of the photometric data in each field, we use FAST \citep{kriek:09} fit the spectral energy distribution (SED) of each galaxy to \citet[BC03]{bc:03} stellar population synthesis models, assuming solar metallicity, a \citet{chabrier:03} Initial Mass Function (IMF), and delayed exponential declining star-formation histories. We assume the best-fit parameters, such as stellar mass, age, and $A_v$, from these fits. We adopt a systematic uncertainty in measured $M_{\star}/L$ of $0.1\unit{dex}$, although we note that in some cases the systematic uncertainties in such measurements could be up to $\sim0.2\unit{dex}$ \citep[e.g.][]{muzzin:09}. Because these stellar mass estimates are based on aperture photometry, corrected by some average aperture-to-total magnitude correction, we rescale stellar masses to reflect the luminosity of the best-fit Sersic profile as:

\begin{equation}
M_{\star} = M/L_{\star,FAST}L_{\mathrm{Tot}}.
\label{eq:mass}
\end{equation}

\noindent In this equation, $M_{\star}$ is the corrected stellar mass, $M/L_{\star,FAST}$ is the stellar mass-to-light ratio estimate from FAST, $L_{\mathrm{Tot}}$ is the total luminosity of the best-fit S\'ersic profile.

Additionally, we use InterRest \citep{taylor:09} to interpolate between observed photometric measurements for each galaxies and calculate rest-frame magnitudes and colors in a number of filters.

\begin{figure*}[!t]
\centering
\subfigure{
\includegraphics[width=0.3\textwidth]{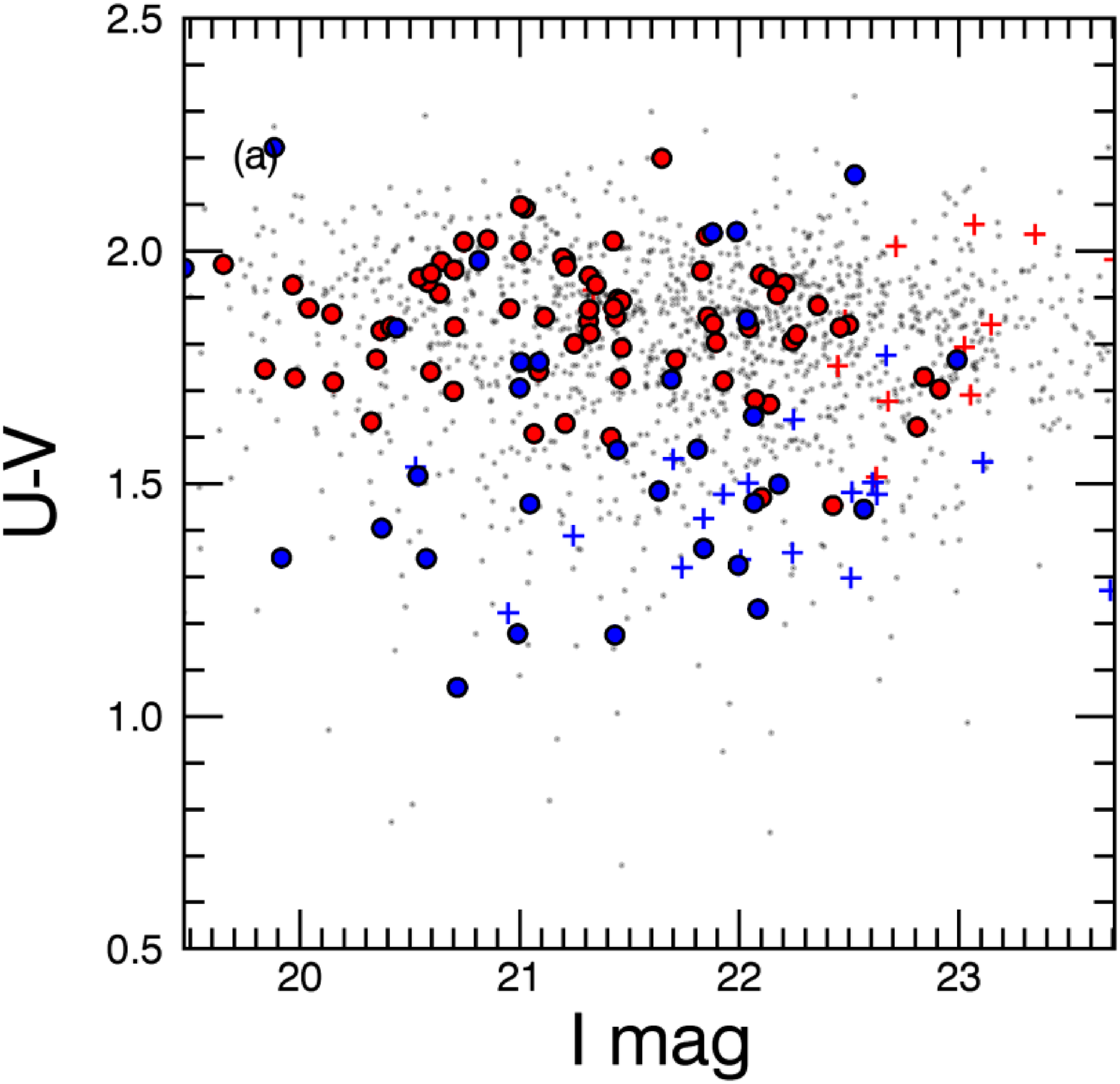}
\label{fig:imagUV}
}
\subfigure{
\includegraphics[width=0.3\textwidth]{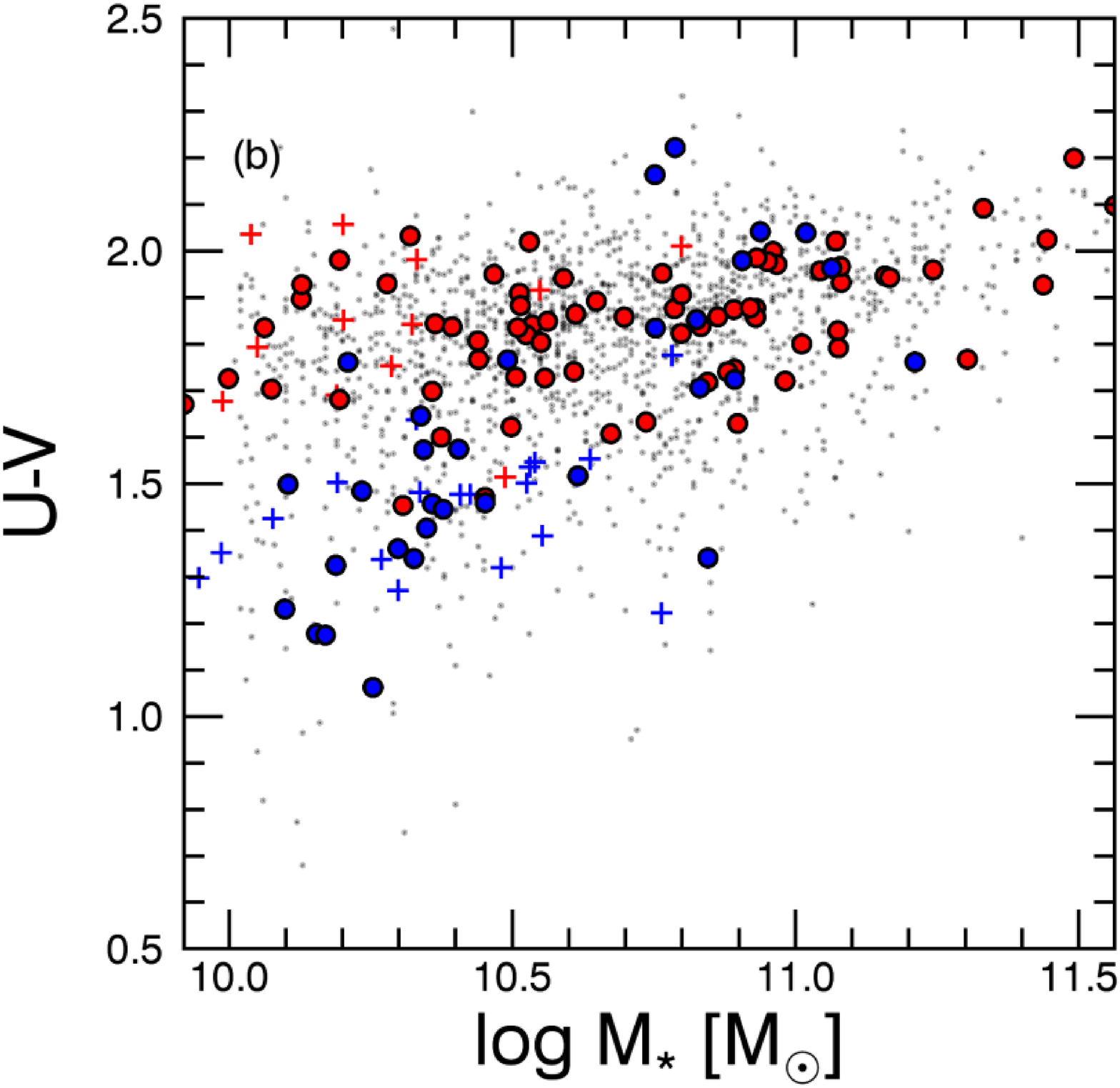}
\label{fig:lmassUV}
}
\subfigure{
\includegraphics[width=0.3\textwidth]{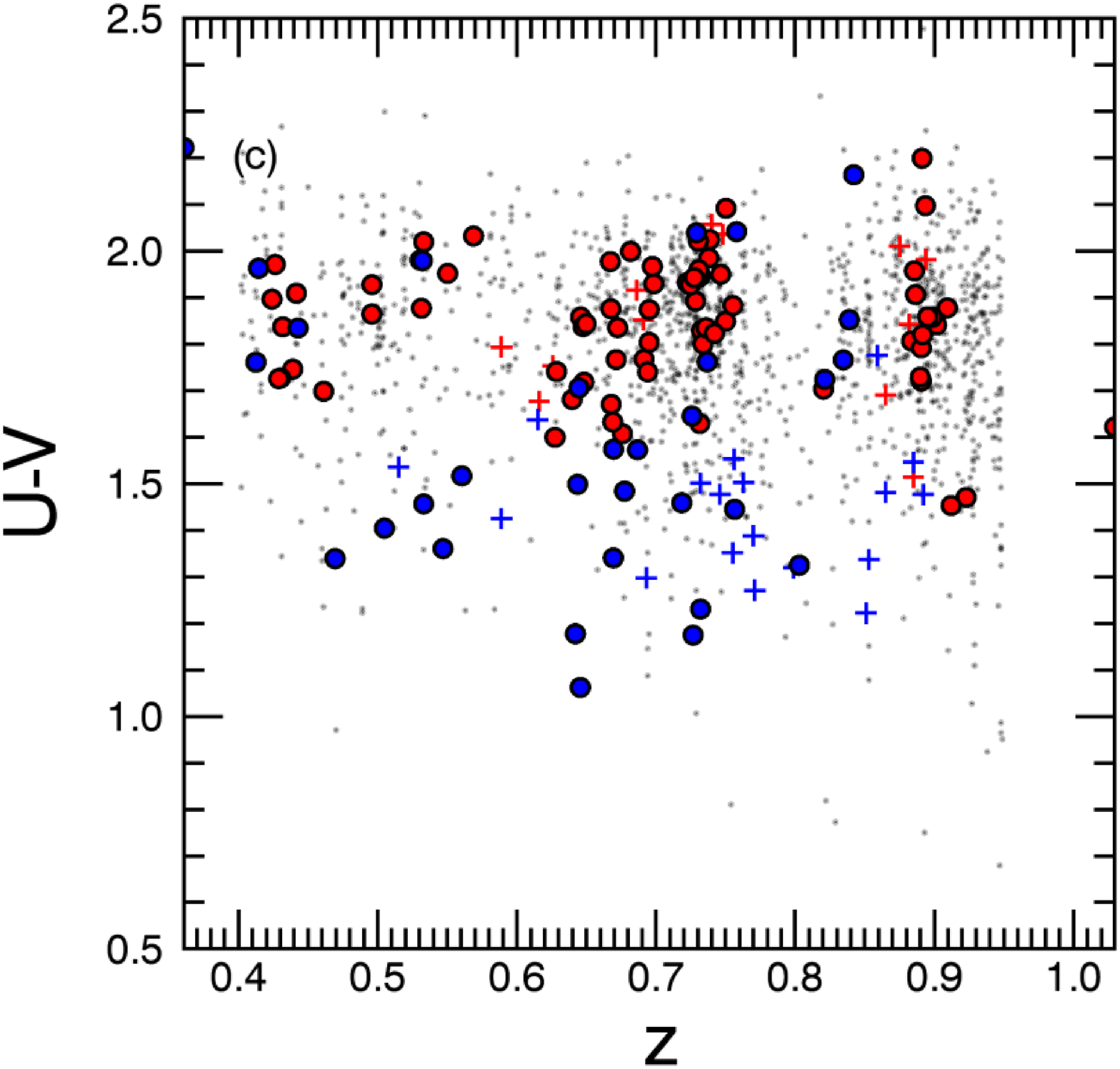}
\label{fig:zUV}
}
\vfill
\subfigure{
\includegraphics[width=0.3\textwidth]{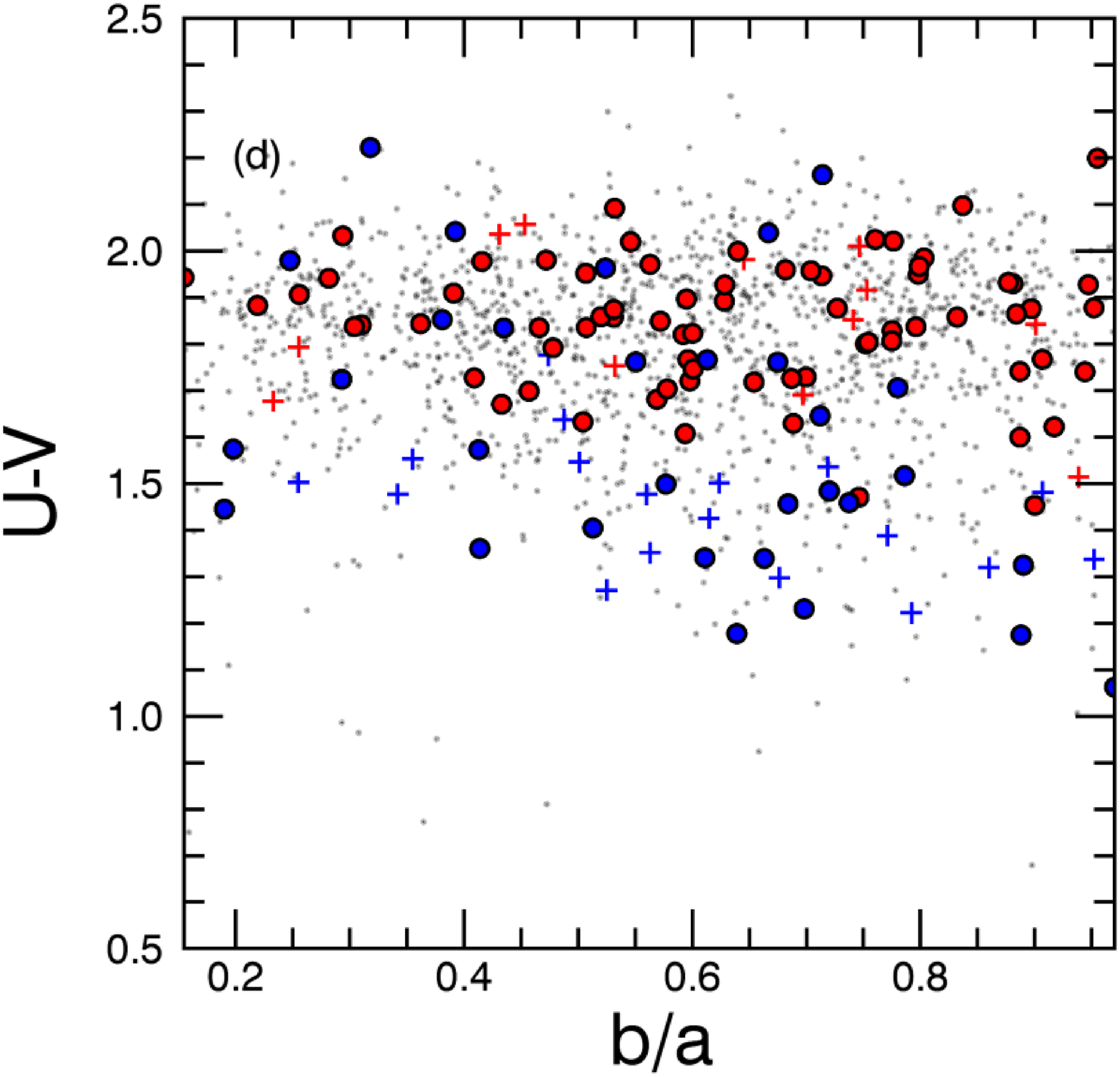}
\label{fig:baUV}
}
\subfigure{
\includegraphics[width=0.3\textwidth]{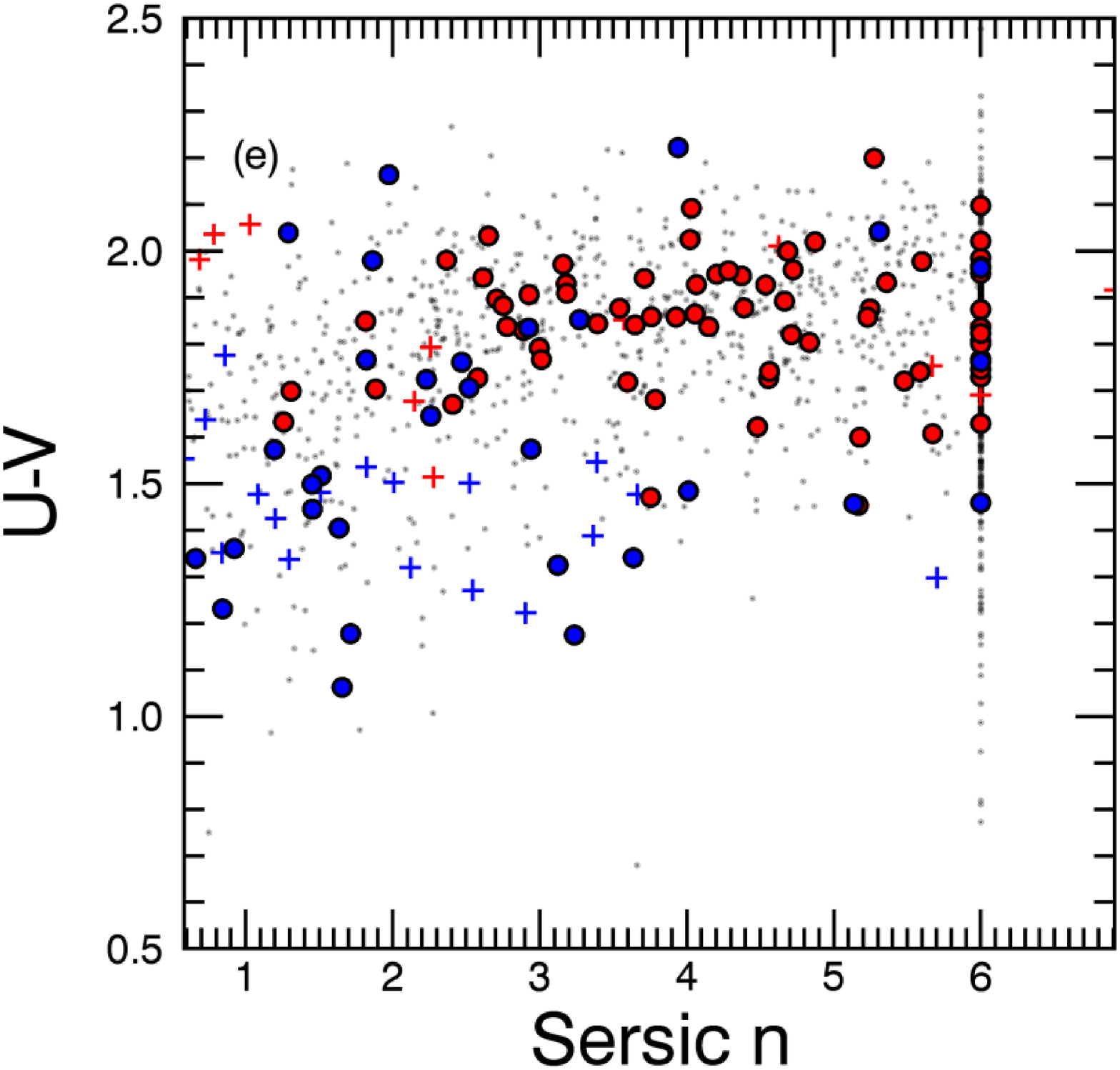}
\label{fig:nUV}
}
\subfigure{
\includegraphics[width=0.3\textwidth]{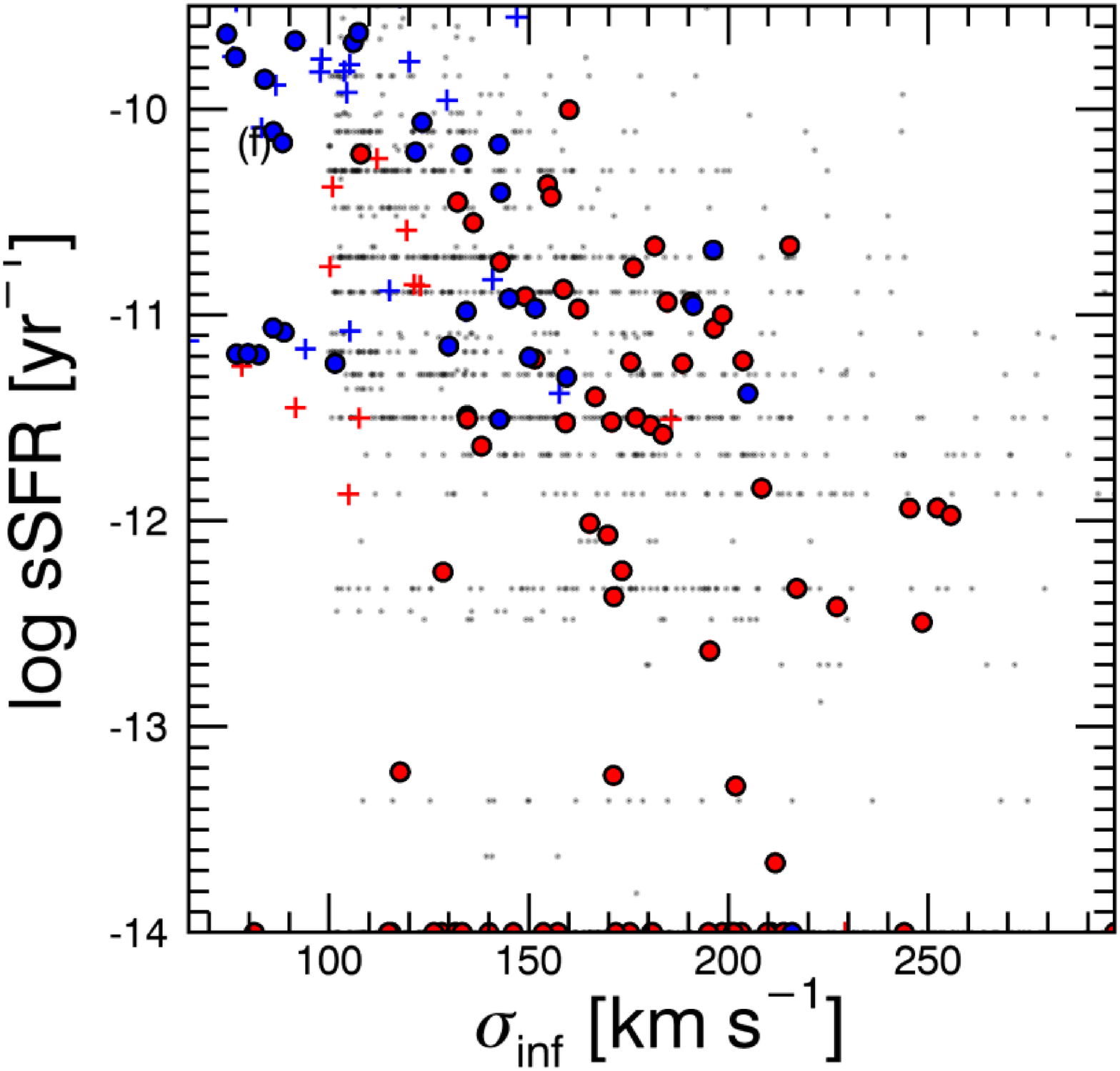}
\label{fig:siginflssfr}
}
\caption{Properties of the $z\sim0.7$ spectroscopic sample (quiescent galaxies in red, star-forming galaxies in blue) relative to $0.4<z_{phot}<0.95$ galaxies in the NMBS-Cosmos field. As in Figure\,\ref{fig:selection}, filled circles represent successful measurements of velocity dispersions (with $<15\%$ statistical error), crosses represent galaxies with spectra that have been excluded from this sample. Blue symbols identify star-forming galaxies, red symbols represent quiescent galaxies, separated using rest-trame U-V and V-J rest frame colors. The spectroscopic sample spans the range of the photometric parent sample in $U-V$ color (panels a-e), I magnitude (panel a), stellar mass (panel b), axis ratio (panel d), Sersic index (panel e), specific star-formation rate and inferred velocity dispersion (panel f). However, at the extremes of the sample (e.g. at high redshift or low mass), the spectroscopic sample is biased relative to a full mass-limited photometric sample.}
\label{fig:completeness}
\end{figure*}

\subsection{Velocity Dispersion Measurements}

Velocity dispersions of stellar absorption features were measured using PPXF \citep{cappellari:04} to fit broadened \citet{bc:03} (BC03) stellar population synthesis models to each spectrum. Possible emission lines (e.g. OII and OIII lines) were excluded from dispersion fitting and spectral regions surrounding the Balmer lines were masked. In order to limit the effect of template mis-match on measured velocity dispersions, we used best-fit synthetic spectrum to the photometry as determined by FAST \citep{kriek:09} and fit only rest-frame $\lambda>4000\unit{\AA}$ to avoid age dependent features such as strong Balmer lines \citep[see e.g.][]{sande:13}. All velocity dispersions were visually inspected and errors were estimated by re-fitting templates added to shuffled residuals from initial fits using PPXF.  Spectra and broadened BC03 templates are included in Figure\,\ref{fig:specs}, ordered by measured velocity dispersion. We compare these error estimates to statistical errors in velocity dispersion calculated by PPXF and verify that they are extremely consistent, with a scatter in relative error of only 0.01.

Measured velocity dispersions are corrected (by adding in quadrature) for the BC03 template resolution, $\sigma=85\unit{km\,s^{-1}}$. Velocity dispersions are then aperture corrected using Equation \ref{eq:apcor} from the 1'' slitwidth to an effective radius. Galaxies with $\geq15\%$ velocity dispersion errors are excluded from the final sample (30 galaxies, 12 quiescent and 18 star-forming), yielding a full sample of 103 galaxies spanning a range of colors and velocity dispersions (see Table \ref{tab:data}).

\subsection{Sample Completeness: Measurement Success Rates for Absorption Line Kinematics}
\label{subsec:completeness}

\begin{figure*}[!hp]
\centering
\includegraphics[height=0.9\textheight]{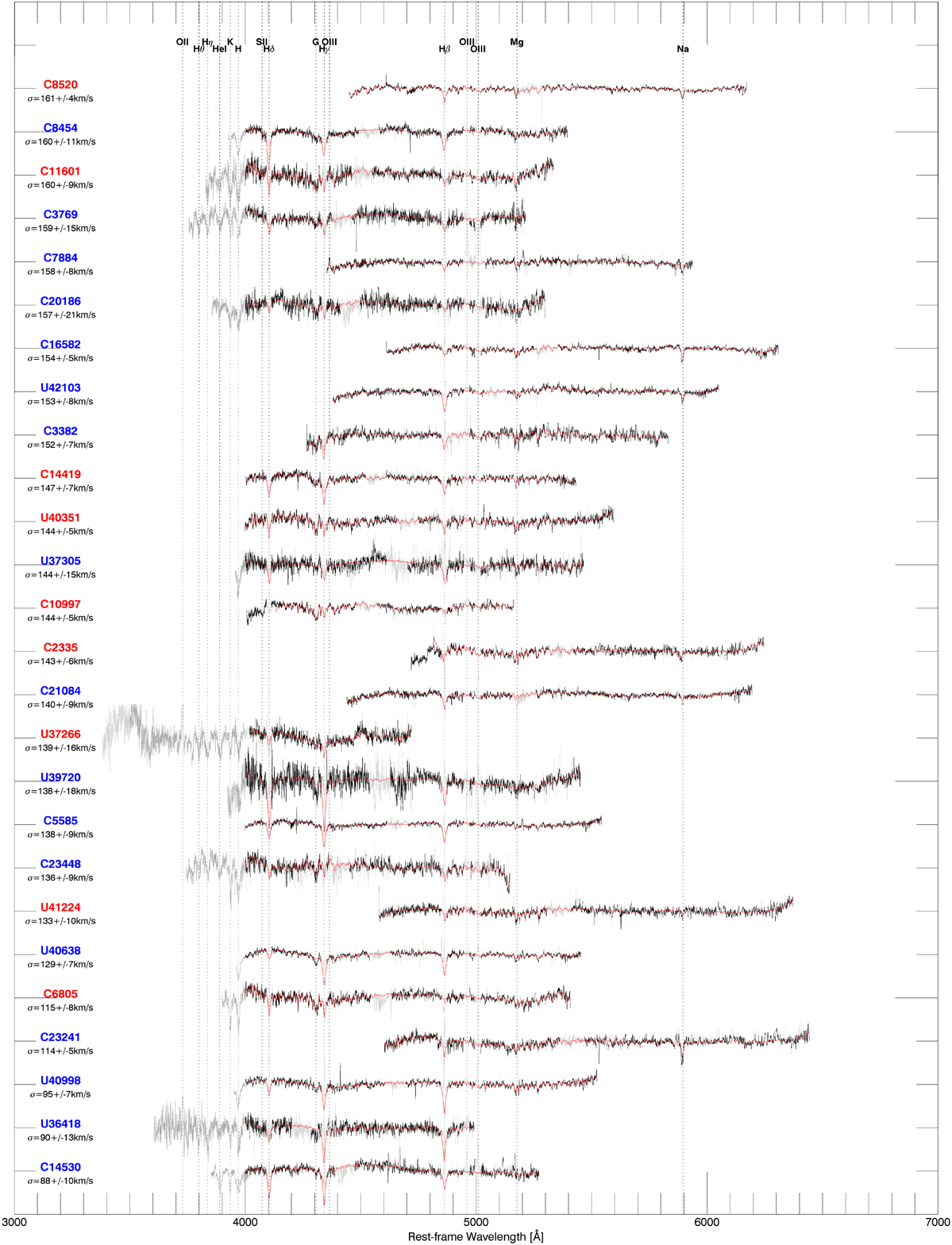}
\caption{Continuum-normalized galaxy spectra, ordered by velocity dispersion and shifted to rest-frame wavelengths. Full spectra are included in gray, regions of the spectra that are included in the dynamical measurements are included in black ($\lambda>4000\unit{\AA}$), and broadened best-fit BC03 templates are red. Key spectral features are labeled at the top of each panel and indicated by dashed vertical lines. Galaxy IDs are labeled left of each spectrum with the color of the label indicating whether a galaxy is star-forming (blue) or quiescent (red).}
\label{fig:specs}
\end{figure*}

\begin{figure*}[!hp]
\centering
\addtocounter{figure}{-1}
\includegraphics[height=0.9\textheight]{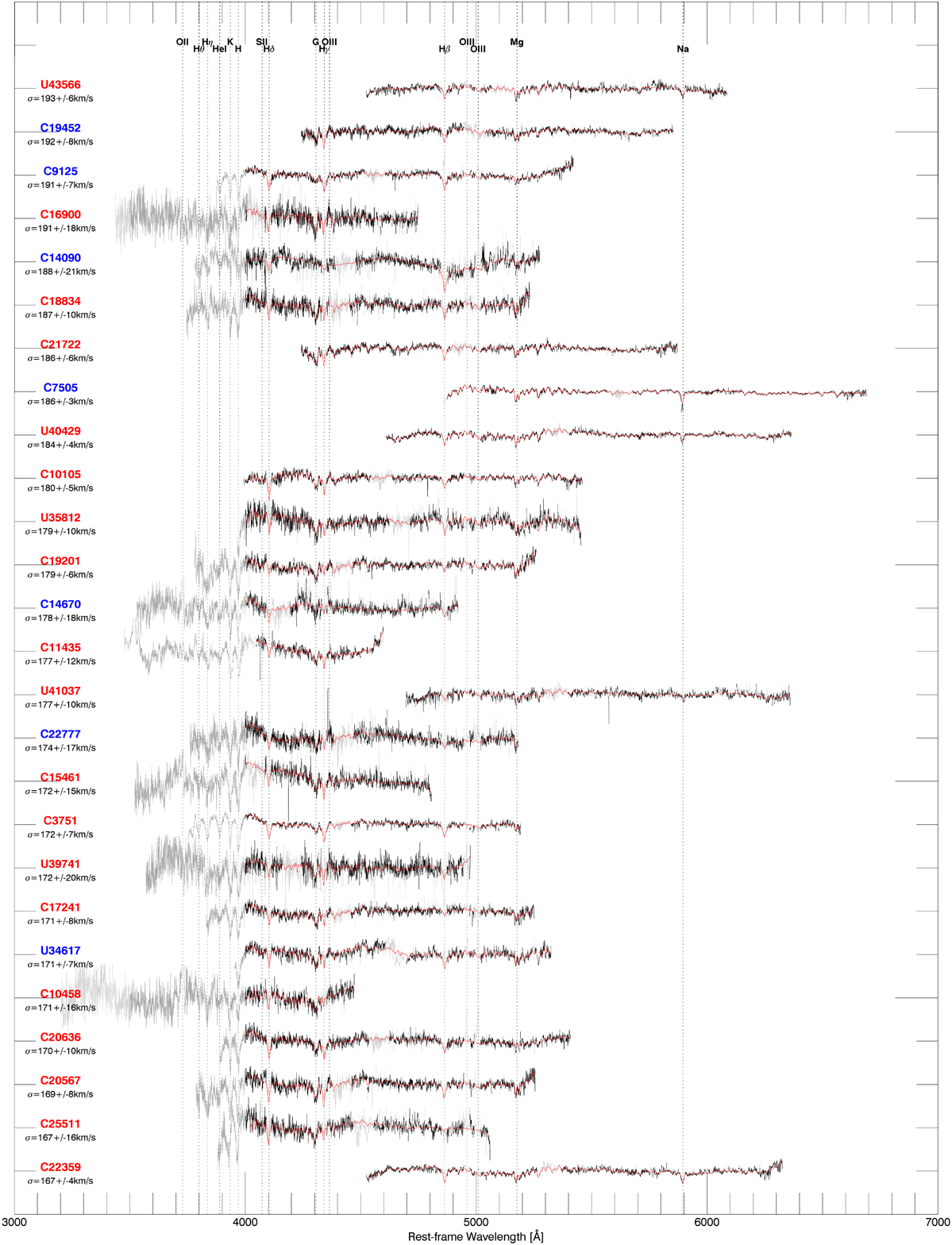}
\caption{DEIMOS spectra of $z\sim0.7$ galaxies, ordered by velocity dispersion -- Continued}
\end{figure*}

\begin{figure*}[!hp]
\centering
\addtocounter{figure}{-1}
\includegraphics[height=0.9\textheight]{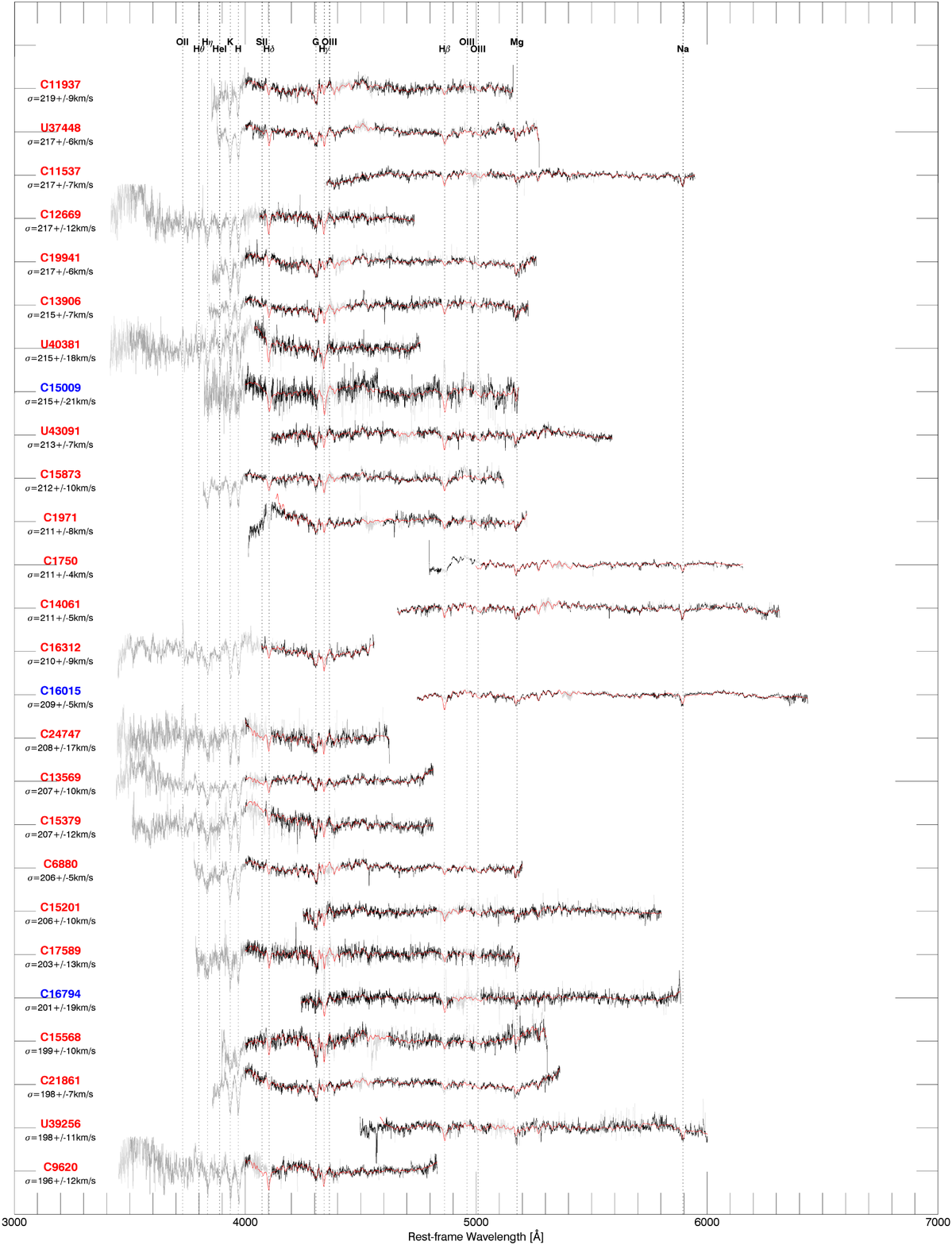}
\caption{DEIMOS spectra of $z\sim0.7$ galaxies, ordered by velocity dispersion -- Continued}
\end{figure*}

\begin{figure*}[!hp]
\centering
\addtocounter{figure}{-1}
\includegraphics[height=0.9\textheight]{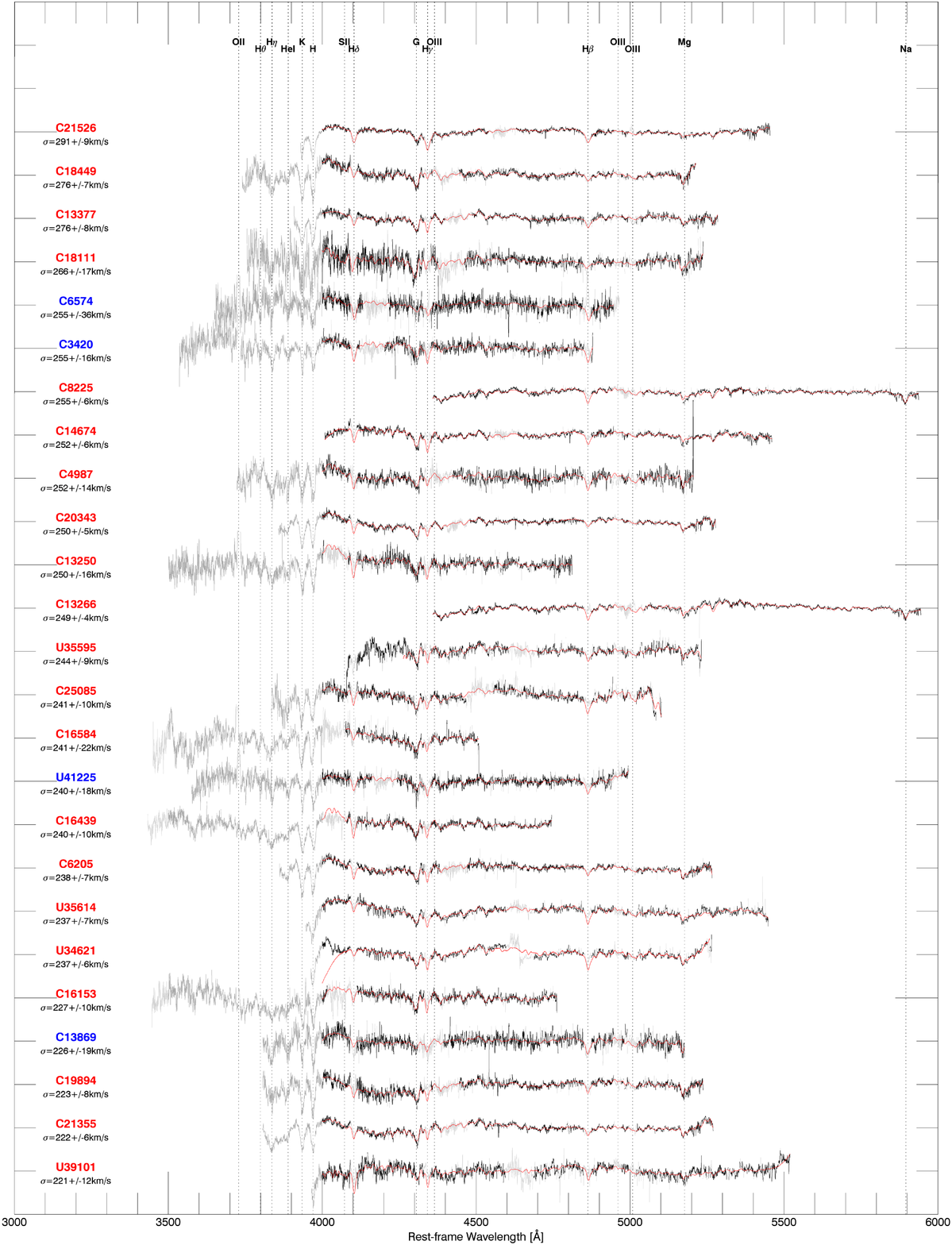}
\caption{DEIMOS spectra of $z\sim0.7$ galaxies, ordered by velocity dispersion -- Continued}
\end{figure*}

This sample of galaxies was selected to span the population of massive ($\sigma\gtrsim100\unit{km\,s^{-1}}$) galaxies at $z\sim0.7$ and therefore the range in stellar populations (or colors), masses, morphologies, and dynamics. Derived properties of the final spectroscopic sample are included in Table \ref{tab:data}. In Figure\,\ref{fig:completeness} we assess the range of the final spectroscopic sample relative to galaxies in the same (photometric) redshift range from the NMBS-Cosmos field (parent population). The parent population is indicated by small gray points; spectroscopic sample is indicated by colored symbols. Our goal is to represent both star-forming and quiescent galaxy populations in this analysis. We distinguish between the two samples using rest-frame $U-V$ and $V-J$ colors \citep[e.g.][]{franx:08,williams:09}, adopting the color-cuts from \citet{whitaker:12a} (Figure\,\ref{fig:selection}). This method has been shown to discriminate between ``red and dead'' galaxies and galaxies which are red and dusty star-formers. In this and subsequent Figures, star-forming galaxies are indicated by blue symbols, quiescent galaxies by red symbols.

For some spectra, we were not able to successfully measure velocity dispersions (indicated by crosses). This is driven partially by S/N of the spectra: these failures appear to be related to a number of correlated galaxy properties. Therefore, the resulting sample of galaxies does not uniformly sample colors, morphologies, and dynamics as described in \S \ref{sect:deimos}. In general star-forming galaxies are more likely to ``fail" in this context. 

The primary factor in failing to measure a velocity dispersion is the observed brightness of the galaxy: faint ($I\gtrsim22$) galaxies are excluded either because they are intrinsically faint or are at the highest redshift end of this survey (Figures \ref{fig:completeness}(a)-(c)). Furthermore, low mass galaxies, as defined based on stellar mass ($\log\,M_{\star}\lesssim10.5$) or inferred dynamics ($\sigma_{inf}\lesssim100-150$\unit{km\,s^{-1}}) are less likely to yield precise velocity dispersions, particularly those included as mask fillers. 75\% of failed measurements of quiescent galaxies, and $\sim40\%$ for star-forming galaxies, have $\sigma_{inf}<100\unit{km\,s^{-1}}$, therefore the sample is much more complete above this limit. Blue star-forming galaxies are also more likely to fail, although this is partially related to the mass bias. Additionally, this sample excludes many galaxies with the highest specific star formation rates (sSFR) (see Figure\,\ref{fig:completeness}(f)). This can be understood as a combination of the aforementioned mass bias against low velocity dispersions and the fact that the youngest stellar population synthesis models have quite weak spectral absorption features (aside from the Balmer lines, which are masked in the dynamical fitting).

Although the existing sample spans a large range in morphologies and inclinations (see images in Figure\,\ref{fig:images}), there also appears to be a bias against face-on disks as many galaxies with rounder shapes ($b/a\gtrsim0.8$ in Figure\,\ref{fig:completeness}(d)) and more disklike profiles (S\'ersic $n\lesssim2.5$ in Figure\,\ref{fig:completeness}(e)) preferentially fail to produce successful dispersion measurements. For galaxies with rotation, such as inclined pure disks, the measured velocity dispersion is a combination of the rotational velocity and intrinsic velocity dispersion \citep[e.g.][]{binney:78}:

\begin{equation}
\sigma_{meas} = \sqrt{\sigma^2 +0.5(V_{rot}\sin\,i)^2}
\label{eq:sigmarot}
\end{equation}

For a face-on inclination ($i=0$), the rotational velocity does not contribute at all to the measured velocity dispersion. For inclined galaxies, measured dispersions will be boosted by the rotational velocities (by broadening the spectral features), which in turn makes them more likely to be included in the spectroscopic sample. The velocity dispersion of an edge-on disk galaxy provides an estimate of the overall dynamics of the galaxy, but just the intrinsic dispersion for a face-on galaxy. Inclination, as probed by projected axis ratios, appears to have a minimal effect on the samples of galaxies presented in this Paper, implying that intrinsic dispersions are high and possibly pointing to the prevalence of bulges in these massive galaxies (see Appendix \ref{sec:inclination}).

Overall this sample is representative for massive galaxies at $z\sim0.7$, but the results of the study will not be as comprehensive for lower mass galaxies or galaxies with high star-formation rates. The former is driven primarily by $S/N$: deeper data would provide higher completeness to lower mass limits. The latter is a limitation to studying absorption line kinematics for very young systems.

\section{The Discrepant Structural and Dynamical Properties of Star-Forming and Quiescent Galaxies}
\label{sec:2d}

\begin{figure*}[!t]
\centering
\subfiguretopcaptrue
\subfigure[][SDSS at $z\sim0$]{
\includegraphics[width=0.45\textwidth]{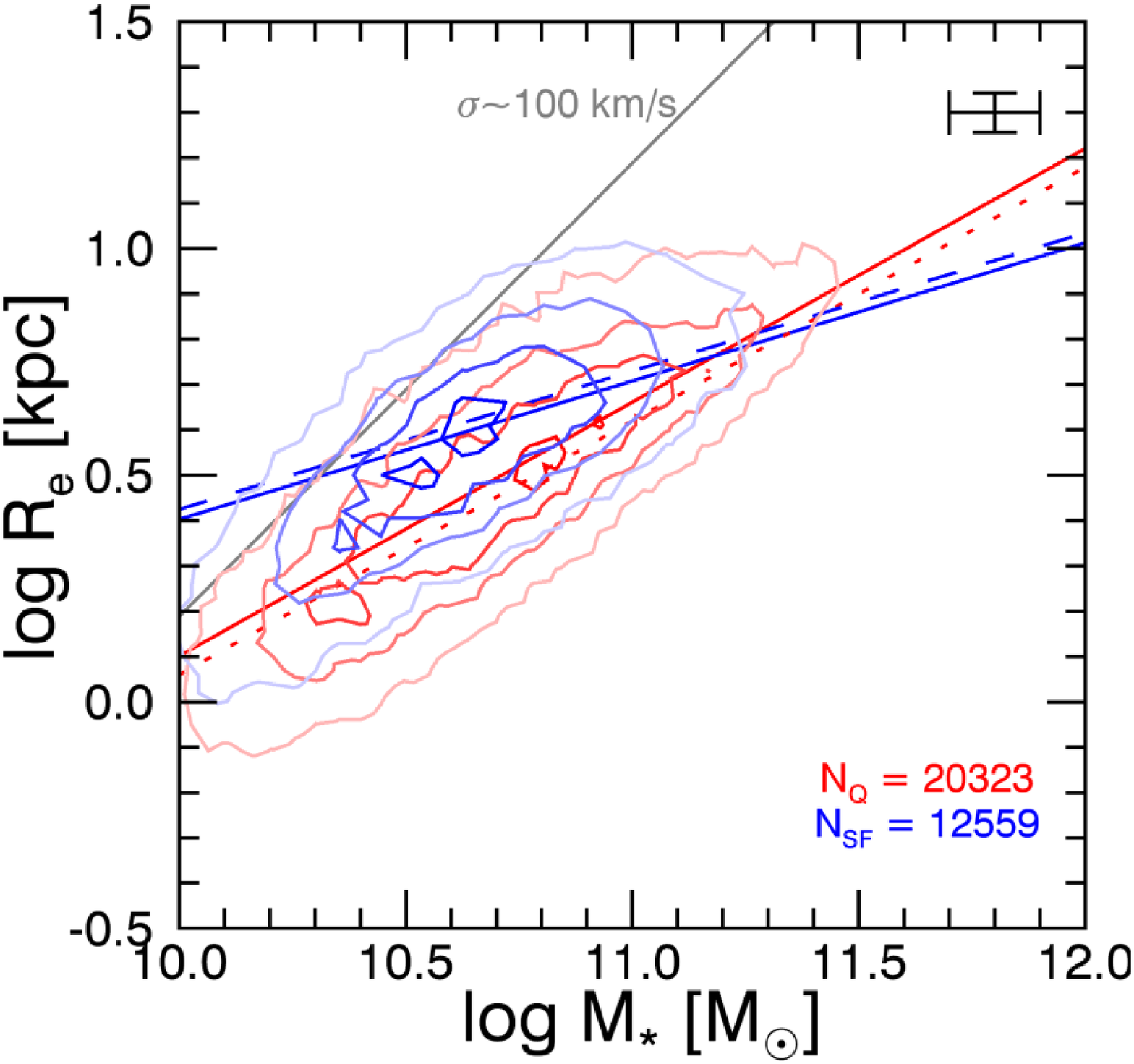}
\label{fig:mass_size_sdss}
}
\subfigure[][DEIMOS at $z\sim0.7$]{
		\includegraphics[width=0.45\textwidth]{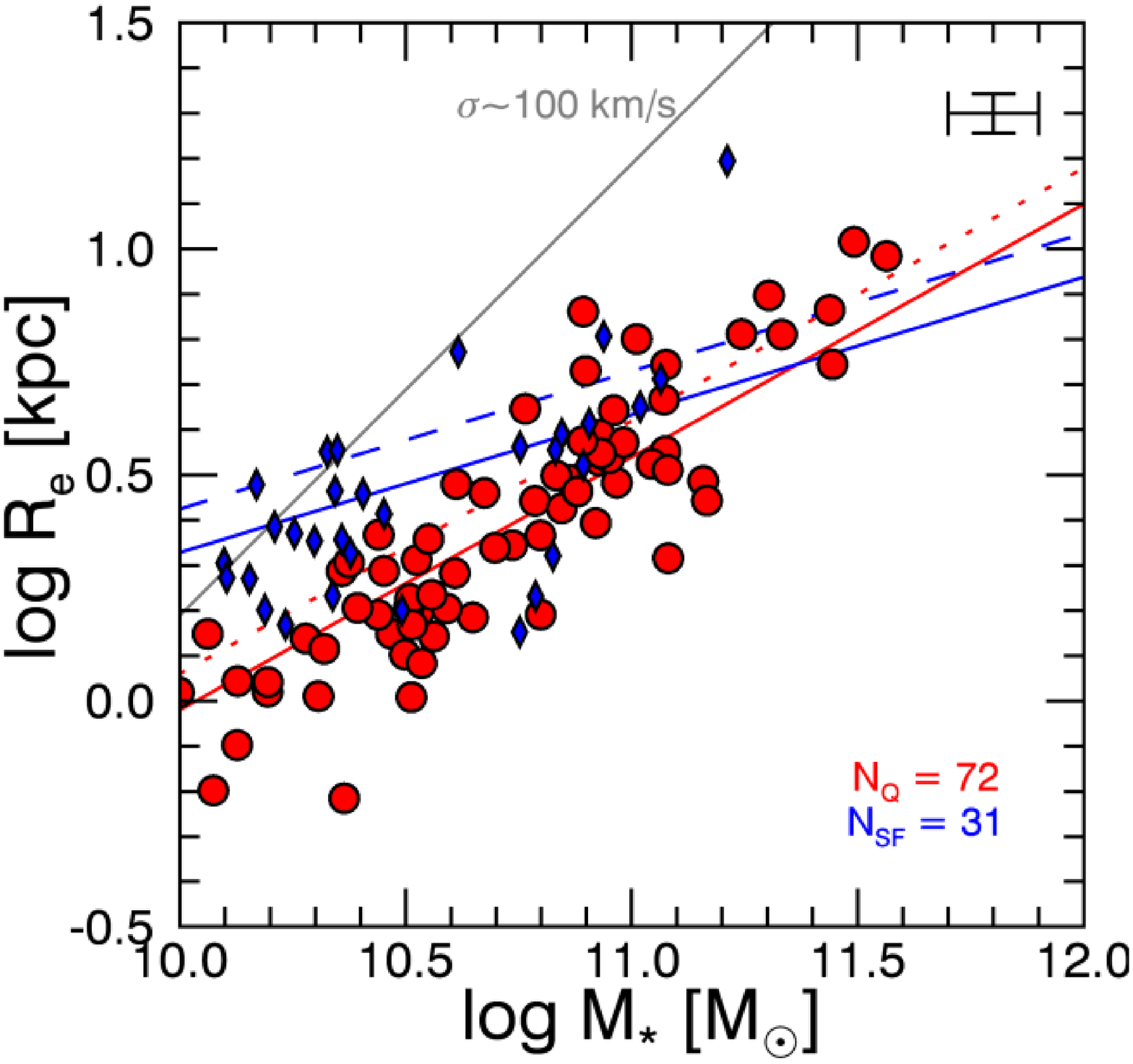}
		\label{fig:mass_size}
}
\caption[]{
Circularized effective radius versus stellar mass for all galaxies in each redshift bin. At fixed mass, star-forming galaxies have larger sizes than quiescent galaxies, both at $z\sim0$ (panel \subref{fig:mass_size_sdss} SDSS sample) and $z\sim0.7$ (panel \subref{fig:mass_size} Deimos sample). Total number of quiescent and star-forming galaxies is indicated in the lower right-hand corner of each panel. Red and blue contours indicate quiescent and star-forming galaxies, respectively, based on K-corrected colors ($g^{0.1}-r^{0.1}$ and $M^{0.1}_r$) in panel \subref{fig:mass_size_sdss}. Red circles and blue diamonds indicate quiescent and star-forming galaxies, respectively, based on rest-frame U-V and V-J colors in panel \subref{fig:mass_size}. The local \citep{shen:03} size-mass relations are indicated for late-type (blue dashed) and early-type
(red dotted) galaxies in each panel.}
\label{fig:mass_size_all}
\end{figure*}

\begin{figure*}[!t]
\centering
\subfiguretopcaptrue
\subfigure[][SDSS at $z\sim0$]{
	\includegraphics[width=0.45\linewidth]{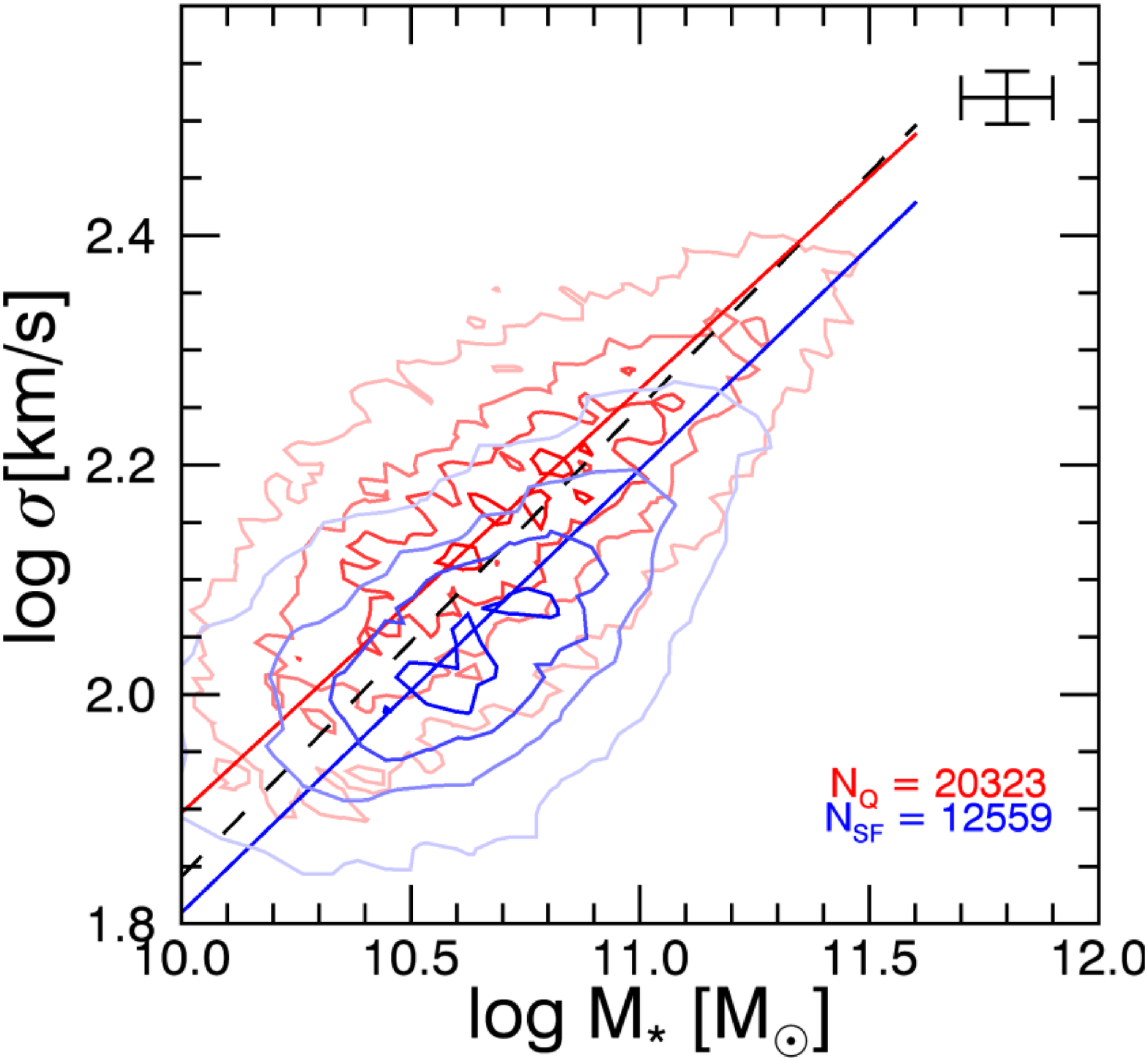}
		\label{fig:mass_sigma_sdss}
}
\subfigure[][DEIMOS at $z\sim0.7$]{
		\includegraphics[width=0.45\linewidth]{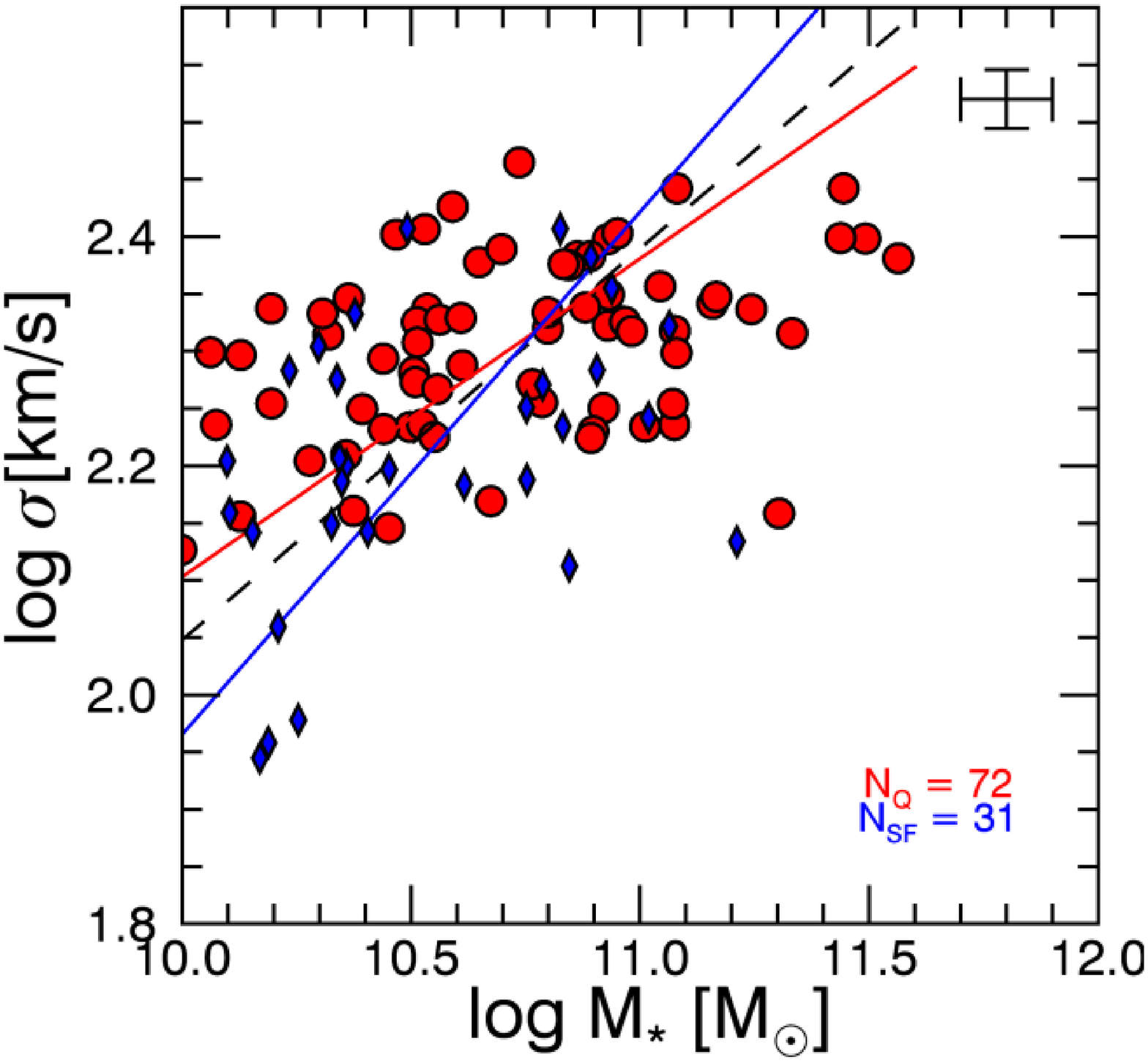}
		\label{fig:mass_sigma}
}
	
\caption[]{
Velocity dispersion versus stellar mass (``mass Faber-Jackson'' relation) for all galaxies at $z\sim0$ (panel \subref{fig:mass_sigma_sdss}) and $z\sim0.7$ (panel \subref{fig:mass_sigma}) (notations and color-coding as in Figure\,\ref{fig:mass_size_all}). Linear fits are included as solid lines for the overall population (black) and separately for star-forming (blue) and quiescent (red) galaxies. At fixed mass, star-forming galaxies have lower velocity dispersions, at each redshift, although the slope of the relation is slightly steeper for star-forming galaxies.
}
\label{fig:mass_sigma_all}
\end{figure*}

\begin{figure*}[!h]
\centering
\includegraphics[width=0.95\textwidth]{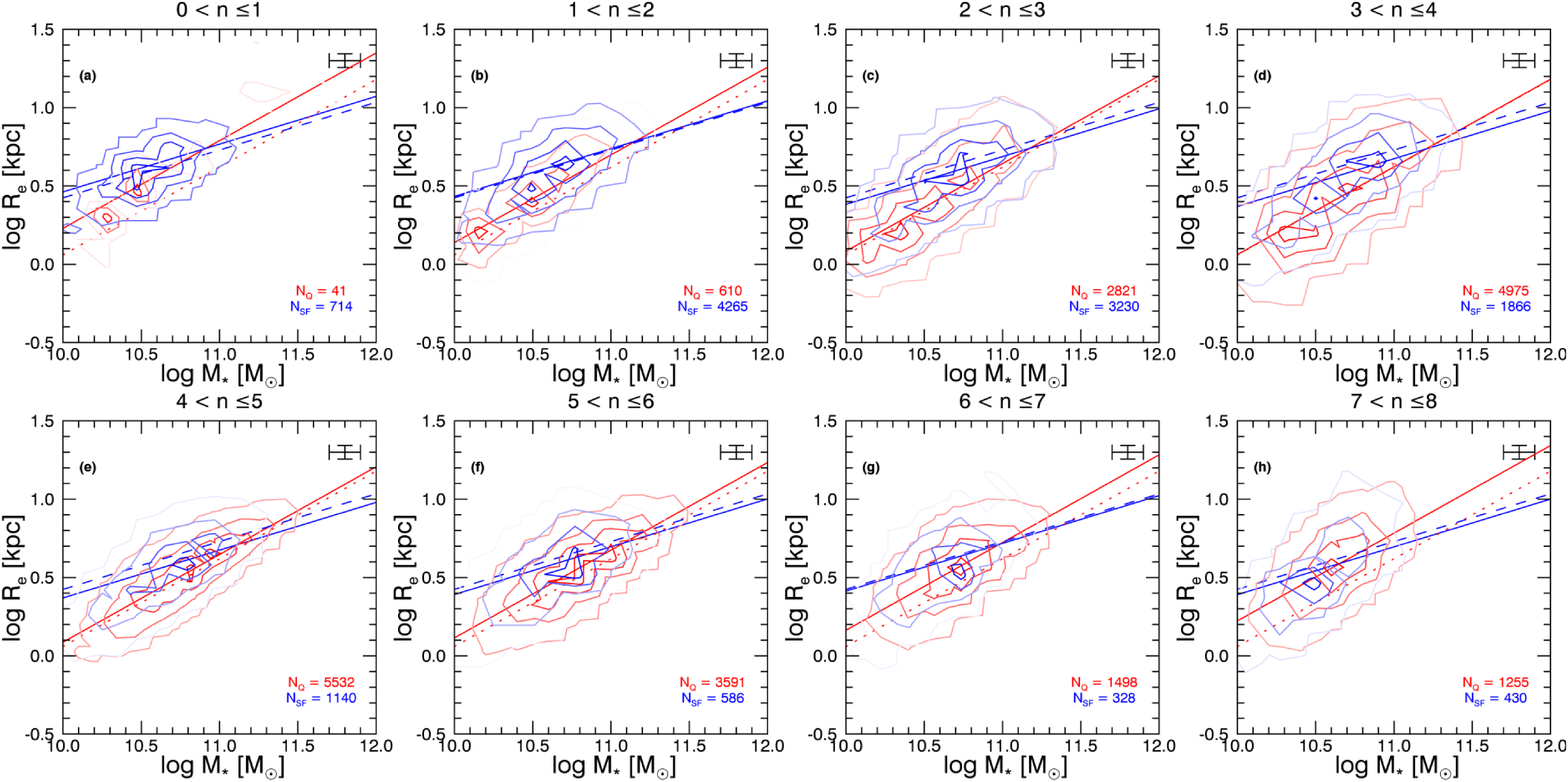}
\caption{Circularized effective radius versus stellar mass for galaxies in the SDSS in bins of best-fit S\'ersic index (n). As in Figure~\ref{fig:mass_size_sdss}, red and blue points indicate quiescent and star-forming galaxies, respectively. The total number of quiescent and star-forming galaxies is indicated in the lower right-hand corner of each panel ($\mathrm{N_{SF}}$ and $\mathrm{N_Q}$). The $z\sim0$ \citep{shen:03} size-mass relations are indicated for late-type (blue dashed line) and early-type (red dotted line) galaxies in each panel. This figure demonstrates that the distinction between star-forming and quiescent galaxies is not merely a cut in profile shape: at fixed mass and fixed Sersic index, star-forming galaxies have larger sizes than quiescent galaxies. They also differ morphologically: galaxies with low S\'ersic indexes tend to be star-forming, those with high S\'ersic indexes tend to be quenched, although this is not an exact delineation.}
\label{fig:mass_size_n_sdss}
\centering
\includegraphics[width=0.8\textwidth]{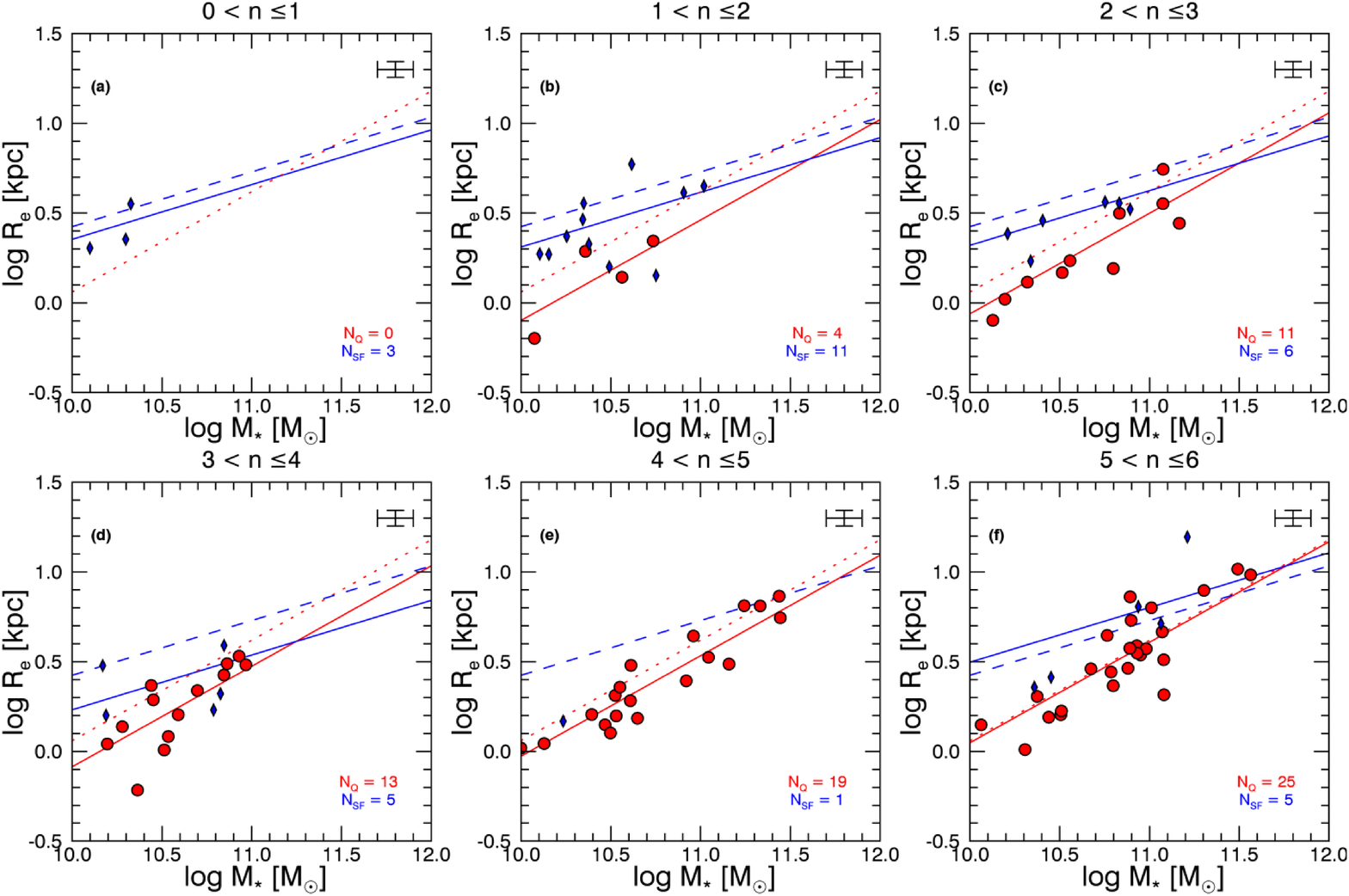}
\caption{Circularized effective radius versus stellar mass for galaxies in the Deimos sample at $z\sim0.7$ in bins of Sersic index (color-coding as in Figure\,\ref{fig:mass_size}). The $z\sim0$ \citep{shen:03} size-mass relations are indicated for late-type (blue dashed line) and early-type (red dotted line) galaxies in each panel. As at lower redshift, star-forming galaxies are generally larger and have lower S\'ersic indices than quiescent galaxies at fixed mass, however the mass-size relations for each population exhibits a large scatter (and therefore this larger samples are required to quantify the relations). }
\label{fig:mass_size_n}
\end{figure*}

In this Section we focus on two traditional scaling relations (size-mass and mass-velocity dispersion, or the mass Faber-Jackson, relation) to highlight the differences between star-forming and quiescent galaxies, both at $z\sim0$ and $z\sim0.7$. These relations are non edge-on projections of the mass fundamental plane, which is the focus of subsequent sections. Although the populations are distinguished based on their rest-frame colors, as a proxy for differences in stellar populations, the separation extends to the overall structural and dynamical properties of each sample of galaxies in this paper.

In Figure\,\ref{fig:mass_size_all} we explore the size-mass relations of the star-forming and quiescent galaxies in the SDSS and at $z\sim0.7$. In Figure\,\ref{fig:mass_size_sdss}, as in the subsequent $z\sim0$ figures in this Paper, the distribution of galaxies in the SDSS is demarcated by a series of contours. Red contours indicate the density of quiescent galaxies, blue contours reflect the density of star-forming galaxies. The relative saturation of the colors are normalized to reflect the density in a given figure. The number of galaxies (quiescent and star-forming) is indicated by $\mathrm{N_Q}$ and $\mathrm{N_{SF}}$ in the bottom right corner of each panel. In Figure\,\ref{fig:mass_size}, and other $z\sim0.7$ figures, points represent individual galaxies. Star-forming galaxies are indicated by blue diamonds, quiescent galaxies by red circles. Again, the number of galaxies in each sample is indicated in the lower right corner of each panel. Finally, average error bars are indicated in the upper right corner of every panel.

The first result is a confirmation that star-forming galaxies are larger, on average, than quiescent galaxies at fixed mass \citep[e.g.][in the SDSS]{shen:03} and \citep[e.g.][at higher redshift]{williams:09,wel:14}. The discrepant normalizations for the two populations holds in both redshift ranges probed by this study. We note that the requirement of reliable velocity dispersion measurements biases this sample against galaxies with low masses and/or large sizes. Therefore, a linear fit to these size-mass relations would be steeper than for purely photometric samples of galaxies. We indicate the approximate size-mass relation of $\sigma\sim100\unit{km\,s^{-1}}$ galaxies (gray solid line), above this line the samples likely suffer from incompleteness. Additionally we include the \citet{shen:03} size-mass relations for early-type (dotted red line) and late-type (dashed blue line) galaxies. By adopting the slopes of these relations, we fit the normalizations to each separate galaxy population (solid red and blue lines). We note that the overall normalizations for this sample differ from the \citet{shen:03} fits, likely due to the differences in \citep{simard:11} size measurements adopted in this work relative to the NYU-VAGC measurements \citep{blanton:05} used in that study. The latter measurements have been shown to underestimate galaxy sizes, by a factor that increases with S\'ersic index \citep[e.g.][]{guo:09}. However, we include these fits mostly for relative comparisons and emphasize the importance of more complete photometric samples to properly measure the size-mass relation.

Figure\,\ref{fig:mass_size} shows the same relation for massive galaxies at $z\sim0.7$. Again, star-forming galaxies lie above quenched galaxies and both samples lie below the \citet{shen:03} relations; in this case this reflects the size evolution of galaxies. A comparison of Figure\,\ref{fig:mass_size_sdss} to Figure\,\ref{fig:mass_size} shows the same thing: both star-forming and quiescent galaxies populations exhibit lower normalizations at $z\sim0.7$ than for galaxies in the SDSS \citep[see e.g.][for more thorough and unbiased study of this evolution to $z\sim3$]{wel:14}. 

In Figures \ref{fig:mass_size_n_sdss} and \ref{fig:mass_size_n}, we present the same size-mass relations split into bins of best-fit S\'ersic index to emphasize the comparison at fixed profile shape.  Panel to panel differences suggest that the sizes of galaxies depend on S\'ersic index in addition to stellar populations and stellar mass. This is particularly obvious for quenched galaxies, however the trends in normalization do not vary linearly with S\'ersic index.  Instead, for galaxies in the SDSS, sizes decrease with S\'ersic index for $n<4$ and then increase with larger S\'ersic indices. A similar trend exists for high $n$ quiescent galaxies in the $z\sim0.7$ sample, but the sample size for star-forming and low-$n$ galaxies too small to assess similar trends in size with S\'ersic index.

Additionally, Figure\,\ref{fig:mass_size_n_sdss} demonstrates that the size difference between star-forming and quenched galaxies exists for the entire population and at fixed S\'ersic index. This is particularly noteworthy and suggests, for example, that a star-forming galaxy whose profile looks like an elliptical galaxy ($n\sim4$) will be larger than a quiescent elliptical galaxy with the same stellar mass. Although such galaxies are less common (star-forming galaxies generally have lower S\'ersic indices), this suggests that the difference between star-forming and quiescent galaxies is less clear than a simple separation between disks and ellipticals. 

Figure\,\ref{fig:mass_size_n} demonstrates that star-forming galaxies in the $z\sim0.7$ sample are also larger and have lower S\'ersic indices than their quiescent counterparts. However, the sample is quite small, so this trend at fixed S\'ersic index is less pronounced, particularly at high values of $n$. The quiescent galaxies also suggest a similar trend of larger size with S\'ersic index, but a larger, less biased sample would be better suited for measuring this effect.  Finally, we emphasize that for any sample of galaxies, mass-size relations have a significant amount of scatter; individual star-forming and quiescent galaxies with exactly the same masses, sizes, and S\'ersic indices are likely to exist. We caution against drawing conclusions from comparisons between small samples of galaxies.

\begin{figure*}[!ht]
\centering
\includegraphics[width=0.95\textwidth]{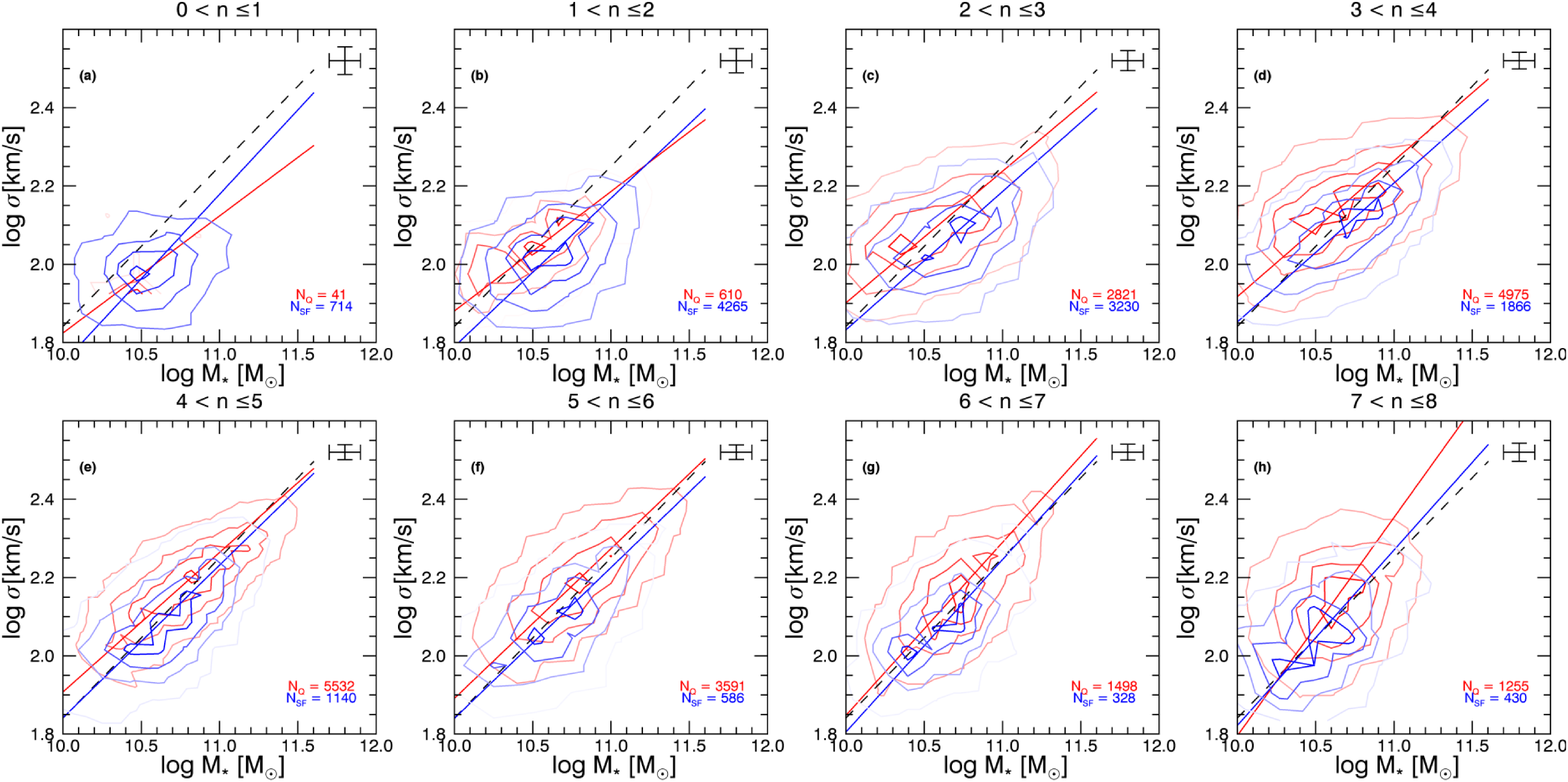}
\caption{Stellar mass Faber-Jackson relation (Velocity dispersion versus stellar mass) for galaxies in the SDSS, split by S\'ersic index (labeled as in Figures \ref{fig:mass_size_sdss} and \ref{fig:mass_sigma_n_sdss}). Overall fit (from Figure \ref{fig:mass_sigma_sdss}) is included as dashed black line, fits to star-forming (blue) and quiescent (red) galaxies at fixed S\'ersic index are shown as solid lines. At fixed mass and S\'ersic index, star-forming and quiescent galaxies lie on different scaling relations, however these relations also vary with $n$. Generally, the normalization of the mass Faber-Jackson relation is lower for star-forming galaxies than for quenched galaxies and the slope of these relations increase with S\'ersic $n$.}
\label{fig:mass_sigma_n_sdss}

\centering
\includegraphics[width=0.8\textwidth]{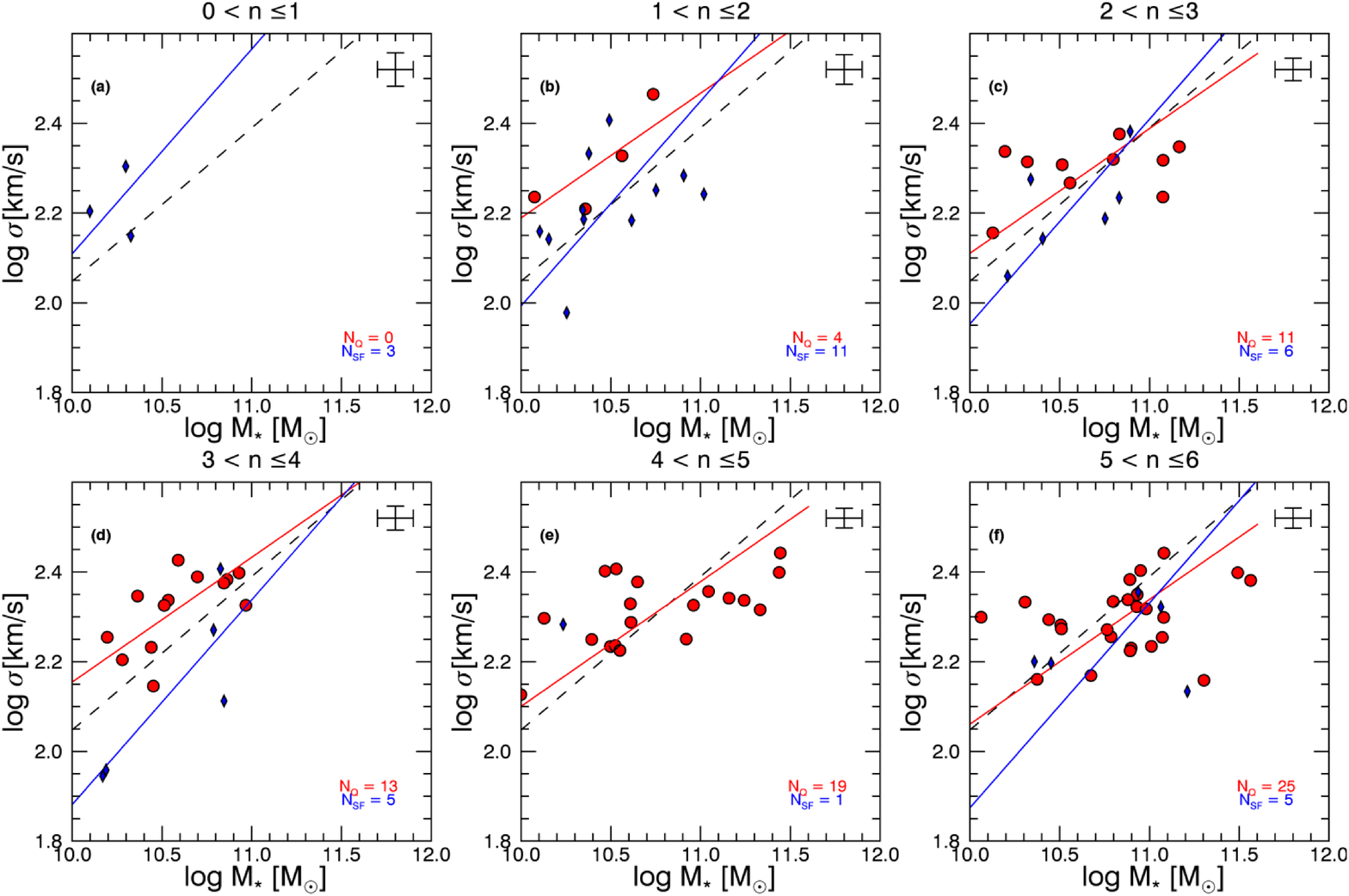}
\caption{Stellar mass Faber-Jackson relation for galaxies in the Deimos sample at $z\sim0.7$ (see also Figure\,\ref{fig:mass_sigma_n_sdss} at $z\sim0$). In each panel the overall mass Faber-Jackson relation at this redshift is included (black dashed line) in addition to fits to the star-forming and quiescent galaxy samples, assuming the slopes from Figure\,\ref{fig:mass_sigma}. Star-forming galaxies generally have lower velocity dispersions than their quiescent counterparts. }
\label{fig:mass_sigma_n}
\end{figure*}

Similarly, we compare the mass-velocity dispersion relations (``stellar mass'' Faber-Jackson relation, or the \cite{faberjackson} relation with luminosity multiplied by $M_{\star}/L$) for star-forming and quiescent galaxies in the full samples (Figure\,\ref{fig:mass_sigma}) and at fixed S\'ersic index (Figures\,\ref{fig:mass_sigma_n_sdss} and \ref{fig:mass_sigma_n}). This projection of the mass fundamental plane has been studied extensively, particularly for early-type galaxies; it has relatively low scatter but is not an edge-on projection of the mass fundamental plane. As discussed in \S \ref{subsec:completeness}, we use aperture-corrected velocity dispersions of star-forming galaxies in this analysis, which is a combination of intrinsic velocity dispersion, rotational velocity, and inclination (see Equation \ref{eq:sigmarot}). The contribution of rotational velocity to this quantity is likely to be more important for star-forming galaxies.

Best-fit slope and normalization of the mass Faber-Jackson relations (Equation \ref{eq:sigma_fj}) for all galaxies and separately for star-forming and quiescent galaxies are reported in Table \ref{tab:massfj}. These fits are calculated by a orthogonal least squares bisector fit, which has been shown to best retrieve the underlying functional relation from astronomical data \citep{isobe:90} and was performed using the IDL task \emph{SIXLIN}, with errors estimated using bootstrap resampling. The slope of the relation ($A_{FJ}(\mathrm{SDSS})=0.41,A_{FJ}(\mathrm{DEIMOS})=0.33$) for the full population of galaxies in both samples falls between the values obtained by the ATLAS-3D team \citep{cappellari:13}, who found that the relation has a mass-dependence: $\sigma\propto M^{0.43}$ for $\sigma_{re}\lesssim140 \unit{km\,s^{-1}}$, $\sigma\propto M^{0.21}$ for larger velocity dispersions. Because the ATLAS-3D sample was comprised of early-type galaxies, comparing to the slopes for the quiescent galaxy populations is more self-consistent, for which the measured slopes also ($A_{FJ}(\mathrm{SDSS})=0.37,A_{FJ}(\mathrm{DEIMOS})=0.28$) fall between the published slopes, given the selection bias towards $\sigma>100\unit{km\,s^{-1}}$.

In the SDSS (Figure\,\ref{fig:mass_sigma_sdss}), the velocity dispersions of quiescent galaxies are generally higher than for star-forming galaxies at fixed mass. In addition, the slope of the best-fit linear relation is slightly steeper for full sample of star-forming galaxies. The trend in relative normalization between the two populations hold at fixed S\'ersic index (Figure\,\ref{fig:mass_sigma_n_sdss}) along with overall variation in normalization with profile shape. This S\'ersic dependence is especially clear for quiescent galaxies, which exhibit steeper slopes and higher normalizations (relative to the overall relation, indicated by a dashed black line) with increasing $n$. Star-forming galaxies with $n\gtrsim1$ exhibit a weaker trend of increasing normalization and increasing slope with S\'ersic index. 

Figure\,\ref{fig:mass_sigma} demonstrates that velocity dispersions are also lower for star-forming galaxies relative to quiescent galaxies in the $z\sim0.7$ sample. The normalization of this relation is offset with respect to the $z\sim0$ relation such that dispersions are higher with redshift. Recently, \citet{sande:13} and \citet{belli:14a} presented a similar trend for samples of quiescent galaxies at $z\sim2$ and $z\sim1.2$ respectively; here we note that the result seems to hold for star-forming galaxies as well. 

As in Figure\,\ref{fig:mass_size_n}, the statistical power of this sample of galaxies breaks down at fixed S\'ersic index (Figure\,\ref{fig:mass_sigma_n}). We include fits in to the two galaxy samples in each panel to guide the eye, assuming the slopes measured from the full subsamples (in Figure\,\ref{fig:mass_sigma}). Although the samples are small, we note that if anything the normalizations of the mass Faber-Jackson relation decrease with S\'ersic index, exhibiting the opposite trend relative to the low-z sample of galaxies. However, we reiterate the need for larger samples to assess this trend robustly.

By comparing star-forming and quiescent galaxy populations in these two dimensional projections of the mass fundamental plane in this section, we emphasize the overall bimodality of the populations of galaxies separated initially based on their stellar populations. From this perspective, the structures (both in size and light profile), stellar dynamics (as measured by absorption line kinematics), and scaling relations between these properties should be treated separately. This would be important for the empirical measurement of galaxy scaling relations and for the theoretical study of the formation and evolution of galaxies which are constrained to follow following those relations. 

\section{Star-Forming and Quiescent Galaxies Lie on the Same Mass Fundamental Plane} \label{sec:massfp}

In contrast with the previous section, we turn our focus to the potential similarity between star-forming and quiescent galaxy populations. Specifically, we present the existence of a mass fundamental plane for star-forming and quiescent galaxy populations alike. The mass fundamental plane is the plane in three-dimensional space between galaxy size (effective radius, $R_e$), velocity dispersion ($\sigma$), and stellar mass surface density ($\Sigma_{\star}\equiv\frac{M_{\star}}{2\pi\,R_e^2}$) \citep[e.g.][]{hyde:09,bezanson:13b}. This plane represents a combination of the two 2-D scaling relations investigated in the previous section. 

\begin{figure*}[!t]
\centering
\subfiguretopcaptrue
\subfigure[][SDSS at $z\sim0$]{
\includegraphics[width=0.45\textwidth]{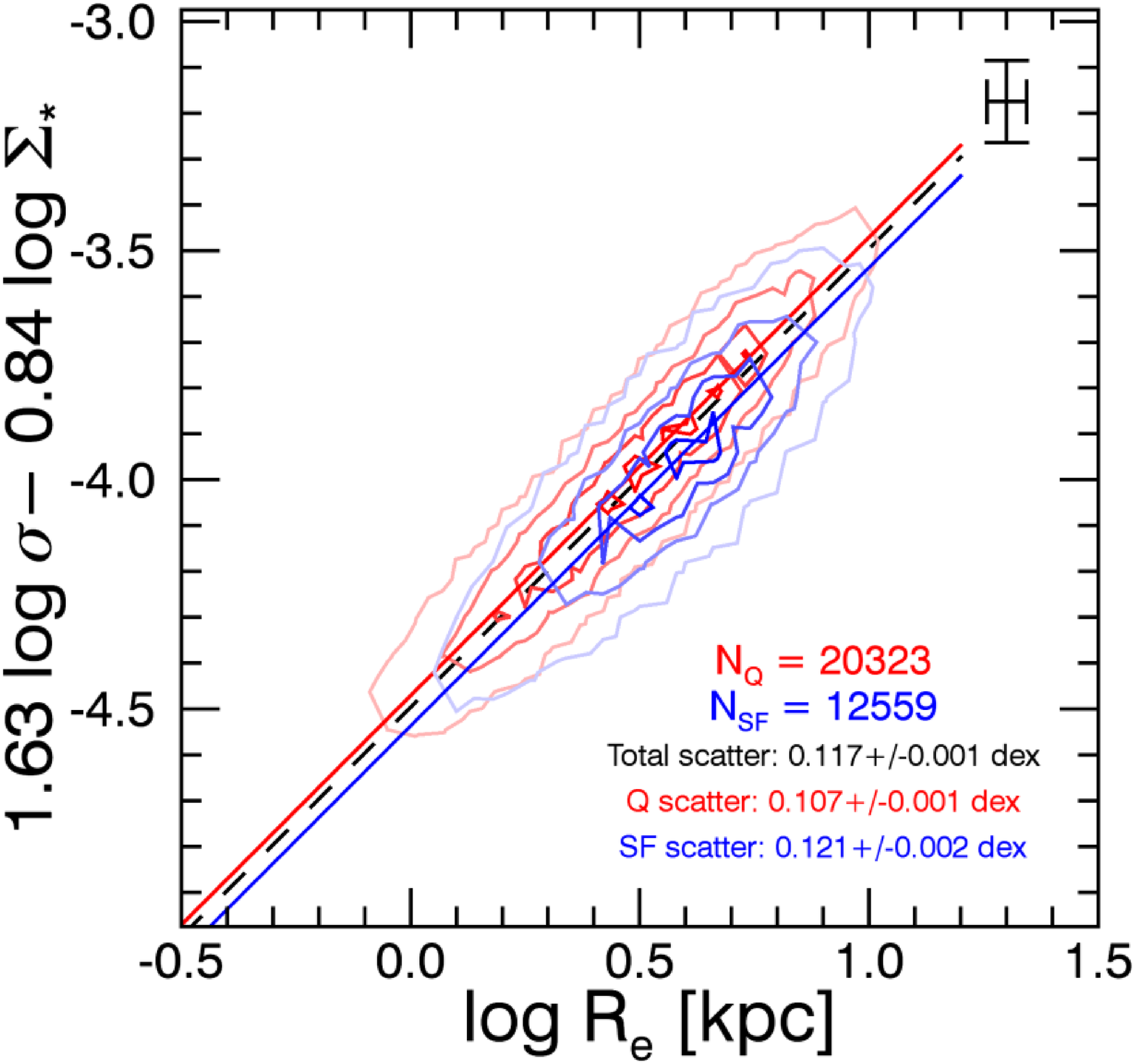}
\label{fig:massfp_sdss}
}
\subfigure[][DEIMOS at $z\sim0.7$]{
\includegraphics[width=0.45\textwidth]{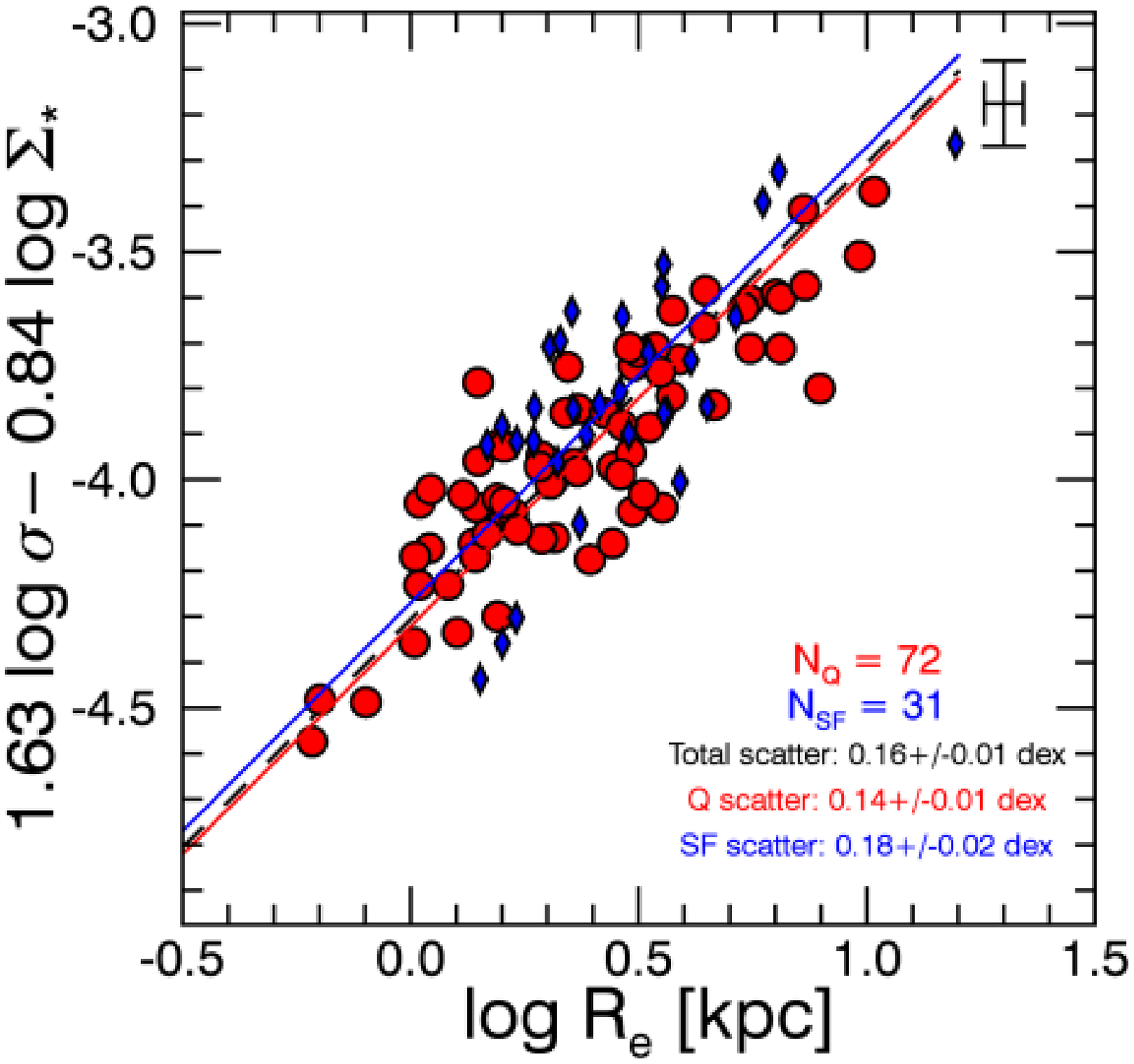}
\label{fig:massfp}
}
\caption{
The mass fundamental plane, or the projected three dimensional surface defined by stellar mass surface density ($\Sigma_*$), velocity dispersion ($\sigma$), and circularized effective radius ($R_e$) of galaxies at $z\sim0$ (panel \subref{fig:massfp_sdss}) and $z\sim0.7$ (panel \subref{fig:massfp}). Distribution of star-forming and quiescent galaxies are red and blue contours (as in e.g. Figures \ref{fig:mass_size_all} and \ref{fig:mass_sigma_all}). Best-fit relations (with fixed slope and varying normalization) are included as lines in each panel. Black dashed lines indicates the best-fit normalization for all galaxies in the sample. Red (and blue) solid lines indicates the best-fit relation to all quiescent (and star-forming) galaxies. The normalization of and scatter about the mass fundamental plane for star-forming and quiescent galaxies is strikingly similar at both redshifts.}
\label{fig:massfp_all}
\end{figure*}

\begin{figure*}[!t]
\centering
\includegraphics[width=0.95\textwidth]{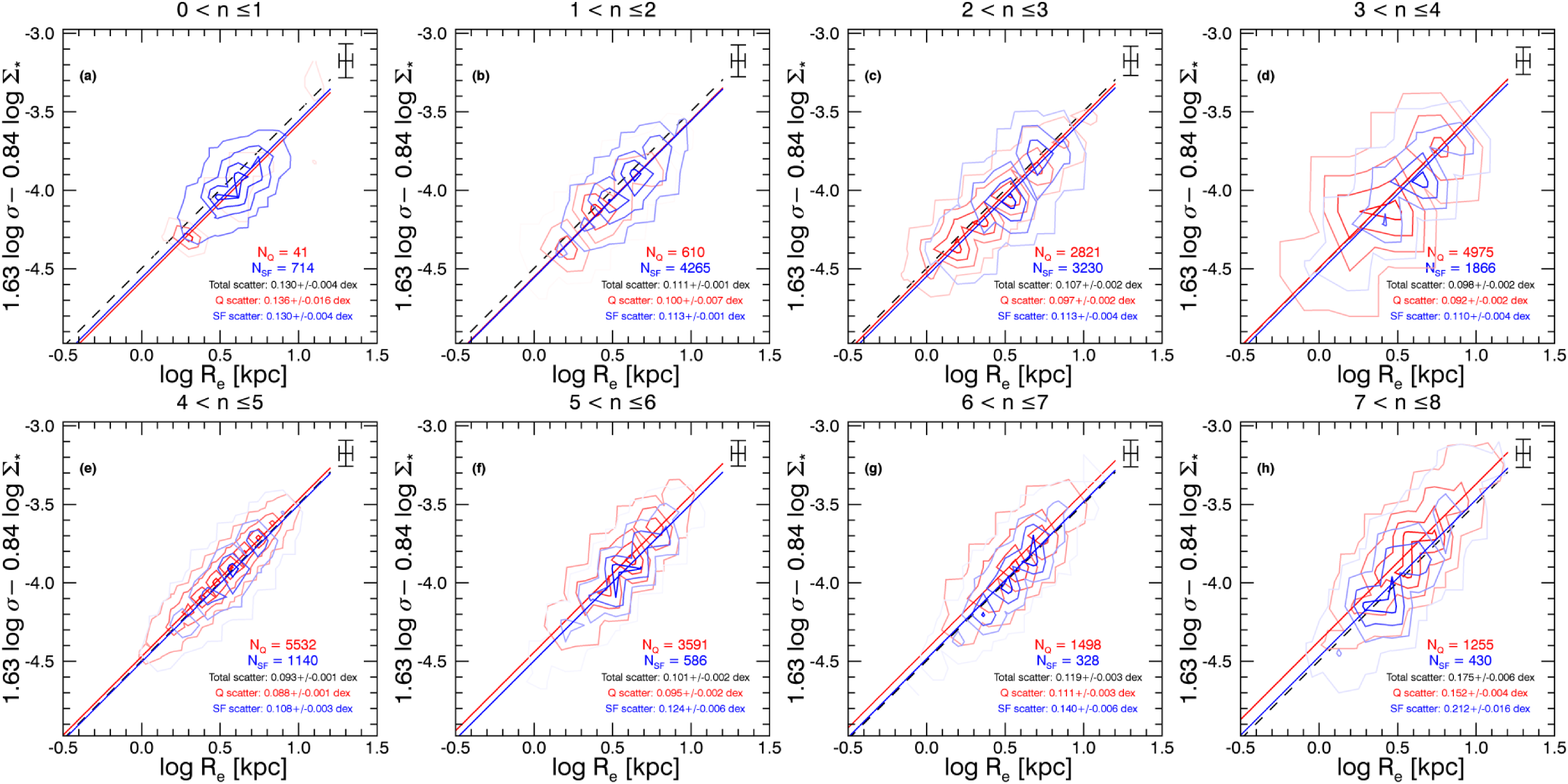}
\caption{The mass fundamental plane for galaxies in the SDSS, now divide into bins of S\'ersic index. Star-forming and quiescent galaxies are red and blue points (as in Figures \ref{fig:mass_size_n_sdss} and \ref{fig:mass_sigma_n_sdss}). Best-fit relations (with fixed slope and varying normalization) are included as lines in each panel. Black dashed lines (same in all panels, see Figure\,\ref{fig:massfp_sdss}) indicates the best-fit normalization for all galaxies in the sample. Red (and blue) solid lines indicates the best-fit relation to quiescent (or star-forming) galaxies in a given S\'ersic bin. The normalization of and scatter about the mass fundamental plane for star-forming and quiescent galaxies is strikingly similar. The effect of structural non-homology are extremely subtle, with normalization varying by $\lesssim 0.1\unit{dex}$ as a function of S\'ersic index.}
\label{fig:massfp_n_sdss}

\centering
\includegraphics[width=0.8\textwidth]{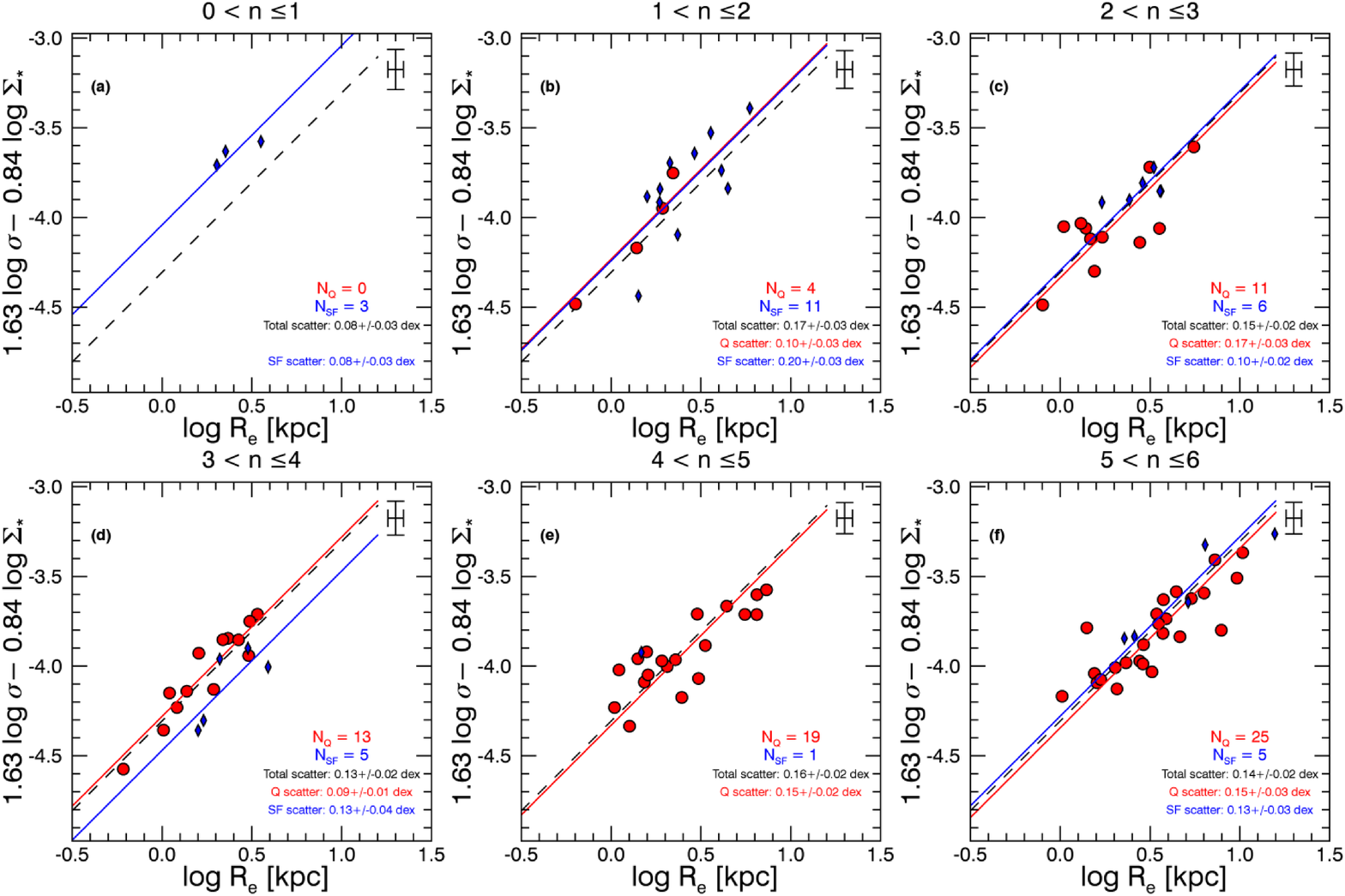}
\caption{The mass fundamental plane for galaxies at $z\sim0.7$. Best-fit relation to all galaxies in the sample is indicated as the dashed black line in all panels, relations for quiescent and star forming galaxies (altogether and in bins of S\'ersic index) shown as solid red and blue lines. Overall normalization of the mass fundamental plane differs from that in the SDSS (see Figure\,\ref{fig:massfp_n_sdss}), however does not vary strongly with quiescence (except perhaps for the small populations of $n<1$ or $n\sim3$ star-forming galaxies) or S\'ersic index.}
\label{fig:massfp_n}
\end{figure*}

Figure\,\ref{fig:massfp_all} shows an edge-on projection of the mass fundamental plane at $z\sim0$ and $z\sim0.7$ for star-forming and quiescent galaxies; Figure\,\ref{fig:massfp_n_sdss} and Figure\,\ref{fig:massfp_n} show the mass fundamental plane separated by Sersic index at each redshift. Although star-forming and quiescent galaxies are discrepant in size and velocity dispersion at fixed mass, Figure\,\ref{fig:massfp_all} demonstrates that the two populations occupy different regions of the same plane. The mass fundamental plane can be written as:

\begin{equation}
\log R_e =  \alpha\log\sigma + \beta\log\Sigma_{\star}+\gamma.
\label{eq:massfp}
\end{equation}

\noindent At $z\sim0$, we adopt the tilt of the mass fundamental plane from \citet{hyde:09} ($[\alpha,\beta]=[1.629,-0.840]$), as measured from a similar sample of galaxies in the SDSS and allow the normalization ($\gamma$) of the mass fundamental plane to vary for each sample.  There is some conflicting evidence in the literature regarding the evolution of the slope of the (luminosity) fundamental plane. \citet{holden:10} demonstrated that there is no evidence for evolution in the tilt since $z\sim1$, however with a larger sample of cluster data, \citet{jorgensen:13} found evidence of evolution in both slope and normalization of the plane with redshift. Fitting the exact functional form of the fundamental plane is an extremely difficult problem as it requires careful modeling of sample selection effects, correlated observational errors, and intrinsic scatter \citep[e.g.][]{jorgensen:96} and is beyond the scope of this paper. Therefore, we assume that the \citet{hyde:09} tilt of the mass fundamental plane also applies at higher redshifts, again allowing the normalization to vary.

We verified that the assumption of a non-evolving tilt does not significantly bias any of our results as follows. First, we performed a simple fit of slope of the mass fundamental plane to the full $z\sim0.7$ galaxy sample using the least trimmed squares algorithm \citep{rousseeuw:06} as implemented by the $LTS\_PLANEFIT$ program \citep{cappellari:13planes} to include observational errors in addition to intrinsic scatter. The plane is fit about each variable; we adopt the averages of each coefficient: $\alpha=1.978\pm0.134$ and $\beta=-0.968\pm0.056$. We note that this is extremely close to the Virial relation ($[\alpha,\beta]=[2,-1]$), and represents a significant evolution from the measured slope at $z\sim0$, but emphasize the potential influence of selection biases on this measurement, which often slice through the plane at non-parallel angles. However, even with this extreme evolution, the decrease in scatter about $\log R_e$ or $\log \sigma$ is minimal and comparable to the bootstrapped error estimates. Therefore we are satisfied to adopt the $z\sim0$ tilt for the mass fundamental plane at all redshifts; in \S \ref{sec:sigma} we include other three-parameter scaling relations.

The best-fit normalization to all galaxies in a given redshift range is shown in each panel of Figure\,\ref{fig:massfp_all} (and later of Figures\,\ref{fig:massfp_n_sdss} and \ref{fig:massfp_n}) as a dashed black line. Normalizations to the separate star-forming and quiescent populations in each panel are included as solid blue and red lines (either for all galaxies, or at fixed S\'ersic index). We calculate the scatter about the fundamental plane as the standard deviation about $\log R_e$, with errors estimated by a 1000 iteration bootstrap simulation. Scatter for the total, quiescent, and star-forming populations are indicated in the lower right corners of each panel.

Figure\,\ref{fig:massfp_sdss} demonstrates clearly that to first order, star-forming and quiescent galaxies lie on nearly the same mass fundamental plane with only very small shifts in normalization ($\sim0.06\unit{dex}$). The scatter about the plane is also very similar for both star-forming ($0.121\unit{dex}$) and quiescent galaxies ($0.107\unit{dex}$). We estimate the contribution of measurement errors to this scatter by monte-carlo simulations within the errors and find an intrinsic scatter of $0.093\unit{dex}$ for star-forming and $0.072\unit{dex}$ for quiescent galaxies.

At fixed S\'ersic index (Figure\,\ref{fig:massfp_n_sdss}), we isolate the effects of structural non-homology on the normalization of the plane. Offsets in normalization from the overall mass fundamental plane are the largest for exponential disklike galaxies (Figure\,\ref{fig:massfp_n_sdss}(a)-(b)), in particular for quenched galaxies. For the rare sample of quiescent disk galaxies, the overall normalization is higher by $\sim0.1\,\unit{dex}$. In general, the normalization of the mass fundamental plane varies as a function of S\'ersic index (and less strongly on stellar population). In this projection, low S\'ersic index galaxies lie below the overall mass fundamental plane and high S\'ersic index galaxies lie above. We fit residuals from the mass fundamental plane as parametrized by:

\begin{equation}
\Delta_{MFP} = A_{nh}+B_{nh}n,
\end{equation}

\noindent and refit the scatter about the non-homology corrected mass fundamental plane. This trend in offset with S\'ersic index is the strongest for quiescent galaxies, with $[A_{nh},B_{nh}] = [-0.099,0.028]$ and including the trend decreases the scatter by $0.009\unit{dex}$ to $0.098$. This is in contrast with the shallower relation, $[A_{nh},B_{nh}] = [-0.083,0.015]$, for star-forming galaxies, which yields only a $0.002\unit{dex}$ decrease in scatter to $0.119\unit{dex}$. The normalization of the overall mass fundamental plane can be thought of as a population-weighted average over the structures of all galaxies; for the entire population the trend,$[A_{nh},B_{nh}] = [-0.100,0.026]$, is similar to that of the quiescent sample and an overall scatter of $0.108\unit{dex}$. We emphasize that these differences are small ($\lesssim0.1\unit{dex}$) and that all galaxies lie on nearly the same plane.

At $z\sim0.7$ the primary conclusion that star-forming and quiescent galaxies lie on roughly the same mass fundamental plane remains (Figure\,\ref{fig:massfp}). However, the details are somewhat different. First, the scatter about these relations is larger than in the SDSS sample. Partially, this is driven by the smaller sample size as reflected by larger errors in the measured scatter (estimated from bootstrap resampling). We assume that measurement errors are very similar for the two samples (e.g. errors in $M_{\star}/L$); implying intrinsic scatter of $\sim0.14\unit{dex}$ for all galaxies and $\sim0.12\unit{dex}$ for quiescent galaxies. This suggests that measurement errors have been underestimated for this sample or that the tightness of the mass fundamental plane decreases with redshift. Measured scatter about the mass fundamental plane is slightly larger for star-forming galaxies, but this is within the errors. Additionally, the overall normalization is slightly higher than at $z\sim0$ ($\sim0.2\unit{dex}$), as shown in \citet{bezanson:13b}. We discuss possible explanations for this effect in \S \ref{sec:discuss}. Furthermore, while there are slight offsets from the mass fundamental plane at fixed S\'ersic index (Figure\,\ref{fig:massfp_n}), these are partially driven by small sample size. We conclude that this sample is insufficient to definitively assess the effects of structural non-homology on the mass fundamental plane at $z>0$.

\begin{deluxetable*}{ccccc}
\tablecaption{Mass Fundamental Plane Normalizations}
\tablehead{
\colhead{sample} & \colhead{S\'ersic range} & 
\colhead{$\gamma$} & \colhead{$\gamma_{Q}$} & \colhead{$\gamma_{SF}$}}
\startdata
SDSS & all & 4.496$\pm$0.001 & 4.471$\pm$0.001 & 4.537$\pm$0.001 \\
SDSS & 0$<n\leq$1 & 4.558$\pm$0.005 & 4.580$\pm$0.021 & 4.557$\pm$0.005 \\
SDSS & 1$<n\leq$2 & 4.556$\pm$0.002 & 4.550$\pm$0.004 & 4.557$\pm$0.002 \\
SDSS & 2$<n\leq$3 & 4.537$\pm$0.001 & 4.523$\pm$0.002 & 4.549$\pm$0.002 \\
SDSS & 3$<n\leq$4 & 4.501$\pm$0.001 & 4.493$\pm$0.001 & 4.524$\pm$0.003 \\
SDSS & 4$<n\leq$5 & 4.475$\pm$0.001 & 4.470$\pm$0.001 & 4.504$\pm$0.003 \\
SDSS & 5$<n\leq$6 & 4.449$\pm$0.002 & 4.442$\pm$0.002 & 4.496$\pm$0.005 \\
SDSS & 6$<n\leq$7 & 4.434$\pm$0.003 & 4.423$\pm$0.003 & 4.484$\pm$0.008 \\
SDSS & 7$<n\leq$8 & 4.396$\pm$0.004 & 4.370$\pm$0.004 & 4.470$\pm$0.010 \\
\\
DEIMOS & all & 4.306$\pm$0.015 & 4.321$\pm$0.017 & 4.270$\pm$0.032 \\
DEIMOS & 0$<n\leq$1 & 4.042$\pm$0.033 & 4.321$\pm$0.017 & 4.042$\pm$0.034 \\
DEIMOS & 1$<n\leq$2 & 4.239$\pm$0.044 & 4.232$\pm$0.045 & 4.241$\pm$0.058 \\
DEIMOS & 2$<n\leq$3 & 4.321$\pm$0.037 & 4.336$\pm$0.052 & 4.295$\pm$0.037 \\
DEIMOS & 3$<n\leq$4 & 4.333$\pm$0.031 & 4.280$\pm$0.025 & 4.470$\pm$0.047 \\
DEIMOS & 4$<n\leq$5 & 4.317$\pm$0.034 & 4.329$\pm$0.034 & 4.092$\pm$0.000 \\
DEIMOS & 5$<n\leq$6 & 4.333$\pm$0.026 & 4.344$\pm$0.029 & 4.279$\pm$0.055 \\
\label{tab:massfp}
\tablecomments{Best-fit normalizations to the mass fundamental plane (see Equation \ref{eq:massfp}). The tilt of the mass fundamental plane is fixed as $[\alpha,\beta]=[1.629,-0.840]$\citep{hyde:09}.}
\end{deluxetable*}

\section{Scatter and the Form of the 3-D Mass Plane} \label{sec:sigma}

Despite the structural and dynamical differences between the populations of star-forming and quiescent galaxies, we have shown that all galaxies follow roughly the same three dimensional relationship between velocity dispersion, size, and stellar mass or stellar mass surface density. In this section, we investigate whether the mass fundamental plane defined in \S \ref{sec:massfp} is the optimal form of the three dimensional surface. Here we consider four scaling relations: the mass Faber-Jackson relation, the mass fundamental plane, the Virial plane, and a Virial relation with a S\'ersic-dependent constant. Measured relations are included in Tables \ref{tab:massfp} and \ref{tab:massfj}. Specifically, we assess the ability of a given relation to predict the velocity dispersion of a galaxy, as measured by the scatter between the inferred and measured dispersions. 

\begin{figure*}[!t]
\centering
\subfigure[][Mass Faber-Jackson]{
\includegraphics[width=0.4\textwidth]{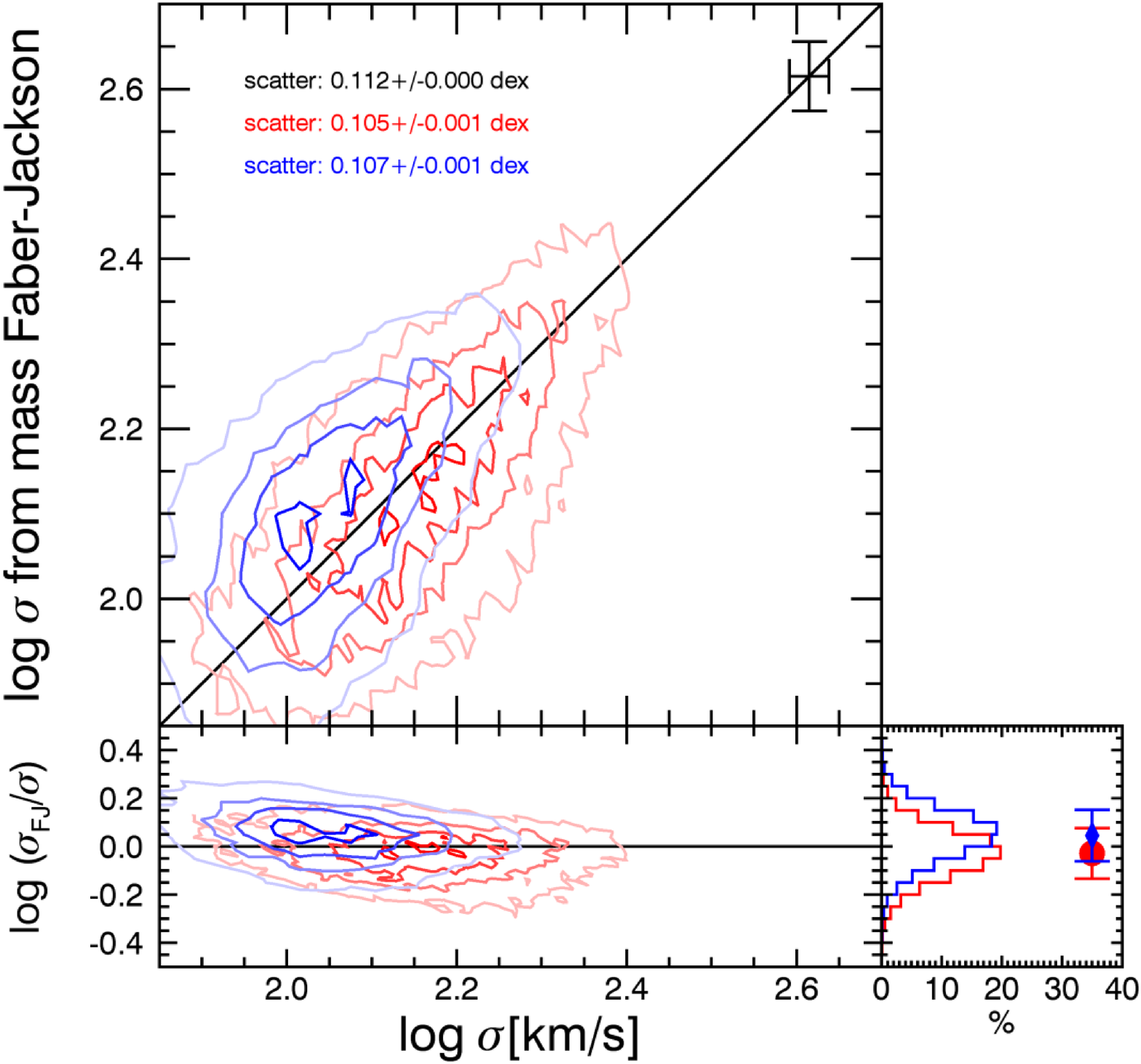}
\label{fig:sigma_sigma_fj_sdss}
}
\subfigure[][Mass Fundamental Plane]{
\includegraphics[width=0.4\textwidth]{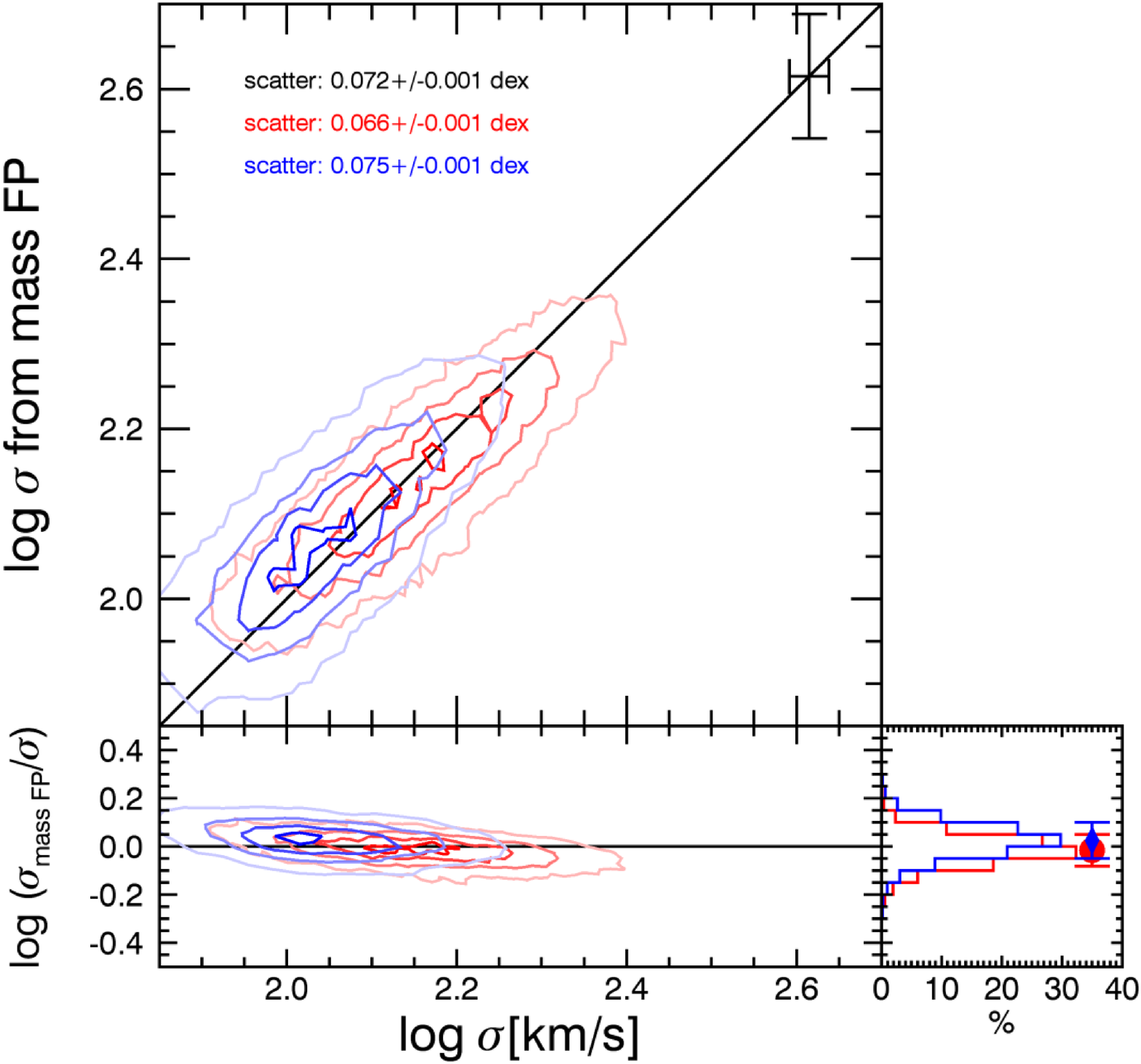}
\label{fig:sigma_sigma_massfp_sdss}
}
\subfigure[][Virial Theorem]{
\includegraphics[width=0.4\textwidth]{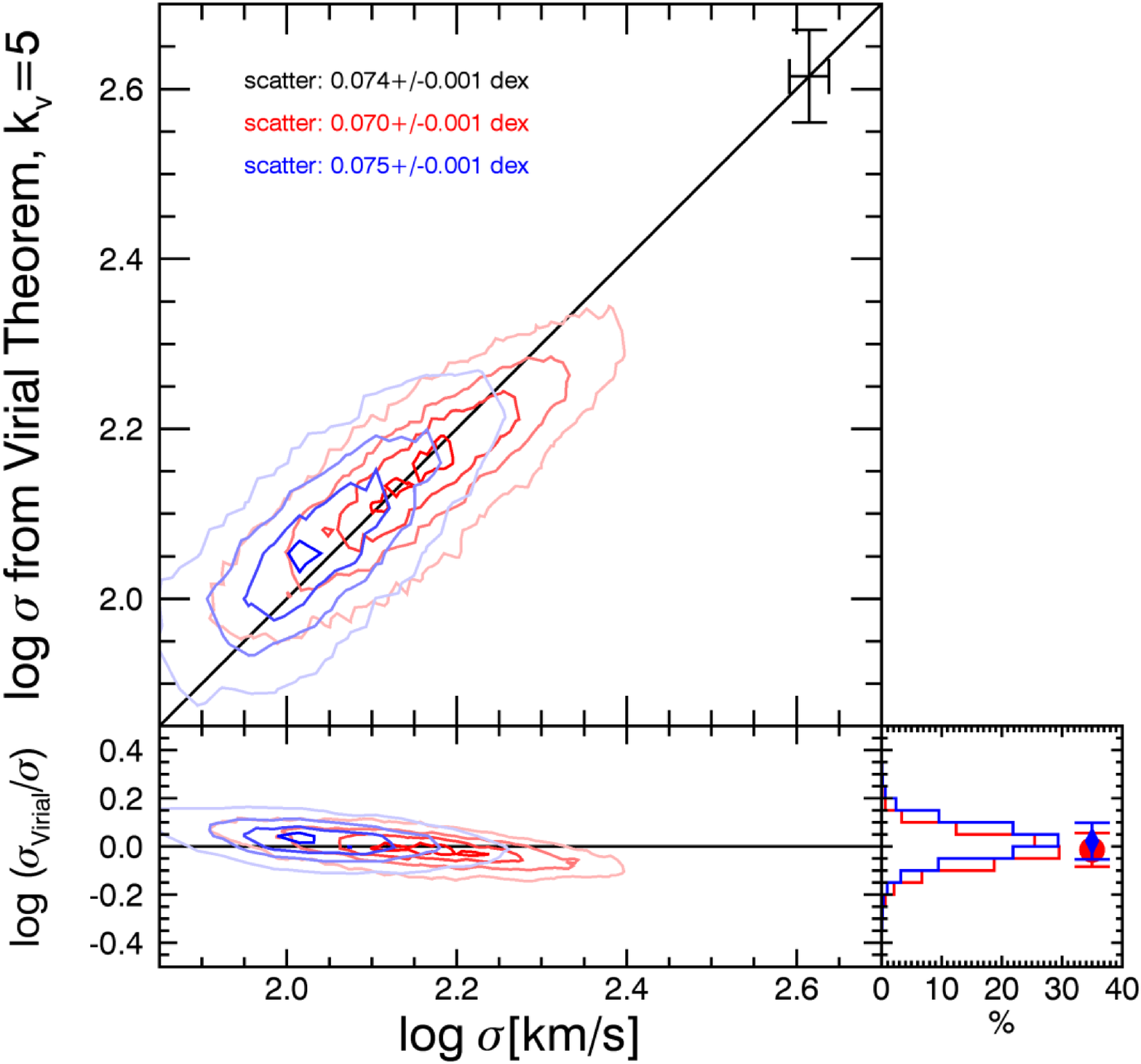}
\label{fig:sigma_sigma_virial_sdss}
}
\subfigure[][Virial Theorem, $k_{\mathrm{V}}(n)$]{
\includegraphics[width=0.4\textwidth]{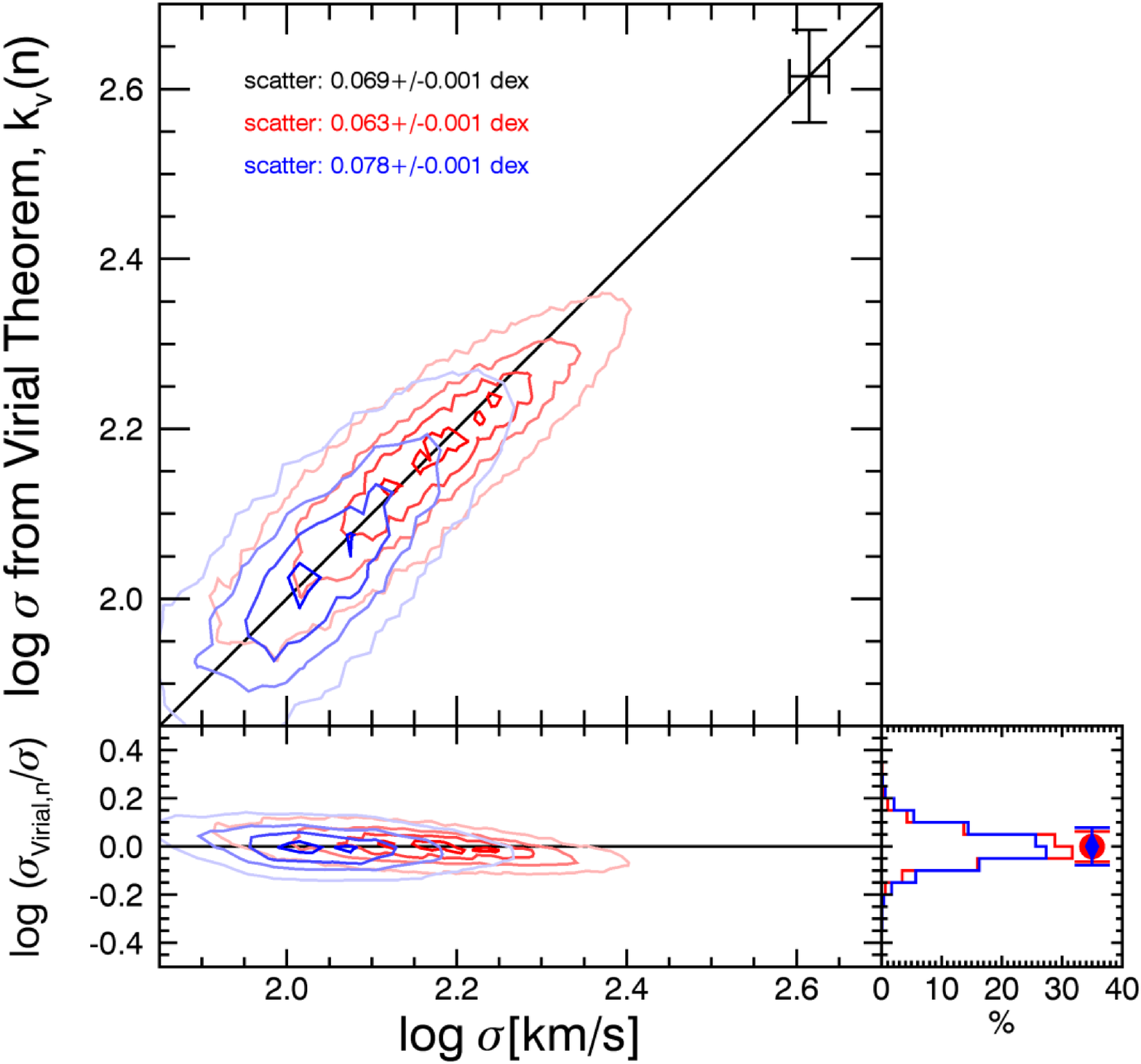}
\label{fig:sigma_sigma_n_sdss}
}
\caption{Comparison of the tightness of scaling relations in the SDSS as revealed by predicted velocity dispersion versus measured velocity dispersion (square panels) and residuals from those relations (bottom panels). \subref{fig:sigma_sigma_fj_sdss} Velocity dispersion predicted by the mass Faber-Jackson relation (see also Figure\,\ref{fig:faberjackson}) versus measured velocity dispersion. \subref{fig:sigma_sigma_massfp_sdss} Velocity dispersion predicted by the mass fundamental plane versus measured velocity dispersion. \subref{fig:sigma_sigma_virial_sdss} Velocity dispersion predicted by the Virial theorem with a single constant (independent of non-homology) versus measured velocity dispersion. \subref{fig:sigma_sigma_n_sdss} Velocity dispersion predicted by the Virial theorem with a S\'ersic-dependent constant versus measured velocity dispersion.   In each case, including galaxy sizes (as in Panels \subref{fig:sigma_sigma_massfp_sdss}-\subref{fig:sigma_sigma_n_sdss}) decreases the scatter in the scaling relations for quiescent and star-forming galaxies alike. However, the scatter is comparable about both Virial planes and the mass fundamental plane.}
\label{fig:sigmasigma_sdss}
\end{figure*}

\begin{figure*}[t]
\centering
\subfigure[][Mass Faber-Jackson]{
\includegraphics[width=0.4\textwidth]{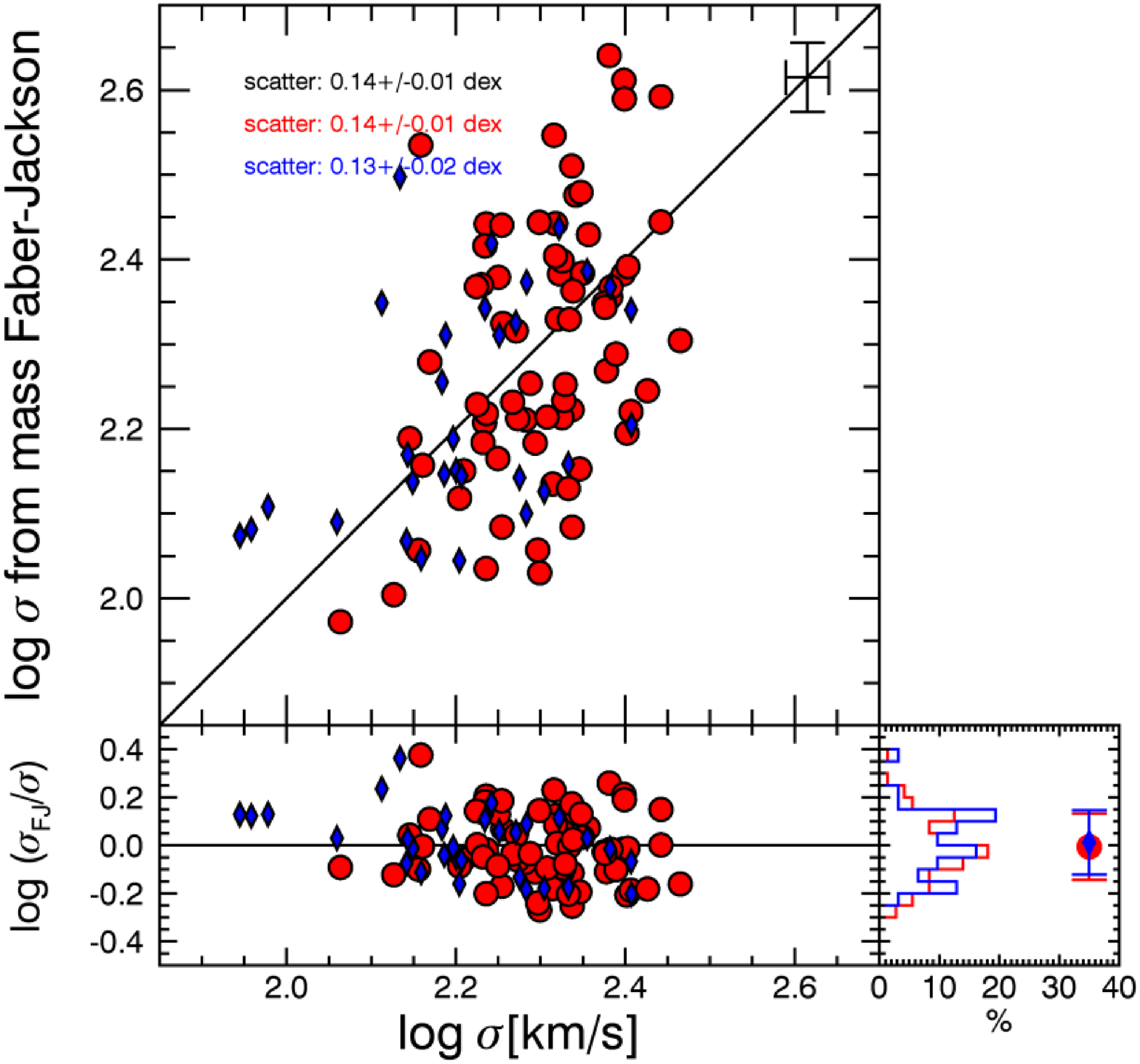}
\label{fig:sigma_sigma_fj}
}
\subfigure[][Mass Fundamental Plane]{
\includegraphics[width=0.4\textwidth]{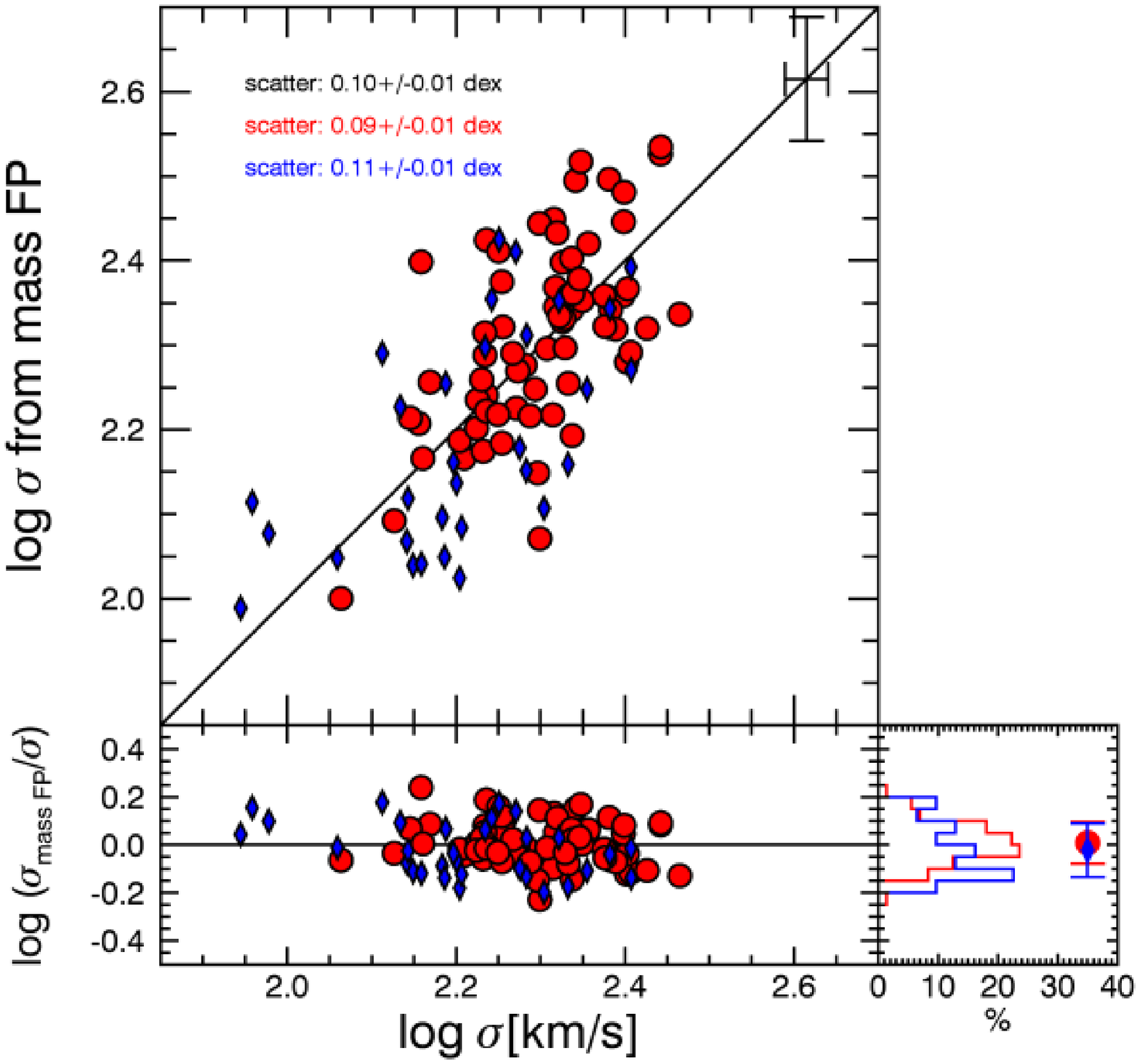}
\label{fig:sigma_sigma_massfp}
}
\subfigure[][Virial Theorem]{
\includegraphics[width=0.4\textwidth]{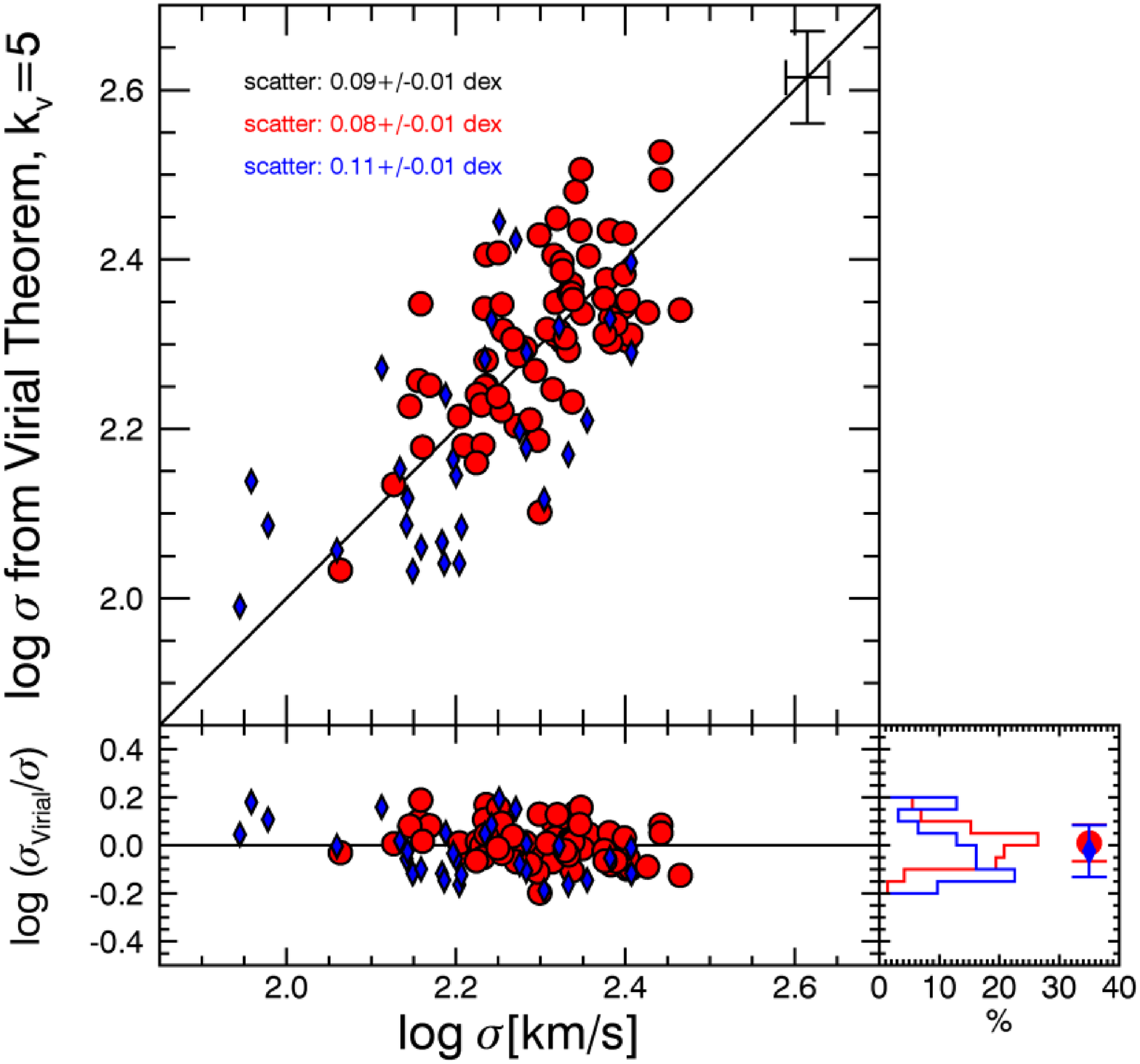}
\label{fig:sigma_sigma_virial}
}
\subfigure[][Virial Theorem, $k_{\mathrm{V}}(n)$]{
\includegraphics[width=0.4\textwidth]{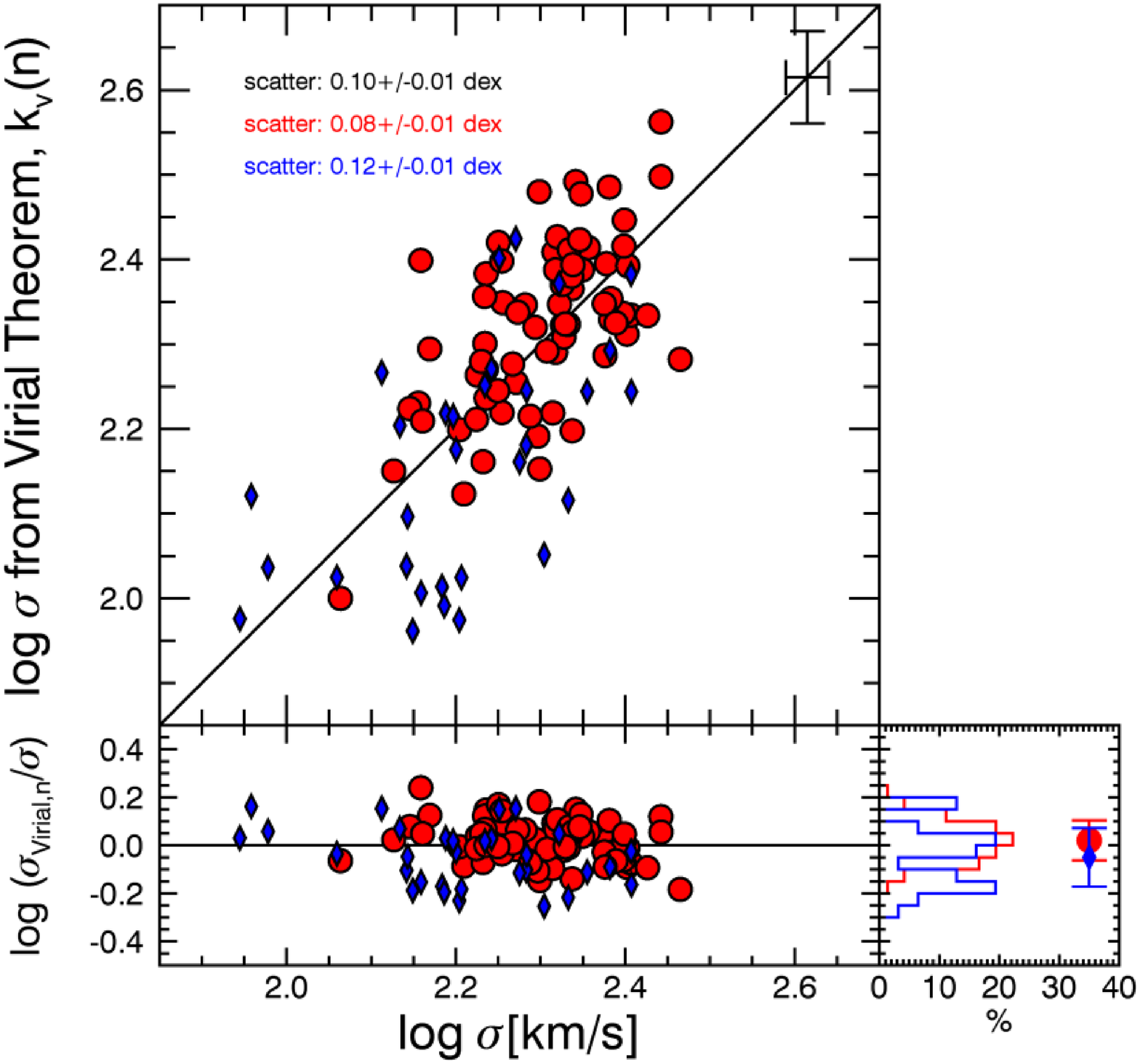}
\label{fig:sigma_sigma_n}
}
\caption{Comparison of the tightness of scaling relations (layout as in Figure\,\ref{fig:sigmasigma_sdss}) at $z\sim0$. Velocity dispersion predicted by \subref{fig:sigma_sigma_fj} mass Faber-Jackson, \subref{fig:sigma_sigma_massfp} the mass fundamental plane \subref{fig:sigma_sigma_virial} Virial theorem with a single constant, \subref{fig:sigma_sigma_n} the Virial theorem with a S\'ersic-dependent constant, and (d) the mass fundamental plane versus measured velocity dispersion. As at $z\sim0$, including galaxy size decreases the scatter in galaxy scaling relations; with very subtle differences between the three possible 3D surfaces.}
\label{fig:sigmasigma}
\end{figure*}

Figure\,\ref{fig:sigmasigma_sdss} shows these comparisons for galaxies in the SDSS sample. Figure\,\ref{fig:sigma_sigma_fj_sdss} shows the velocity dispersion predicted from the mass Faber-Jackson relation (shown in Figure\,\ref{fig:mass_sigma_sdss}):

\begin{equation}
\log\sigma_{\mathrm{inf,FJ}} = \mathrm{A_{FJ}}\log\,M_{\star}+\mathrm{B_{FJ}}
\label{eq:sigma_fj}
\end{equation}

\noindent versus measured velocity dispersion (large panel), residuals as a function of velocity dispersion (lower panel), and a histogram of residuals for star-forming and quiescent galaxies (right) separately. This and the following relations assume that velocity dispersions are measured in $\unit{km\,s^{-1}}$, sizes are measured in $\unit{kpc}$, and stellar masses are quoted in $M_{\odot}$. The mean and standard deviations of the residuals for star-forming and quiescent galaxies are indicated as blue/red points with error-bars to the right of the histograms. Scatter for all galaxies and quiescent/star-forming populations in black, red, and blue text in the upper left corner.

The scatter in this relation is highest of the four tested scaling relations, with a scatter in velocity dispersion of $\sim0.11\unit{dex}$ for all populations. Furthermore, the residuals exhibit additional correlations in the bottom panel, with offset median values for star-forming and quiescent galaxies. We note that separately measuring the mass Faber-Jackson relation for star-forming and quiescent galaxies reduces the scatter only minimally (by $0.01\unit{dex}$), although it does remove residual correlations (see Appendix \ref{sec:SFQmassfj}).

\begin{deluxetable*}{cccccccc}
\tablecaption{Mass Faber-Jackson Parameters}
\tablehead{
\colhead{sample} & \colhead{$\mathrm{A_{FJ}}$} & \colhead{$\mathrm{B_{FJ}}$} & \colhead{$\mathrm{A^Q_{FJ}}$} & \colhead{$\mathrm{B^Q_{FJ}}$} & \colhead{$\mathrm{A^{SF}_{FJ}}$} & \colhead{$\mathrm{B^{SF}_{FJ}}$}
}
\startdata
SDSS & 0.41$\pm$0.00 & -2.25$\pm$0.02 & 0.37$\pm$0.00 & -1.80$\pm$0.02 & 0.39$\pm$0.00 & -2.06$\pm$0.03 \\
\vspace{-0.1cm}
DEIMOS & 0.34$\pm$0.04 & -1.37$\pm$0.43 & 0.28$\pm$0.06 & -0.68$\pm$0.62 & 0.46$\pm$0.11 & -2.60$\pm$1.11 \\
\label{tab:massfj}
\tablecomments{Best fit mass Faber-Jackson relations at $z\sim0$ and $z\sim0.7$ (see Equation \ref{eq:sigma_fj}).}
\end{deluxetable*}

Figure\,\ref{fig:sigma_sigma_massfp_sdss} has the same layout as Figure\,\ref{fig:sigma_sigma_fj_sdss}, but compares the velocity dispersion predicted by the mass fundamental plane:

\begin{equation}
\log\sigma_{\mathrm{inf,massFP}} = (\log R_e - \beta \log \Sigma_{\star} - \gamma)/\alpha
\label{eq:sigma_mfp}
\end{equation}

\noindent to the measured velocity dispersion (measured normalizations $\gamma$ are provided in Table \ref{tab:massfp}). The scatter about this relation is lower ($0.07\unit{dex}$), though there seems to be a slight residual trend, possibly indicating a slight tilt relative to the \citet{hyde:09} mass fundamental plane.

Another possible relation to connect the dynamical and structural properties of galaxies is the Virial theorem, which states that:

\begin{equation}
M_{\star} = \frac{M_{\star}}{M_{dyn}}\frac{kR_e\sigma^2}{G}.
\end{equation}

\noindent The constant in this equation, $k$, can be estimated from analytical models or measured empirically. Based on the observed dynamical and structural properties of local elliptical galaxies $k\approx5$ \citep{cappellari:06}. Figure\,\ref{fig:sigma_sigma_virial_sdss}, compares measured dispersion to the velocity dispersion predicted by the Virial theorem:

\begin{equation}
\log\sigma_{\mathrm{inf,V}} = 0.5\left(\log M_{\star} - \log R_e\right) + C_{\mathrm{V}}
\label{eq:sigma_vir}
\end{equation}

\noindent in which the constant is a combination of the gravitational constant, $G$, the Virial constant and a normalization as $C_V = 0.5\log \left( \frac{G}{(M_{\star}/M_{dyn})(5)}\right)$. See Appendix \ref{sec:mass_mass} for a discussion of the relationship between dynamical and stellar mass. We fit the normalization and find $C_{\mathrm{V}}=2.848$ at $z\sim0$, $C_{\mathrm{V}}=-2.956$ at $z\sim0.7$. This relation yields a similar overall scatter to the mass fundamental plane ($0.07\unit{dex}$), although this is slightly higher for star-forming galaxies ($0.08\unit{dex}$).

Dynamical models of galaxies with varied stellar distributions predict a variation of the Virial constant; for example with S\'ersic index \citep[e.g.][]{ciotti:91,bertin:02,cappellari:06}. Given the range in galaxy morphology probed by this study, this could be important. Figure\,\ref{fig:sigma_sigma_n_sdss} compares measured to inferred velocity dispersion calculated with the Virial theorem, but including an analytically derived Virial constant which depends on S\'ersic index:

\begin{equation}
\log \sigma_{\mathrm{inf,V}}(n) = 0.5\left(\log M_{\star} - \log R_e\right)+C_{\mathrm{V(n)}},
\label{eq:sigma_vir_n}
\end{equation}
\noindent in which the constant $C_{V(n)} =0.5\log \left( \frac{G}{M_{\star}/M_{dyn}}\right) -0.5\log{k_{\mathrm{V}}(n)}$. The S\'ersic-dependent constant, $k_{\mathrm{V}}(n)\approx 8.87-0.831n+0.0241n^2$ \citep{cappellari:06}, is derived analytically by solving the spherically symmetric Jeans equation with the assumption that mass follows a S\'ersic profile and all orbits are isotropic.  We allow the normalization to vary to account for differences in $\frac{M_{\star}}{M_{dyn}}$, finding $C_{V(n)} (z\sim0) = -2.539-0.5\log{k_{\mathrm{V}}(n)}$ and $C_{V(n)} (z\sim0.7) = -2.646-0.5\log{k_{\mathrm{V}}(n)}$. Again the overall scatter is the same ($0.07\unit{dex}$), which is an average between a tightened fit for quiescent galaxies ($0.06\unit{dex}$) and an increase in scatter for star-forming galaxies ($0.08\unit{dex}$).

The overall scatter in the latter three relations is comparable: $\sim0.07\unit{dex}$ for the full galaxy population in each case, indicating the importance of including size measurements in these scaling relations. Although the scatter for the quiescent galaxy population decreases slightly when a S\'ersic-dependent Virial constant is included ($\sim0.06\unit{dex}$), and trends in the residuals differ only subtly, all three surfaces adequately describe the relationship between galaxy structures and dynamics for star-forming and quiescent galaxies. This is not the case for scatter about $\log R_e$ (see Appendix \ref{sec:scatter_re}), in which the mass fundamental plane exhibits much lower scatter, however this may be driven by correlated measurement errors.

Figure\,\ref{fig:sigmasigma} demonstrates that similar results hold at $z\sim0.7$. Again the scatter derived from the mass Faber-Jackson is larger (Figure\,\ref{fig:sigma_sigma_fj}) than for the mass fundamental plane (Figure\,\ref{fig:sigma_sigma_massfp}) or the Virial theorem, both without (Figure\,\ref{fig:sigma_sigma_virial}) and with (Figure\,\ref{fig:sigma_sigma_n}) a S\'ersic-dependent constant. The overall scatter in the three latter relations is roughly the same, although it is larger for star-forming galaxies, in contrast with the $z\sim0$ SDSS sample. 
\begin{deluxetable*}{ccccccc}[!t]
\tabletypesize{\footnotesize}
\tablecaption{Measured and Intrinsic Scatter in Velocity Dispersions Inferred from Scaling Relations}
\tablehead{
\colhead{Relation} & \colhead{rms} & \colhead{Intrinsic rms} & \colhead{rms (Q)} & \colhead{Intrinsic rms (Q)} & \colhead{rms (SF)} & \colhead{Intrinsic rms (SF)} }
\startdata
\cutinhead{SDSS: $z\sim0$}
Faber-Jackson & 0.112 $\pm$ 0.000 & 0.101 & 0.105 $\pm$ 0.001 & 0.095 & 0.107 $\pm$ 0.001 & 0.094 \\
Mass Fundamental Plane & 0.072 $\pm$ 0.001 & 0.055 & 0.066 $\pm$ 0.001 & 0.049 & 0.075 $\pm$ 0.001 & 0.057 \\
Virial Theorem ($k_V=5$) & 0.074 $\pm$ 0.001 & 0.064 & 0.070 $\pm$ 0.001 & 0.061 & 0.075 $\pm$ 0.001 & 0.063 \\
Virial Theorem ($k_{\mathrm{V}}(n)$) & 0.069 $\pm$ 0.001 & 0.058 & 0.063 $\pm$ 0.001 & 0.053 & 0.078 $\pm$ 0.001 & 0.066 \\
\cutinhead{DEIMOS: $z\sim0.7$}
Faber-Jackson & 0.14 $\pm$ 0.01 & 0.13 & 0.14 $\pm$ 0.01 & 0.13 & 0.13 $\pm$ 0.02 & 0.12 \\
Mass Fundamental Plane & 0.10 $\pm$ 0.01 & 0.08 & 0.09 $\pm$ 0.01 & 0.07 & 0.11 $\pm$ 0.01 & 0.10 \\
Virial Theorem ($k_V=5$) & 0.09 $\pm$ 0.01 & 0.08 & 0.08 $\pm$ 0.01 & 0.07 & 0.11 $\pm$ 0.01 & 0.10 \\
Virial Theorem ($k_{\mathrm{V}}(n)$) & 0.10 $\pm$ 0.01 & 0.09 & 0.08 $\pm$ 0.01 & 0.07 & 0.12 $\pm$ 0.01 & 0.11 \\
\label{tab:scatter}

\tablecomments{Measured and intrinsic scatter in velocity dispersion from various scaling relations at $z\sim0.7$, for all galaxies and then separately for quiescent (Q) and star-forming (SF) galaxies. Inferred velocity dispersions calculated using the following equations: 
Faber-Jackson from Eq. \ref{eq:sigma_fj}, Mass Fundamental Plane from Eq. \ref{eq:sigma_mfp}, Virial Theorem with constant from Eq. \ref{eq:sigma_vir} and with S\'ersic-dependent constant from Eq. \ref{eq:sigma_vir_n}.}
\end{deluxetable*}

\section{Discussion and Conclusions}
\label{sec:discuss}

We have shown that despite the structural and dynamical differences between star-forming and quiescent galaxies, all massive ($\sigma\gtrsim100\unit{km\,s^{-1}}$) galaxies lie within the same three dimensional phase space when structures (size and profile shape) and stellar dynamics (velocity dispersion) are included concurrently. The data do not suggest a significant preference for any of the specific 3D surfaces examined in this work: the mass fundamental plane, the Virial plane, or the space defined by the Virial theorem with a S\'ersic-dependent constant. 

\citet{cappellari:13} pointed out that the existence of a mass fundamental plane (or ``mass plane'') for elliptical galaxies does not strongly constrain galaxy formation models, it is simply a statement that galaxies are in virial equilibrium.  However, in the context of this work, the fact that star-forming and quiescent galaxies both lie on mass fundamental planes and the normalization is so similar could have important implications for galaxy formation models. 

The inclusion of star-forming and quiescent galaxies in the same scaling relations can provide interesting constraints on galaxy formation models and can be useful for observational astronomy at high redshifts, however it is not a new technique. \citet{taylor:10} demonstrated that the stellar masses of galaxies of all types were directly correlated with dynamical masses derived from sizes, velocity dispersions, and S\'ersic indices, using comparable galaxies from the SDSS. This work is also similar to studies of the Fundamental Manifold \citep[e.g.][]{zaritsky:06,zaritsky:07,zaritsky:08}. In particular, \citet{zaritsky:08} demonstrated that both disks and spheroids across a wide range of masses lie on the same plane in velocity, size, and surface brightness space when allowing for variation in mass-to-light ratios. In that work, the velocity (V) was comprised either of rotational velocity or intrinsic velocity dispersion; by measuring the line-of-sight, projected velocity dispersion as in this work, $\sigma$ is nearly the same quantity as $V^2\equiv0.5V_r^2+\sigma^2$ in the Fundamental Manifold formalism. In this Paper, we verify the existence of a similar mass fundamental plane in which mass-to-light ratios are derived from stellar population synthesis modeling alone.  Furthermore, we extend these conclusions to higher redshift and propose that such unified scaling relations are not just an interesting property of galaxies, but and important and flexible tool to study galaxies through cosmic time.

Although this study shows that quenched and star-forming galaxies exist within the same 3D space, they still populate different regions of that space. Therefore, the process(es) which are responsible for driving the transition or quenching of galaxies from one population may alter galaxy structures and dynamics and must do so within the allowed space defined by the mass fundamental plane. It has been suggested that star-forming galaxies at high redshift are structurally similar to local elliptical galaxies, even though they are larger at a fixed epoch \citep[e.g.][]{franx:08,wel:14}. The evolution of galaxies in this plane is certainly due to a combination of such simple, passive evolution and stronger structural evolution, likely driven by galaxy merging. Another potentially important constraint is the weak dependence of the structural and dynamical scaling relations on profile shape, as quantified by S\'ersic index. The shape of the mass Fundamental Plane and the projections explored in this Paper, the mass-size and Faber-Jackson relations, vary subtly with S\'ersic n. Therefore, models of passive evolution must also compare detailed structures, not just sizes, of galaxy populations at different redshifts. With a larger sample of galaxy dynamics across galaxy populations and time, one could compare the structures, both sizes and profile shapes, and dynamics of high redshift disk galaxies to quenched descendent galaxies at later times and assess the relative importance of various physical processes.

To first order the mass fundamental plane for quiescent galaxies does not evolve strongly since $z\sim2$ \citep{bezanson:13b}, however the normalization of the plane is slightly different between the two epochs probed by this study ($\sim 0.2 \unit{dex}$). This could be due to a combination of observational and/or physical reasons. First, dynamics of galaxies are observed within somewhat different spectroscopic apertures. Although aperture corrections are made based on empirical velocity dispersion profiles and the corrections themselves are generally small ($\sim 4\%$ in the SDSS and $\sim 2\%$ at $z\sim0.7$), these relations are found for local ellipticals and could vary with redshift or morphological type. Furthermore, $M_{\star}/L$ are measured within different physical apertures between the two redshift slices and color gradients could introduce offsets in the measured mass fundamental plane. Additionally, it could be related to the distribution of S\'ersic indices of the two samples: the $z\sim0.7$ sample of galaxies has a larger fraction of galaxies with high S\'ersic indices $n>4$. In the SDSS we showed that the normalization of the mass fundamental plane varies subtly with S\'ersic index relative to the population-weighted average (Figure \ref{fig:massfp_n_sdss}), bringing high-$n$ galaxies closer to the normalization of the $z\sim0.7$ mass fundamental plane.   

Perhaps the more physically interesting case would involve slight evolution in the density profiles in the central regions of galaxies: either in stars or dark matter.  As galaxies grow through cosmic time, the more massive ones appear to be growing in an ``inside-out'' manner \citep[e.g.][]{bezanson:09,hopkins:09cores}. This translates to an increase in S\'ersic indices \citep[e.g.][]{dokkum:10,patel:13a} with time for massive galaxies. In Figures \ref{fig:massfp_n_sdss} and \ref{fig:massfp_n} we demonstrated the slight dependence of the mass fundamental plane normalization on structure (quantified by S\'ersic index). Different distributions of stellar density profile shapes imply slightly different dynamics.

Additionally, this could be due to a variation in the dark matter fraction in the inner regions of galaxies. Indications of evolution in $M_{DM}/M_{\star}(<R_e)$ with redshift have been suggested theoretically \citep[][]{hopkins:09scaling}, based on simulations \citep[][]{hilz:12a}, and empirically \citep[e.g.][]{sande:13} as the stellar components undergo significant growth with time. Only with a clear understanding of the relative growth of dark matter and stellar components and the shape of the combined gravitational potential could we assess the resulting impact on the evolution of the normalization of the mass fundamental plane. 

Potentially the most interesting implication of this work in the context of observational studies of galaxies at high-redshift is that it suggests that given the measured stellar mass and size of a galaxy, regardless of its light profile shape, whether it is a disk or spheroidal, and whether it is forming stars or not, we can predict its stellar dynamics. These dynamics can be used to create a census of galaxies through cosmic time as a function of their velocity dispersions \citep[see e.g.][]{bezanson:11,bezanson:12}. Flexibility with respect to galaxy morphology is particularly important in the context of building evidence that massive galaxies were more disklike, based on axis ratio distributions \citep[e.g.][]{wel:11,weinzirl:11,bruce:12,chevance:12,chang:13} and S\'ersic indices \citep[e.g.][]{dokkum:10,patel:13a}. 

Furthermore, when following populations of galaxies through cosmic time, dynamics are likely to be the most stable property of individual galaxies with time \citep[see e.g.][]{bezanson:11,bezanson:12, belli:14a,belli:14b}. Under this assumption, the ability to infer the dynamics of a galaxy from less observationally expensive data is invaluable. As an example, there is a growing body of literature following the detection and verification of compact quiescent galaxies at $z\gtrsim1.5$ \citep[e.g.][]{daddi:05,trujillo:06,dokkumnic:08,szomoru:10,sande:13} and discussion of possible formation mechanisms, particularly the identification of star-forming progenitors \citep[e.g.][]{barro:13a,barro:13b,patel:13a,toft:14}. Therefore, the best way to connect progenitors and these descendants may be by their dynamics. Ideally dynamics would be directly measured using deep spectroscopy, however this is extremely challenging even in optimal cases. At $z\gtrsim2$, stellar absorption line kinematics have only been measured directly for roughly a dozen quiescent galaxies \citep[e.g.][]{dokkumnature:09,sande:11,onodera:12,sande:13,toft:12,belli:14b} and only indirectly from gas dynamics for star-forming or sub-mm galaxies \citep[e.g.][]{tacconi:08,toft:14,barro:14,nelson:14}. If the existence of a common mass fundamental plane extends to these early times, it implies that we can efficiently connect star-forming and quenched galaxies using inferred dynamics with relative ease from excellent, and existing, HST data from the CANDELS \citep{candels,candelsb} and 3DHST \citep{3dhst,skelton3dhst} surveys. 

Finally, star-forming and quenched galaxies lie on a common mass fundamental plane in both of the samples presented in this work, however only the SDSS sample is large enough to probe the relations as a function of morphology. Although there are hints that non-homology affects the normalization of the mass fundamental plane differently as a function of redshift, only with a much larger sample of galaxies, including a full range of star-forming and quiescent disks and spheroidals, could we begin to assess this evolution and evaluate the importance of such subtle effects as morphological and baryonic to dark matter profile evolution through cosmic time. 

\vspace{1 cm}

The authors wish to acknowledge the funding agencies, institutional support, and observatories that supported the collection and analysis of the data included in this work without with this project would not have been possible. We also wish to thank Rik Williams and Ryan Quadri for providing access to the most recent UDS photometric catalog.
 
Support for this work was provided by NASA through Hubble Fellowship grant \#HF-51318.01-A awarded by the Space Telescope Science Institute, which is operated by the Association of Universities for Research in Astronomy, Inc., for NASA, under contract NAS 5-26555. 

Some of the data presented herein were obtained at the W.M. Keck Observatory, which is operated as a scientific partnership among the California Institute of Technology, the University of California and the National Aeronautics and Space Administration. The Observatory was made possible by the generous financial support of the W.M. Keck Foundation. The authors wish to recognize and acknowledge the very significant cultural role and reverence that the summit of Mauna Kea has always had within the indigenous Hawaiian community.  We are most fortunate to have the opportunity to conduct observations from this mountain.  The analysis pipeline used to reduce the DEIMOS data was developed at UC Berkeley with support from NSF grant AST-0071048. 

Funding for the SDSS and SDSS-II has been provided by the Alfred P. Sloan Foundation, the Participating Institutions, the National Science Foundation, the U.S. Department of Energy, the National Aeronautics and Space Administration, the Japanese Monbukagakusho, the Max Planck Society, and the Higher Education Funding Council for England. The SDSS Web Site is http://www.sdss.org/. The SDSS is managed by the Astrophysical Research Consortium for the Participating Institutions. The Participating Institutions are the American Museum of Natural History, Astrophysical Institute Potsdam, University of Basel, University of Cambridge, Case Western Reserve University, University of Chicago, Drexel University, Fermilab, the Institute for Advanced Study, the Japan Participation Group, Johns Hopkins University, the Joint Institute for Nuclear Astrophysics, the Kavli Institute for Particle Astrophysics and Cosmology, the Korean Scientist Group, the Chinese Academy of Sciences (LAMOST), Los Alamos National Laboratory, the Max-Planck-Institute for Astronomy (MPIA), the Max-Planck-Institute for Astrophysics (MPA), New Mexico State University, Ohio State University, University of Pittsburgh, University of Portsmouth, Princeton University, the United States Naval Observatory, and the University of Washington.

This work is based partially on observations taken by the 3D-HST Treasury Program (GO 12177 and 12328) with the NASA/ESA HST, which is operated by the Association of Universities for Research in Astronomy, Inc., under NASA contract NAS5-26555.

\appendix

\section{The Effects of Inclination on Measured Velocity Dispersion} \label{sec:inclination}

\begin{figure*}[t]
\centering
\includegraphics[width=0.45\textwidth]{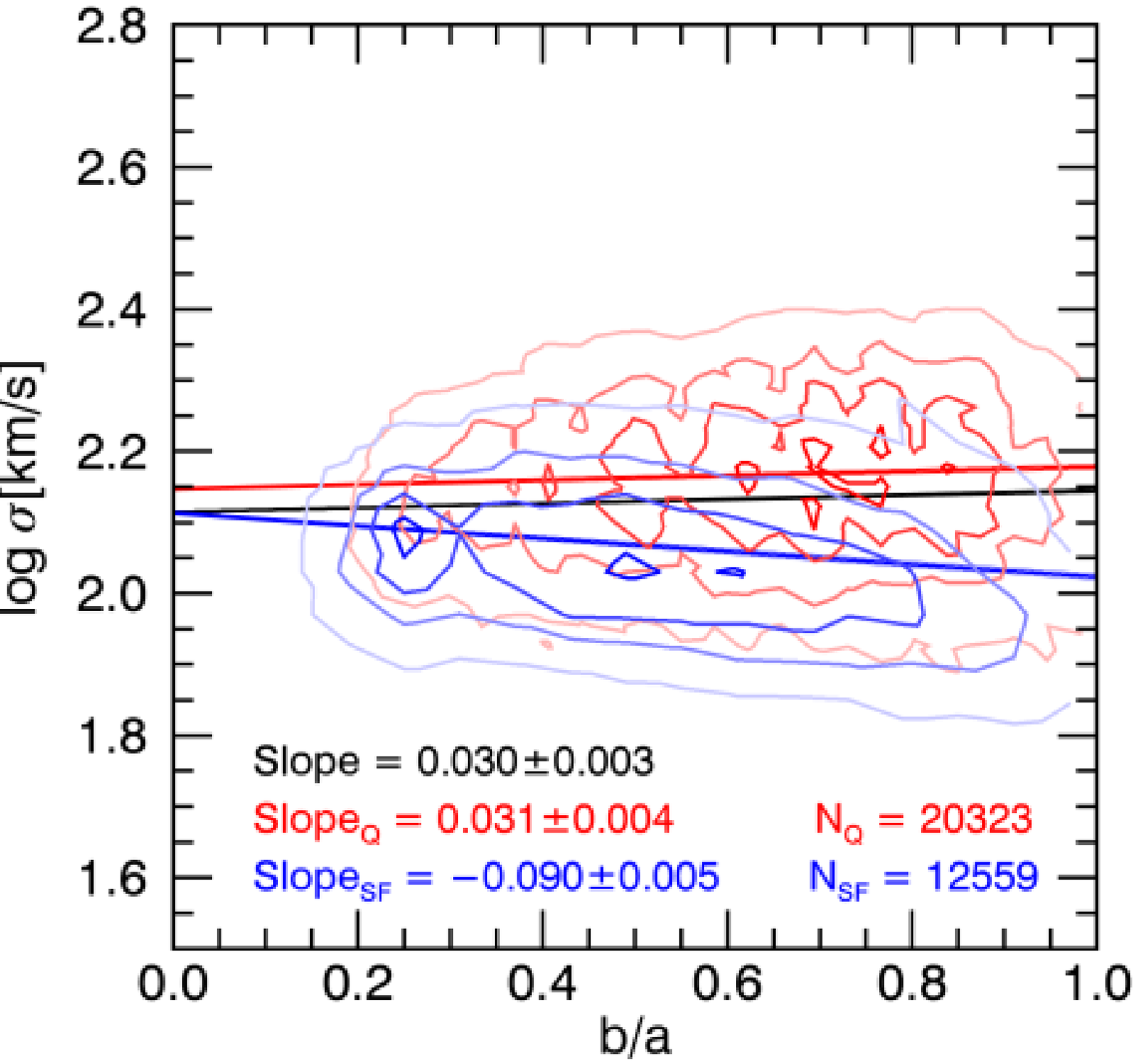}
\includegraphics[width=0.45\textwidth]{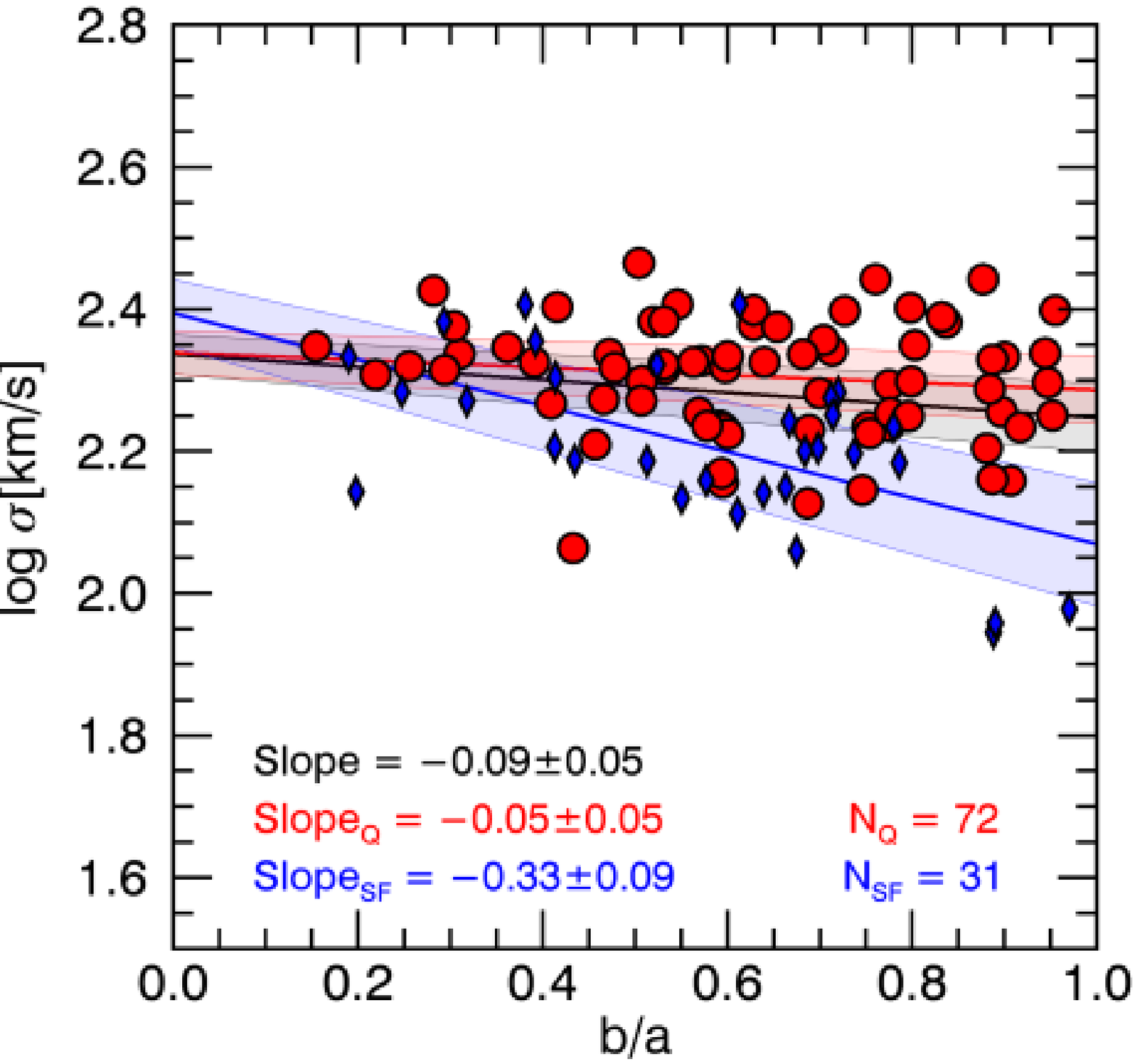}
\caption{Velocity dispersion vs. axis ratio ($b/a$) for galaxies in the SDSS (left panel) and at $z\sim0$ (right panel). In the left panel, the distribution of quiescent galaxies and star-forming galaxies are indicated by red and blue contours, in the right panel by red and blue points. Linear fits are included as black (all galaxies), red (quiescent galaxies), and blue (star-forming galaxies) solid lines. Quiescent galaxies do not exhibit trends as a function of projected axis ratio, while rounder (higher $b/a$) galaxies have slightly lower measured velocity dispersions at both redshifts. This trend is slightly stronger at $z\sim0.7$ (right). These data suggest that inclination contributes subtlety to the scatter in scaling relations of star-forming galaxies.}
\label{fig:sigma_ba}
\end{figure*}

\begin{figure*}[t]
\centering
\includegraphics[width=0.95\textwidth]{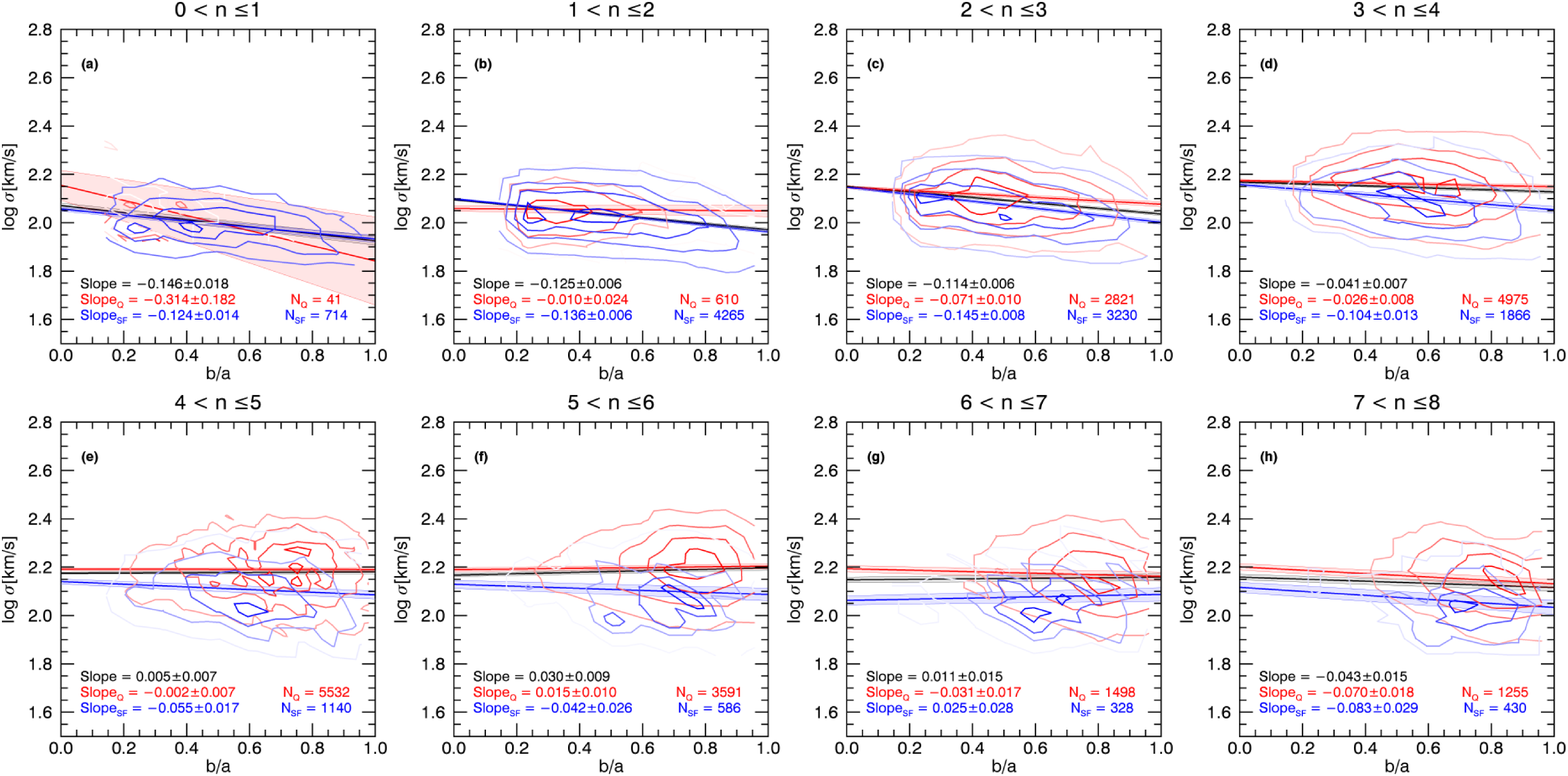}
\caption{Velocity dispersion vs. axis ratio ($b/a$) for galaxies in the SDSS, separated by best-fit S\'{e}rsic index. Annotations and symbols are as in Figure \ref{fig:sigma_ba}. For galaxies in the SDSS, it is apparent that the weak trends in velocity dispersion with axis ratio are important only for galaxies with more disklike ($n\lesssim4$), particularly for those that are star-forming.}
\label{fig:sigma_ba_n_sdss}
\end{figure*}

In \S \ref{subsec:completeness} we introduced a theoretical dependence of the measured velocity dispersion of a disk galaxy on observed inclination (Equation \ref{eq:sigmarot}).  In this formulation, measured velocity dispersion will be a combination of intrinsic dispersion and inclination corrected rotational velocity.  However, increasing intrinsic velocity dispersion, potentially due to the presence of a significant bulge component, would minimize the importance of inclination on the measured velocity dispersions. Although the inclination of a galaxy cannot be directly measured from the available data, we can use the projected axis ratio as a proxy measurement. In the following figures, we assess trends in measured velocity dispersion as a function of axis ratio, and therefore inclination, for the entire population of galaxies and separated by stellar populations and morphologies.

Figure \ref{fig:sigma_ba}, shows trends in measured velocity dispersion as a function of axis ratio for galaxies in the SDSS (left panel) and at $z\sim0.7$ (right panel). As in the previous figures, quiescent galaxies are indicated by red contours (SDSS) or circles ($z\sim0.7$); star-forming galaxies are indicated by blue contours or symbols. Linear fits to the full population (black), quiescent (red), and star-forming (blue) galaxies are included as solid lines. Errors in the linear fits are estimated via bootstrap resampling of each sample and the $1\sigma$ range of the fits are indicated by the shaded regions. Slopes of the lines are indicated in the lower left corner of each panel. At both redshifts, there is little to no statistically significant trend in velocity dispersion with projected axis ratio for quiescent galaxies. However, there is a weak trend at each redshift for star-forming galaxies such that velocity dispersion decreases slightly for rounder galaxies. This effect is small, suggesting that although inclination does impact measured velocity dispersions, it is a very weak effect. In the SDSS, the mean change in velocity dispersion across the full range in axis ratio (from $b/a\sim0.2$ to $b/a\sim1$) implies a discrepancy of only $0.07\unit{dex}$ or a factor of $1.18$ in velocity dispersion. This effect appears to be stronger at $z\sim0.7$, at which the same change in $b/a$ would imply $\sim0.26\unit{dex}$ difference or factor of $\sim1.8$ in velocity dispersion. However, we note that the potential incompleteness of this sample for rounder galaxies may impact the measurement of this trend. Furthermore, the method of measuring galaxy dynamics differs subtly between the two samples: in the SDSS, the velocity dispersion is measured within the $3''$ fiber, whereas the $z\sim0.7$ spectra are integrated across a $1''$ slit. At $z\sim0.06$, $3''$ corresponds to a physical scale of $\sim3.5\unit{kpc}$, compared to $\sim7\unit{kpc}$ of the $1''$ slit at $z\sim0.7$. Additionally, given the size evolution of galaxies, this larger physical aperture corresponds to an even larger fraction of the galaxies that are being probed in the higher redshift sample. This suggests within the centers of galaxies, such as those probed by the SDSS fibers, the dynamics are more dominated by central bulges. However, at higher redshift, when the dynamics are probed in larger apertures, the effects of inclination will be stronger and could contribute to the scatter in such dynamical scaling relations.

In Figure \ref{fig:sigma_ba_n_sdss}, we evaluate the dependence of this trend on structural parameters (in the SDSS only), splitting galaxies into bins of S\'ersic index. Color-coding, notations, and symbols are as in Figure \ref{fig:sigma_ba}. In these panels, we see that the trends with axis ratio are stronger for more disklike galaxies (at lower S\'ersic n), for both star-forming and quiescent galaxies. Above $n\sim4$, the trends all but disappear. This is partially due to the fact that these disky galaxies likely have more oblate structures and therefore axis ratio will be a more sensitive probe of inclination. Additionally, one would expect these galaxies to have more rotational support (higher $v/\sigma$), which determines the contribution of inclination to Equation \ref{eq:sigmarot}. Although this paper demonstrates that the scaling relations of massive galaxies can be determined without the inclusion of an inclination correction, we conclude that the inclination of galaxies with disklike morphologies will contribute a small amount of scatter to the relations. This may become increasingly important for higher redshift studies, as the effect will likely increase with the physical scale of the spectroscopic aperture relative to the decreasing sizes of evolving galaxy populations.

\section{Does the mass Faber-Jackson relation tighten for separate galaxy populations?} \label{sec:SFQmassfj}

The stellar mass Faber-Jackson relation for star-forming and quiescent galaxies differs in normalization and slope, therefore the measured relation for the overall population of galaxies at a given redshift will be an average of the two relations. In \S 5, we demonstrated that the scatter about the Faber-Jackson relation is greater than the scatter in relations which incorporate galaxy sizes. Locally, the Fundamental Plane for elliptical galaxies exhibits less scatter than its projection, the Faber-Jackson relation, however in this Appendix we investigate whether some of this scatter is due to adopting a uniform definition of the Faber-Jackson relation for both galaxy populations. 

\begin{figure*}[t]
\centering
\includegraphics[width=0.32\textwidth]{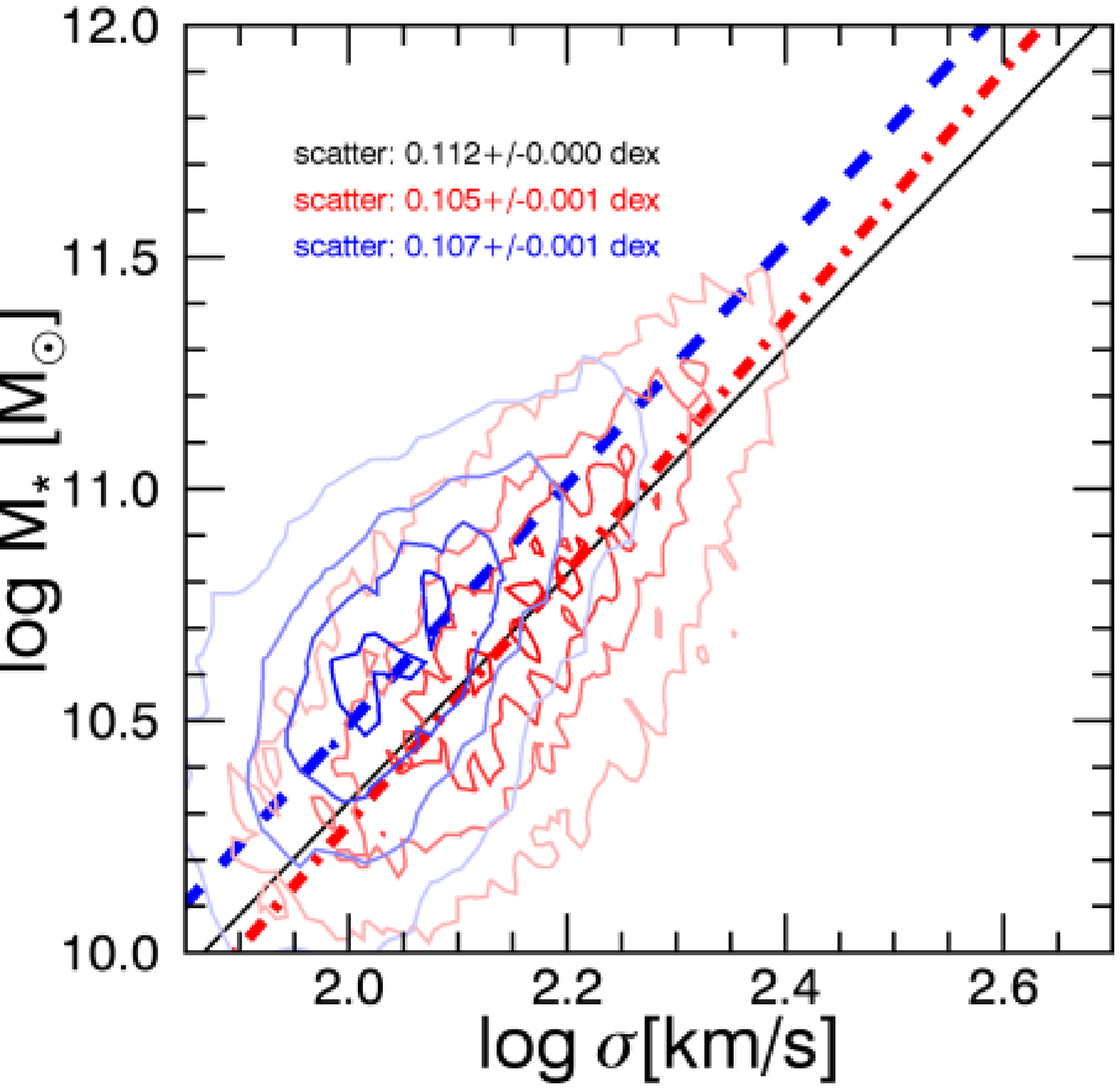}
\includegraphics[width=0.32\textwidth]{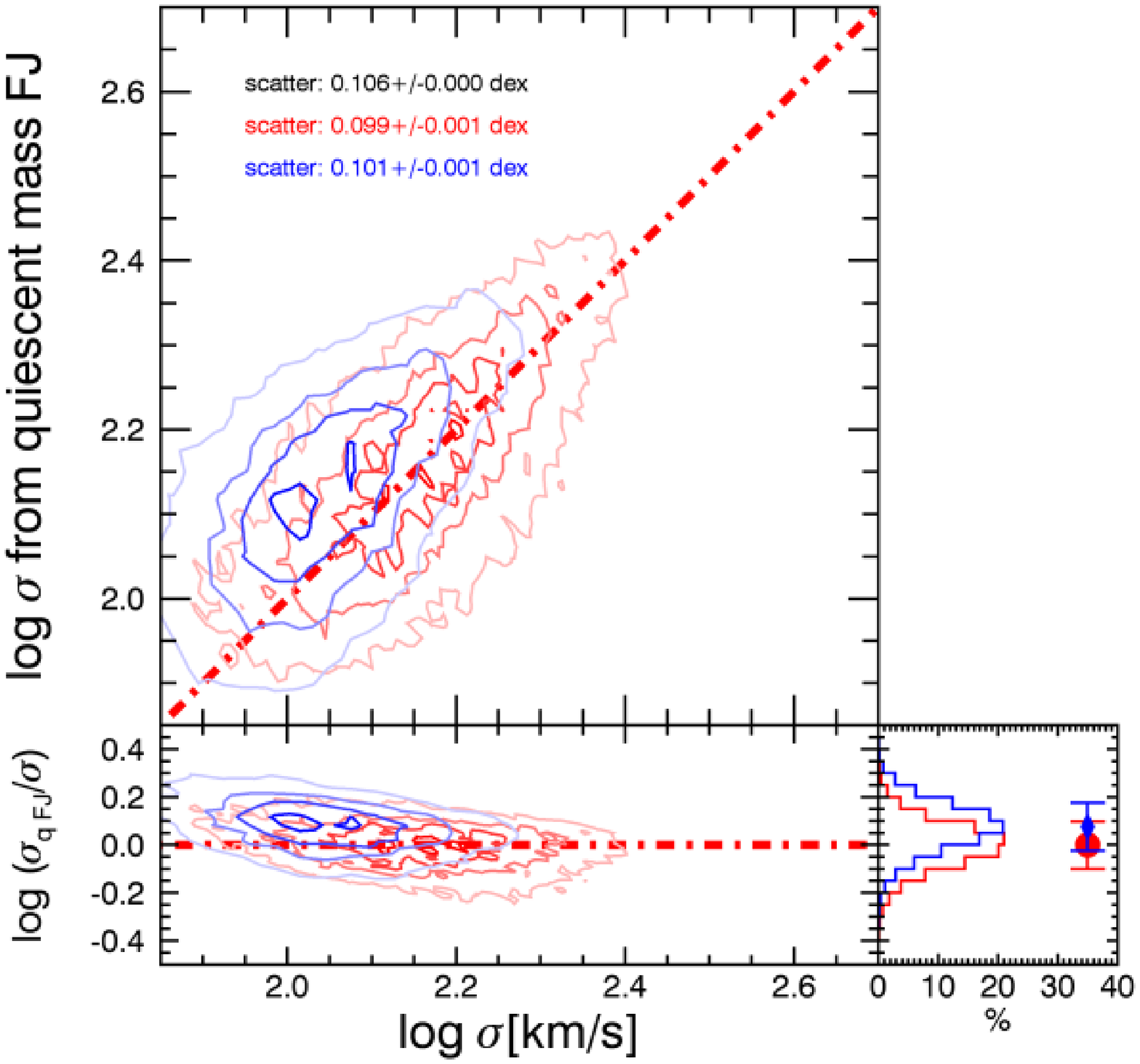}
\includegraphics[width=0.32\textwidth]{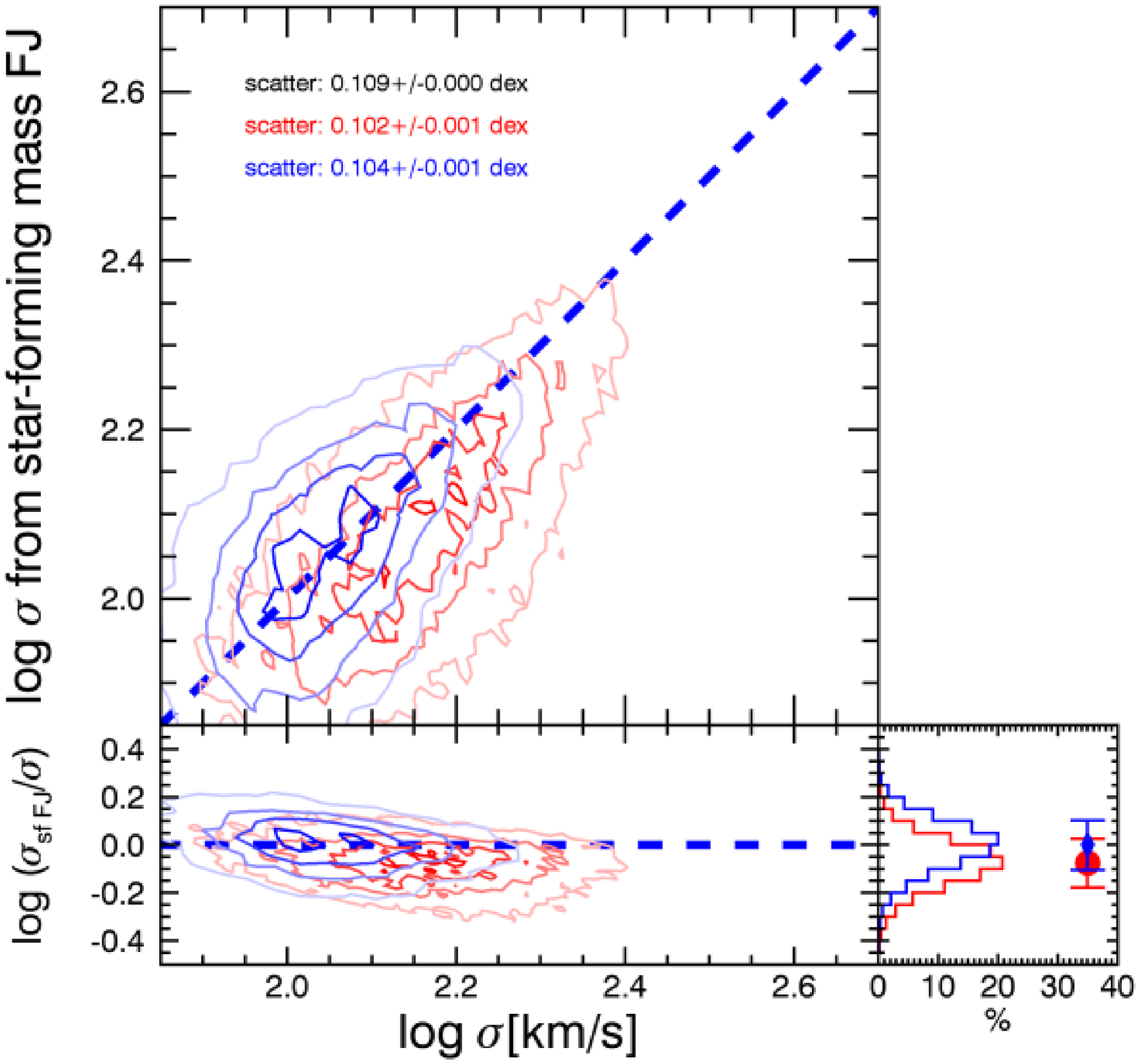}
\caption{Mass Faber-Jackson relation (Left Panel) in the SDSS. Best-fit relation is measured for all galaxies (solid black line), but differs slightly in slope and normalization for star-forming (dashed blue line) and quiescent (dot dashed red line) galaxies. Expanding upon Figure\,\ref{fig:sigmasigma_sdss}, we demonstrate that the scatter in the velocity dispersion predicted by the mass Faber-Jackson relation versus measured velocity dispersion does not decrease by adopting the best-fit relation to star-forming (middle panel) or quiescent (right panel) galaxies.}
\label{fig:faberjackson}
\end{figure*}

Figure\, \ref{fig:faberjackson}a shows the Faber-Jackson relation for galaxies in the SDSS. The black line indicates the overall scaling relation (used in Figure\, \ref{fig:sigmasigma_sdss}a, the dashed blue and red lines indicate the fits to the star-forming and quiescent populations. Scatter about the velocity dispersion predicted by these relations are shown in the center (quiescent) and right (star-forming) panels. Scatter for all galaxies (in black), quiescent galaxies (red), and star-forming galaxies (blue) is indicated in the upper left corner of each panel. It is clear that although using the individual relations reduces the scatter in the mass Faber-Jackson relation very slightly ($\sim0.01\unit{dex}$), the three dimensional planes exhibit far lower scatter for star-forming and quiescent galaxies alike.

\section{Three Parameter Scaling Relations: Scatter about $\log R_e$}\label{sec:scatter_re}

\begin{figure*}[!t]
\centering
\includegraphics[width=0.32\textwidth]{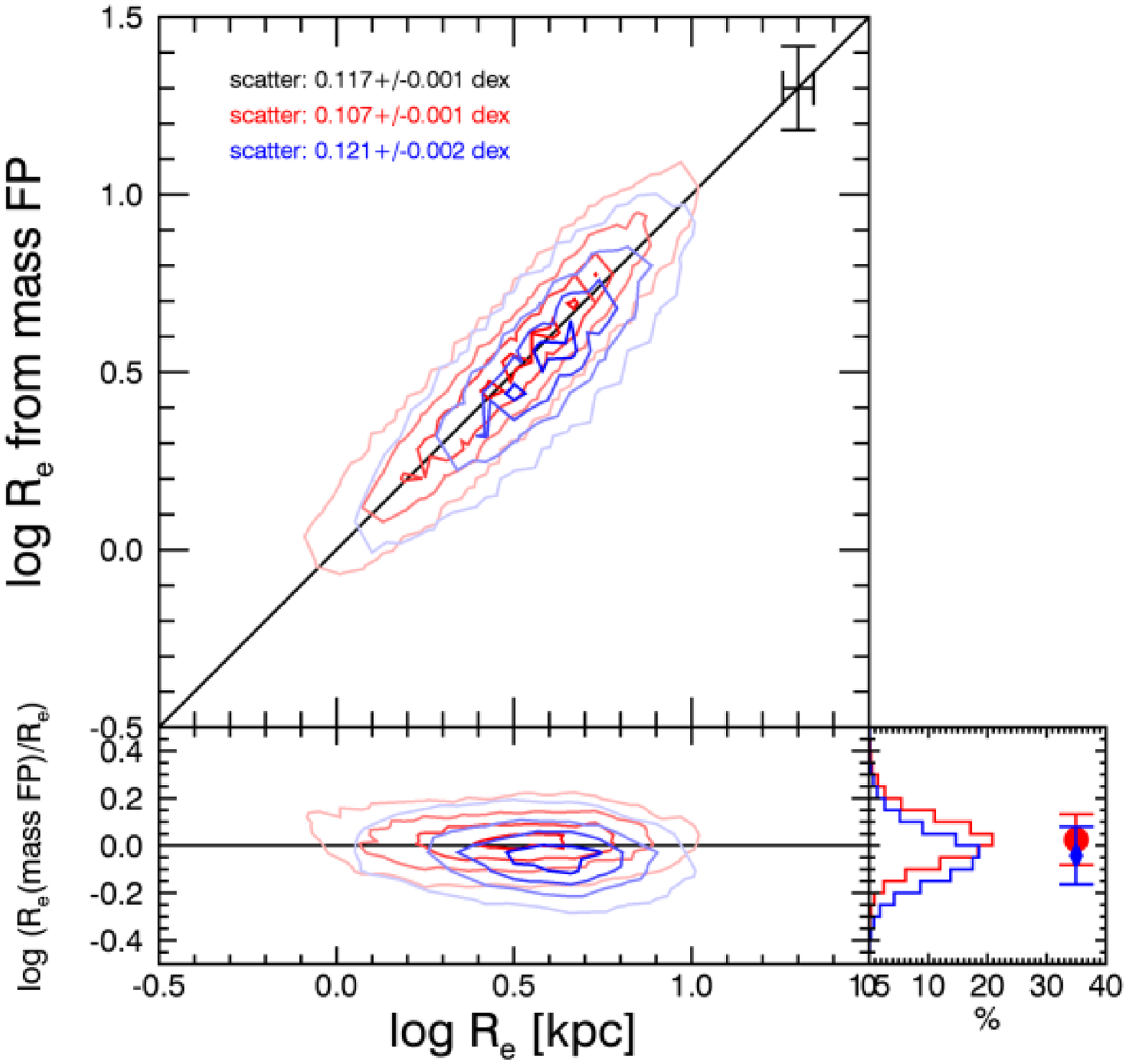}
\includegraphics[width=0.32\textwidth]{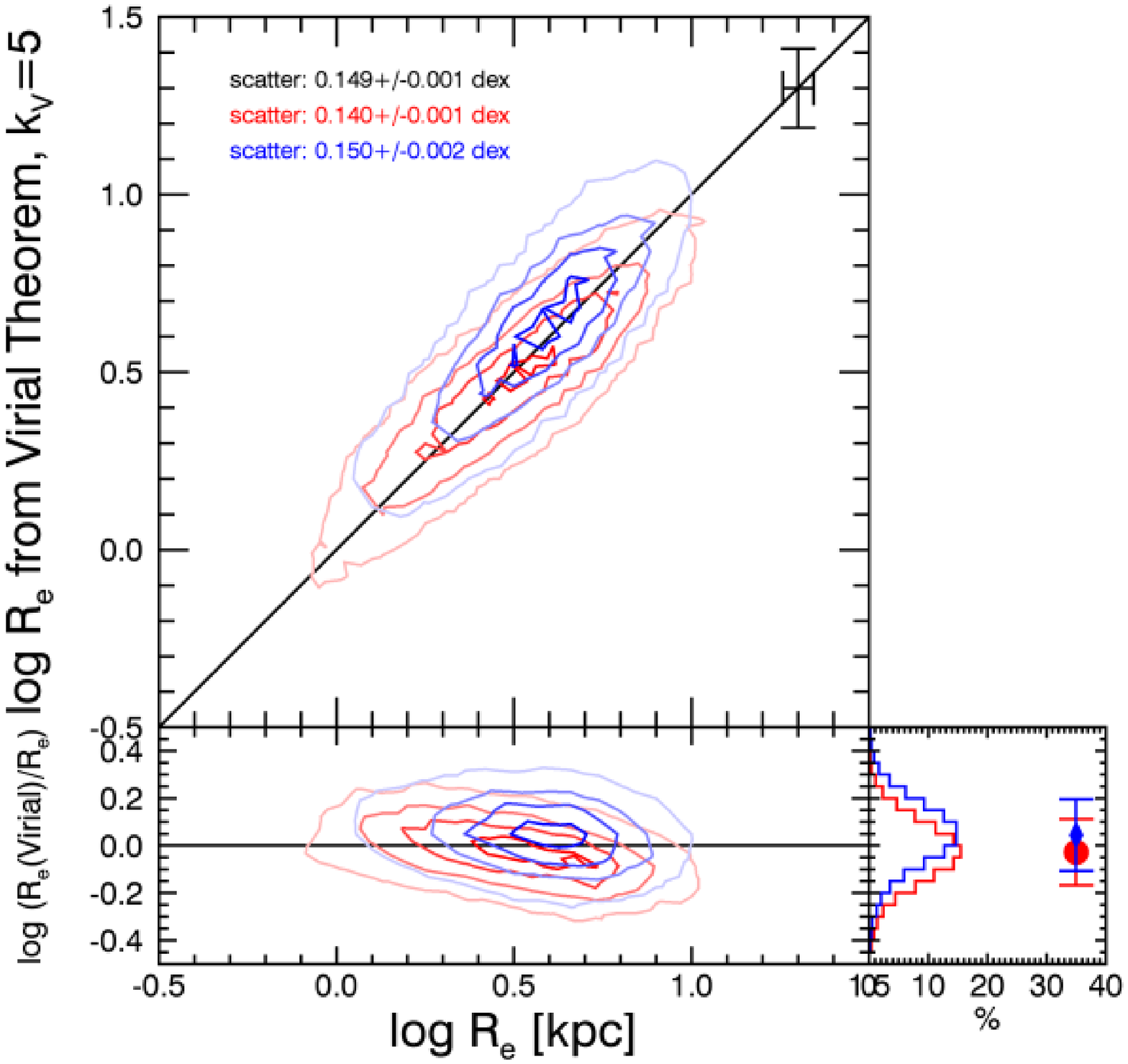}
\includegraphics[width=0.32\textwidth]{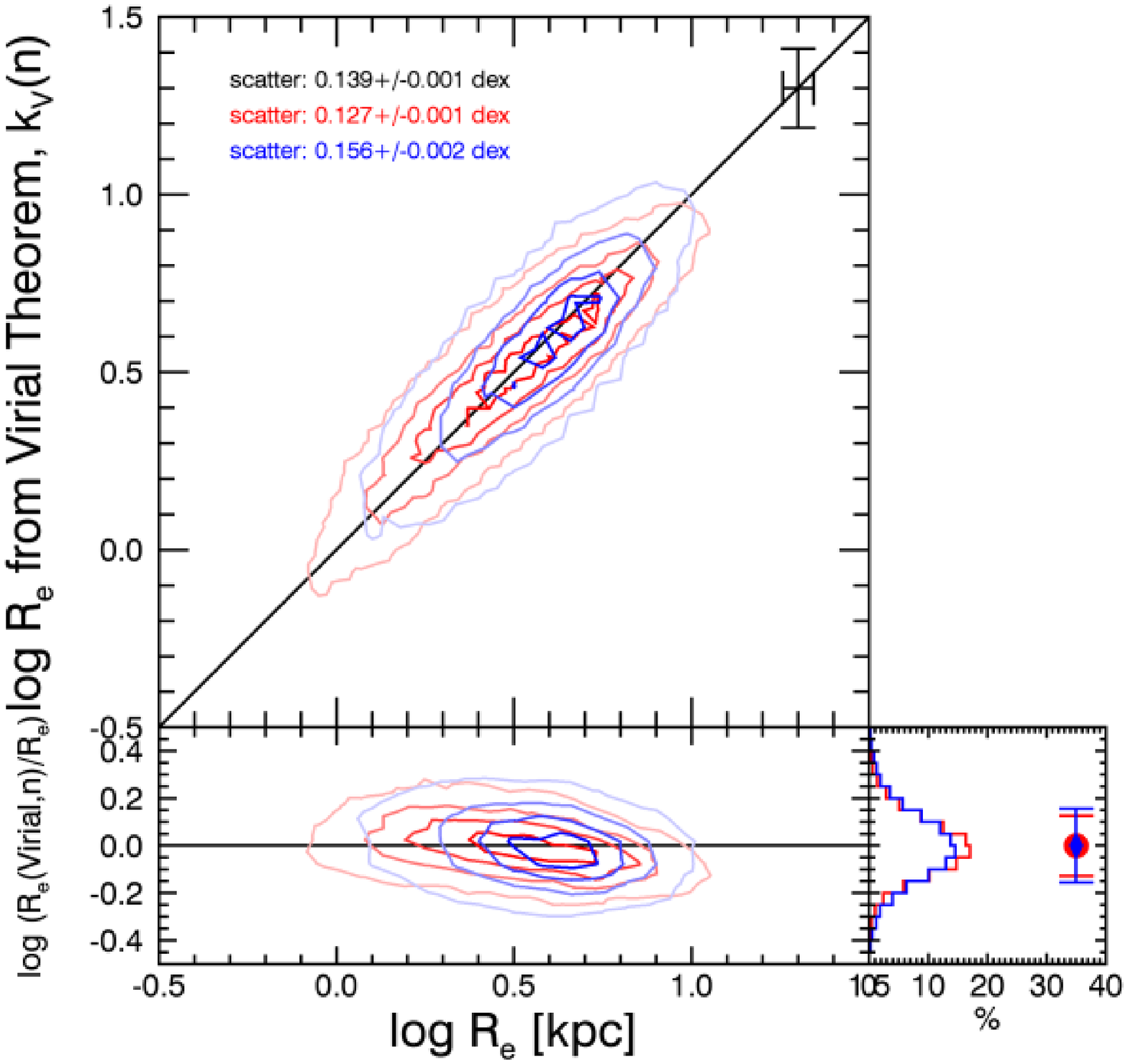}
\caption{Scatter in $\log R_e$ in the SDSS from (left) the mass fundamental plane (center) the Virial theorem with a fixed constant and (right) with a constant that depends on S\'ersic index. The vertical axis in each is determined by the equations reported in Table \ref{tab:re_re}.}
\label{fig:re_re_sdss}
\end{figure*}

\begin{figure*}[!h]
\centering
\includegraphics[width=0.32\textwidth]{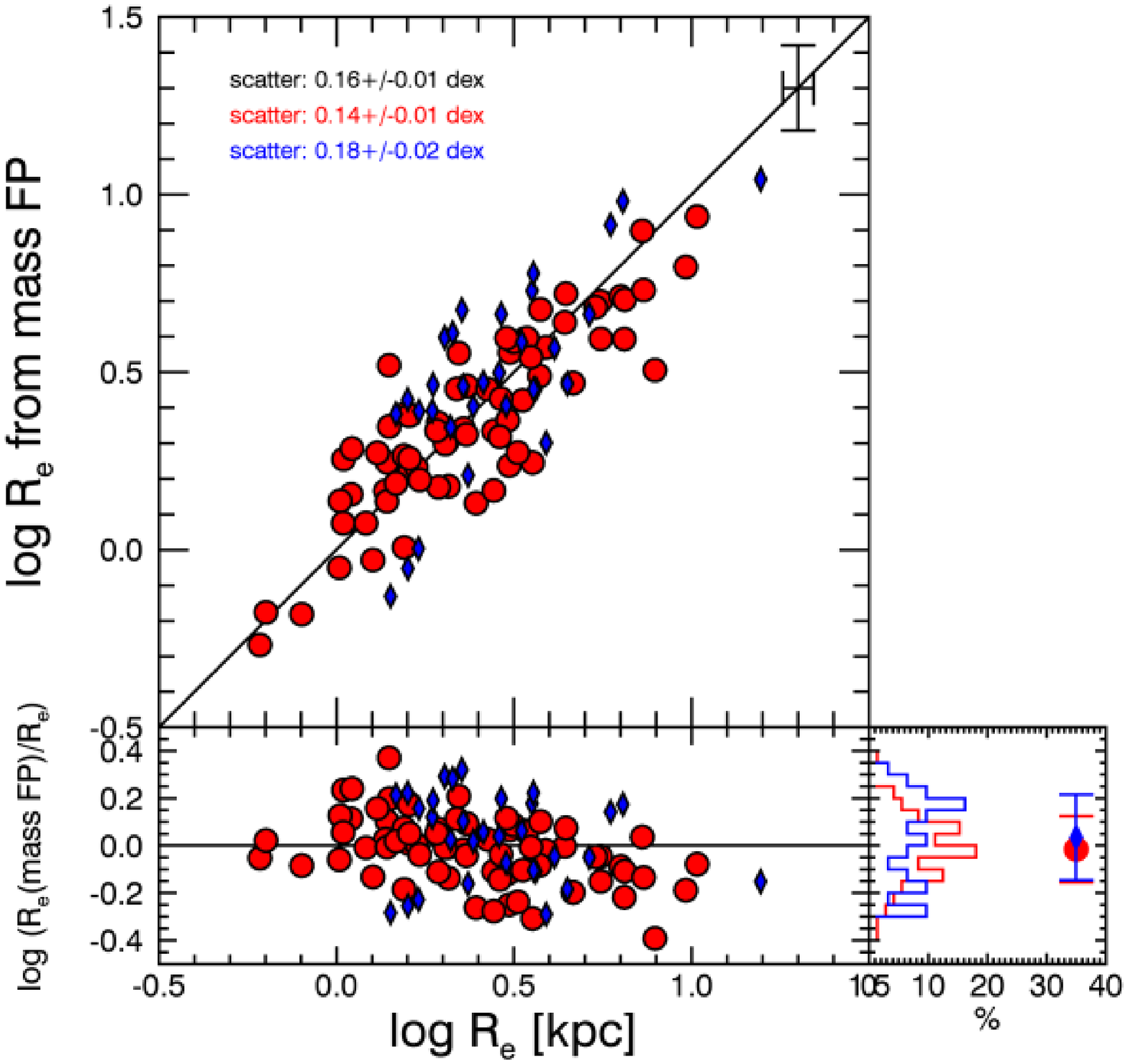}
\includegraphics[width=0.32\textwidth]{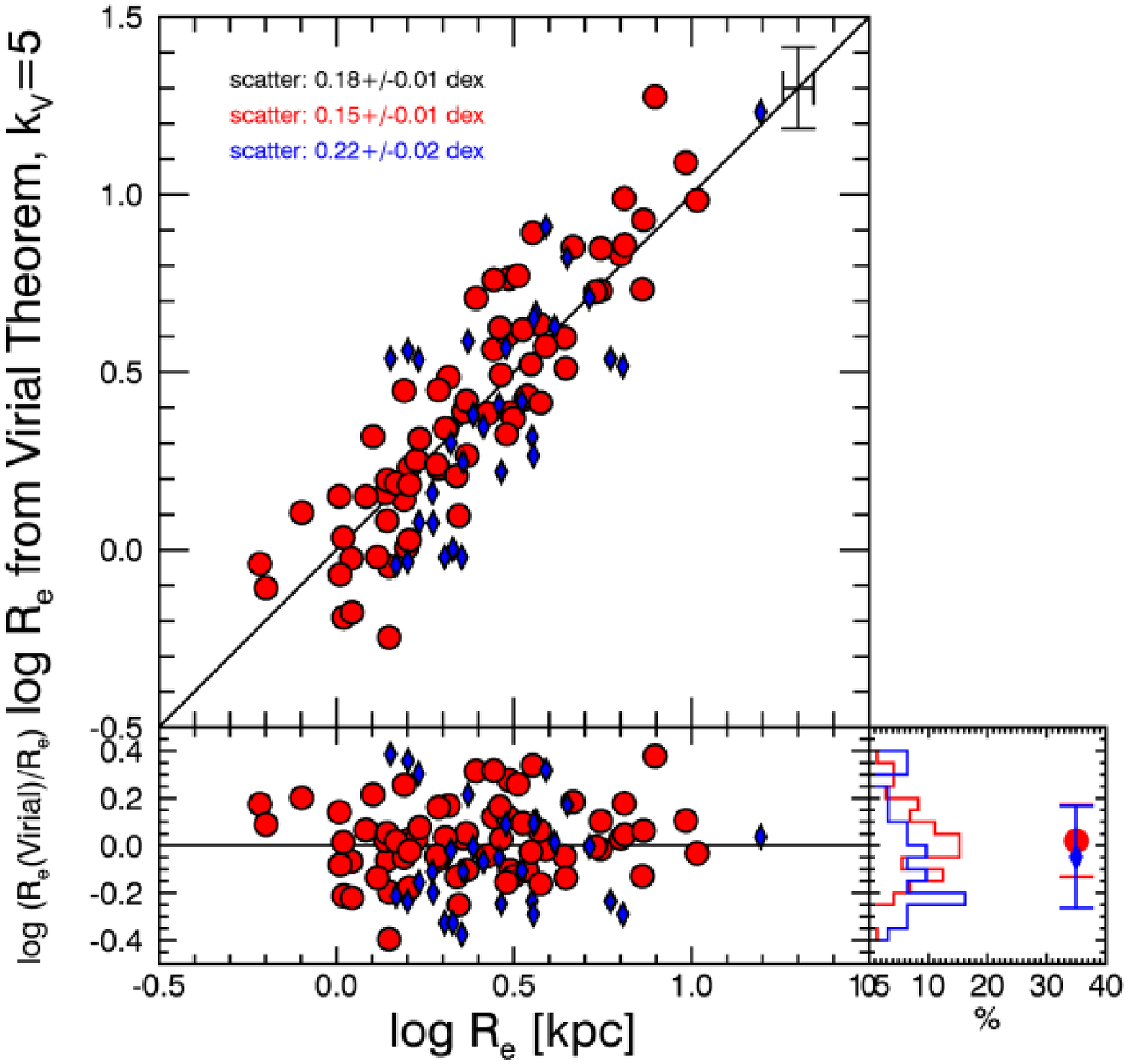}
\includegraphics[width=0.32\textwidth]{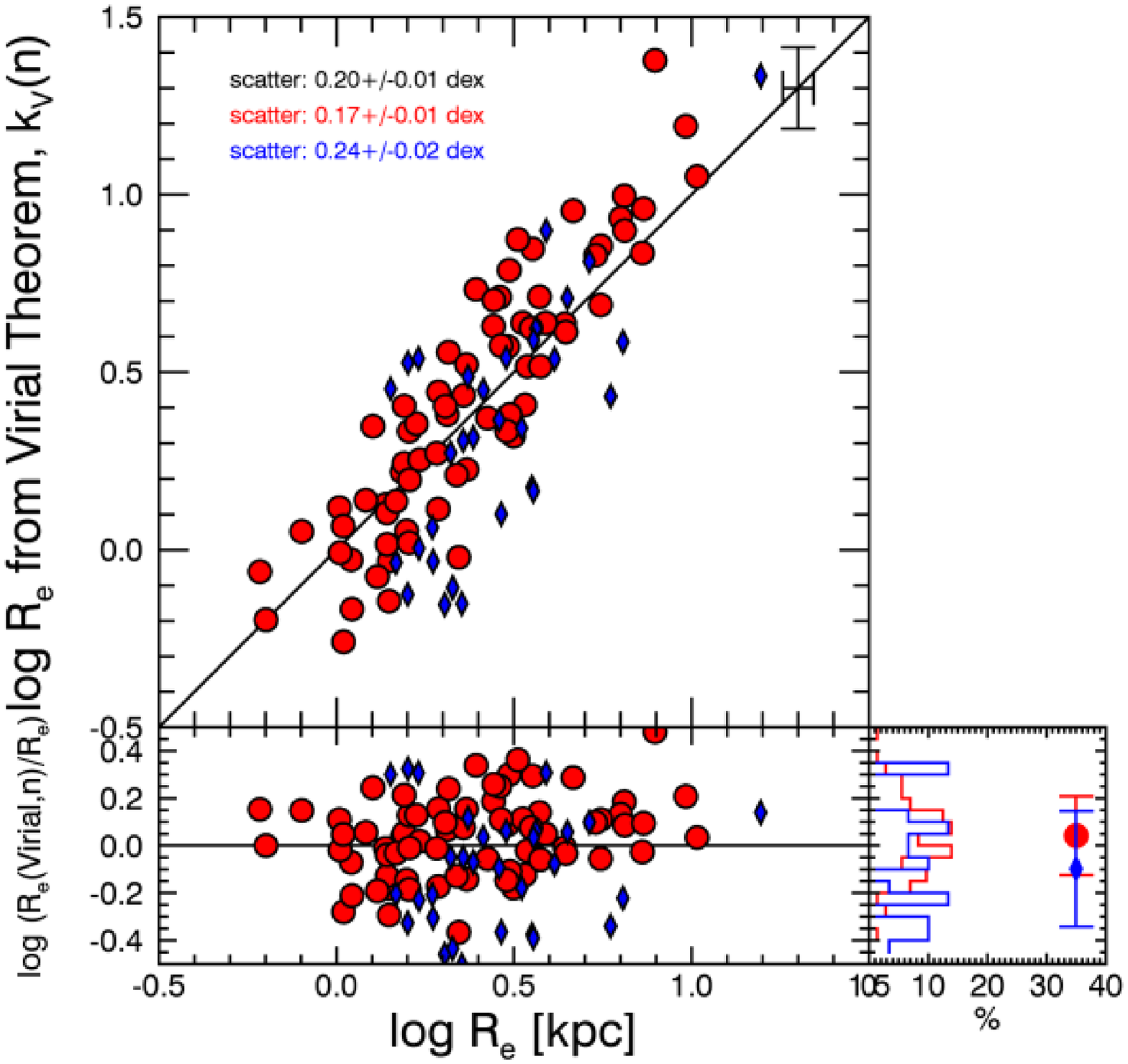}
\caption{Scatter in $\log R_e$ at $z\sim0.7$ from the mass fundamental plane  (left panel) and the Vrial theorem with a fixed constant (center panel) and a S\'ersic-dependent content (right panel).}
\label{fig:re_re}
\end{figure*}

In this Appendix, we investigate the scatter in the mass fundamental plane and the Virial relations in an additional projection (see also Figures \ref{fig:massfp}, \ref{fig:massfp_n_sdss}, and \ref{fig:massfp_n}).  For each relation, we calculate the scatter about $\log R_e$. The slopes of the scaling relations are held fixed, but in each case the normalization is allowed to vary, although this will not affect the scatter. Best-fit normalizations, measured and intrinsic scatter, as determined through Monte-Carlo simulations, are included in Table \ref{tab:re_re} . Figure \ref{fig:re_re_sdss} illustrates the scatter about $\log R_e$ in the SDSS. The left panel includes sizes inferred by the mass fundamental plane, the center panel from the Virial theorem with a fixed constant and the right panel with a S\'ersic-dependent constant. Figure \ref{fig:re_re} follows the same layout, but includes galaxies in the DEIMOS sample.  Unlike the scatter in velocity dispersion (see Figures \ref{fig:sigmasigma_sdss} and \ref{fig:sigmasigma}), the scatter in the mass fundamental plane is significantly tighter than that of the Virial relations. However, we expect this to be due, at least in part, to correlated errors in sizes and stellar mass surface densities. In this projection, the difference in scatter between star-forming and quiescent populations is also significant, $0.01-0.03\unit{dex}$ at $z\sim0$ and up to $\sim0.07\unit{dex}$ at $z\sim0.7$, particularly when using $k_{\mathrm{V}}(n)$. 

\begin{deluxetable*}{lcccccc}
\tabletypesize{\scriptsize}
\tablecaption{Measured and Intrinsic Scatter in $\log R_e$ from Scaling Relations}
\tablehead{
\colhead{Relation} & \colhead{rms} & \colhead{$\mathrm{rms_{int}}$} & \colhead{rms (Q)} & \colhead{$\mathrm{rms_{int}}$ (Q)} & \colhead{rms (SF)} & \colhead{$\mathrm{rms_{int}}$ (SF)} }
\startdata
\cutinhead{SDSS: $z\sim0$}
Mass Fundamental Plane & 0.117 $\pm$ 0.001 & 0.088 & 0.107 $\pm$ 0.001 & 0.074 & 0.121 $\pm$ 0.002 & 0.094 \\
$\log R_e = 1.63\log\sigma - 0.84\log\Sigma_{\star} + 4.496$ \\ 
Virial Theorem ($k_V=5$) & 0.149 $\pm$ 0.001 & 0.130 & 0.140 $\pm$ 0.001 & 0.121 & 0.150 $\pm$ 0.002 & 0.132 \\
$\log R_e = \log M_{\star} - 2.0\log\sigma - 5.695$ \\
Virial Theorem ($k_{\mathrm{V}}(n)$) & 0.139 $\pm$ 0.001 & 0.118 & 0.127 $\pm$ 0.001 & 0.104 & 0.156 $\pm$ 0.002 & 0.138 \\
$\log R_e = \log M_{\star} - 2.0\log\sigma + \log k_{\mathrm{V}}(n) - 5.078$ \\
\cutinhead{DEIMOS: $z\sim0.7$}
Mass Fundamental Plane & 0.16 $\pm$ 0.01 & 0.13 & 0.14 $\pm$ 0.01 & 0.12 & 0.18 $\pm$ 0.02 & 0.16 \\
$\log R_e = 1.63\log\sigma - 0.84\log\Sigma_{\star} + 4.306$ \\ 
Virial Theorem ($k_V=5$) & 0.18 $\pm$ 0.01 & 0.16 & 0.15 $\pm$ 0.01 & 0.13 & 0.22 $\pm$ 0.02 & 0.20 \\
$\log R_e = \log M_{\star} - 2.0\log\sigma - 5.911$ \\
Virial Theorem ($k_{\mathrm{V}}(n)$) & 0.20 $\pm$ 0.01 & 0.19 & 0.17 $\pm$ 0.01 & 0.15 & 0.24 $\pm$ 0.02 & 0.23 \\
$\log R_e = \log M_{\star} - 2.0\log\sigma + \log k_{\mathrm{V}}(n) - 5.292$ \\
\label{tab:re_re}

\tablecomments{Measured and intrinsic scatter in $\log R_e$ from various scaling relations, for all galaxies and then separately for quiescent (Q) and star-forming (SF) galaxies. Quantitative relations are included in the table and the S\'ersic-dependent Virial constant from \citep{cappellari:06} is approximated by $k_{\mathrm{V}}(n) = 8.87-0.831n+0.0241n^2)$ as in Eq. \ref{eq:sigma_vir_n}.}
\end{deluxetable*}

\section{Virial Estimates of Dynamical Mass}\label{sec:mass_mass}

\begin{figure*}[t]
\centering
\includegraphics[width=0.4\textwidth]{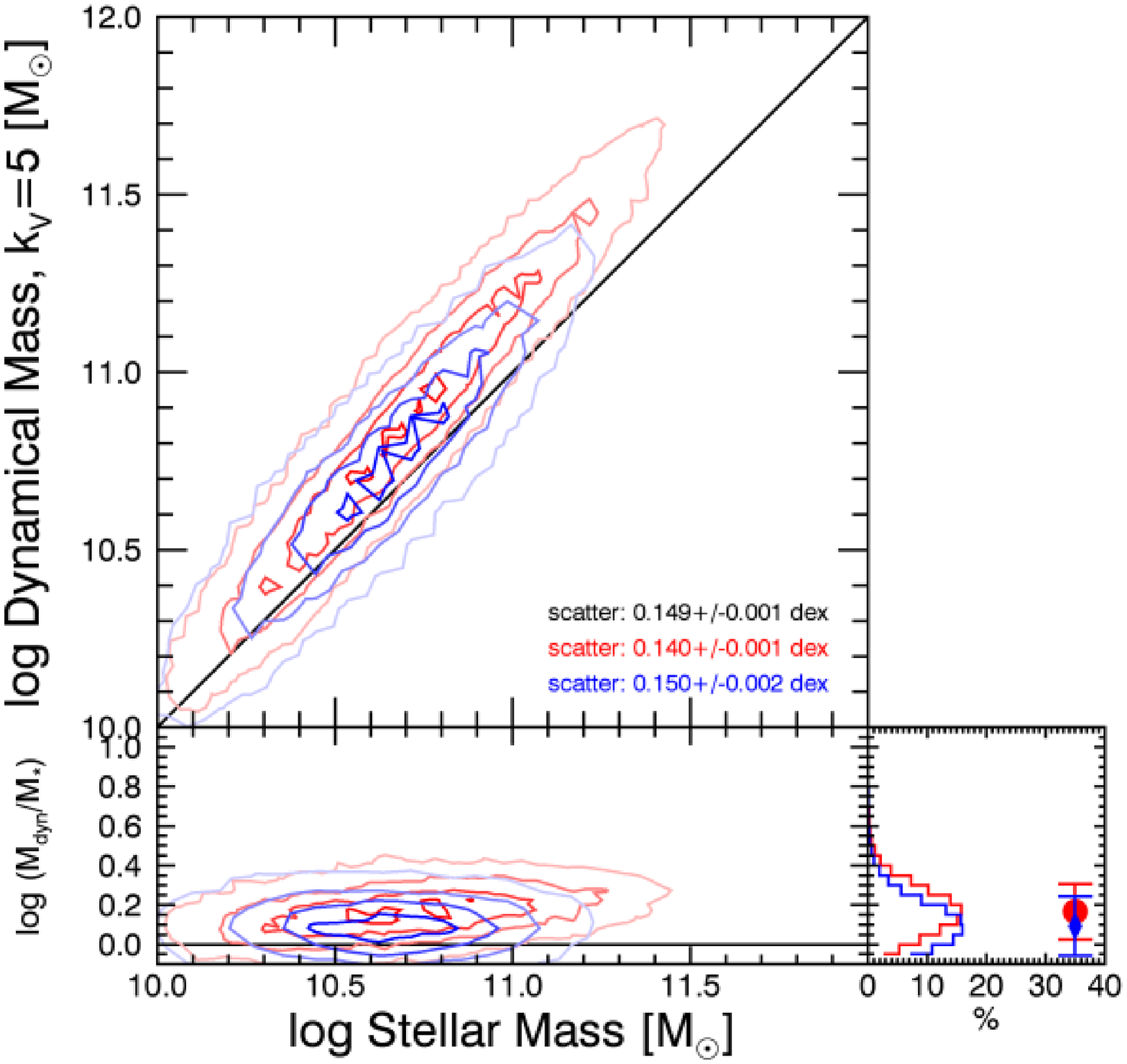}
\includegraphics[width=0.4\textwidth]{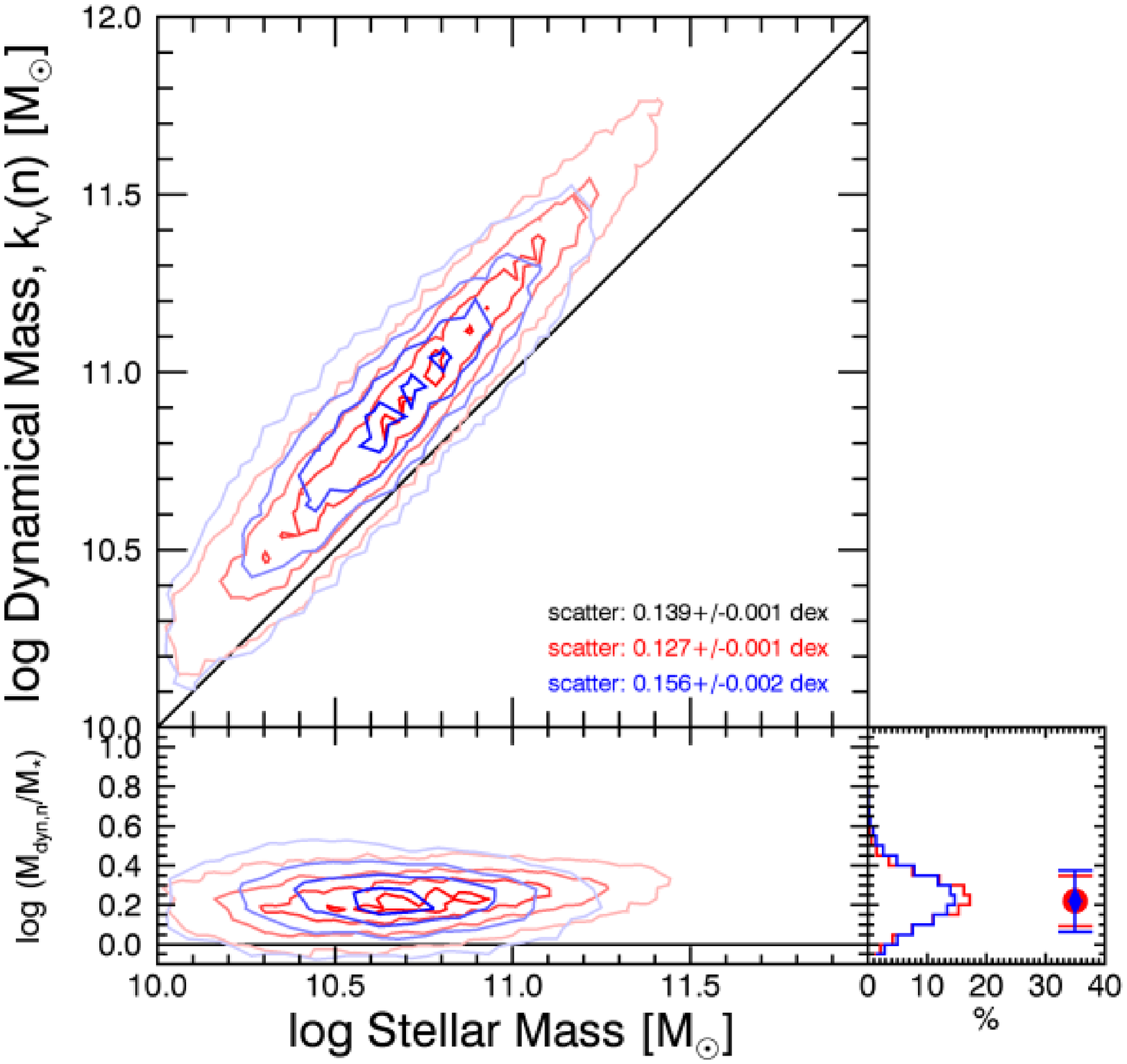}
\caption{Dynamical and stellar masses of galaxies in the SDSS from (left) the Virial theorem with a fixed constant and (right) with a constant that depends on S\'ersic index.}
\label{fig:mass_mass_sdss}
\end{figure*}

\begin{figure*}[h]
\centering
\includegraphics[width=0.4\textwidth]{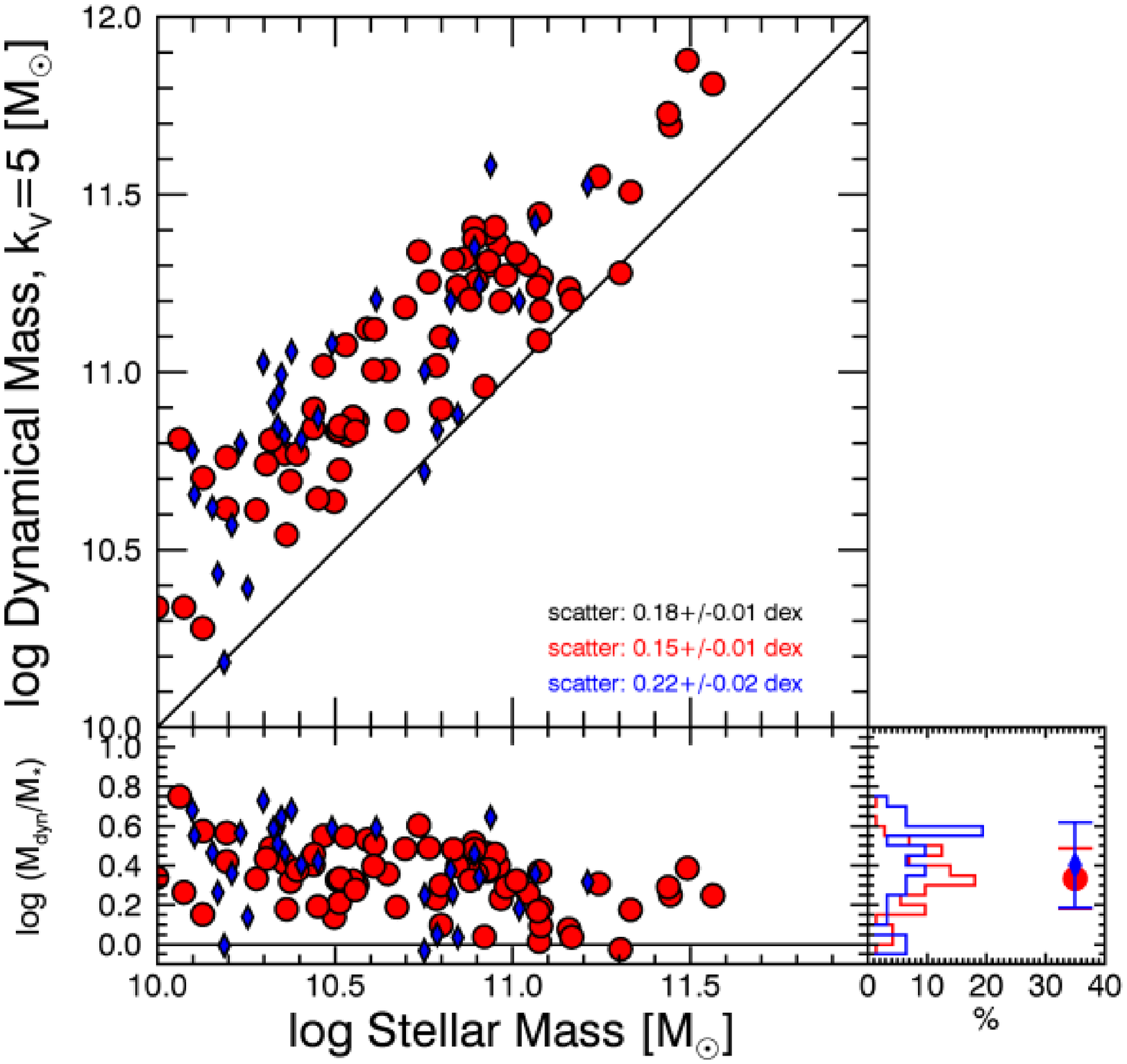}
\includegraphics[width=0.4\textwidth]{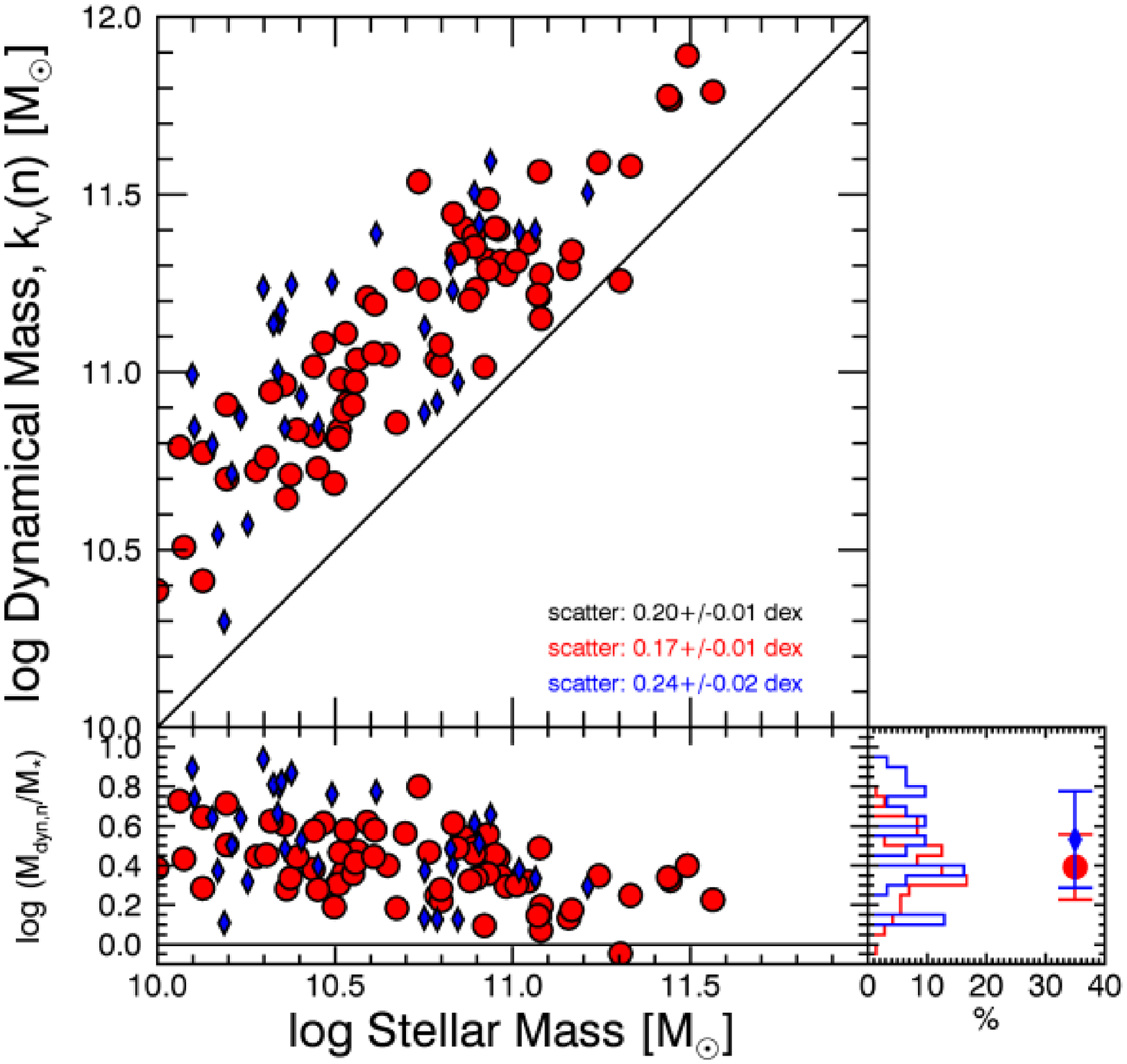}
\caption{Dynamical and stellar masses of galaxies at $z\sim0.7$ from (left) the Virial theorem with a fixed constant and (right) with a constant that depends on S\'ersic index.}
\label{fig:mass_mass}
\end{figure*}

Throughout this Paper, we mention Virial estimates of galaxy dynamical masses. Although all such measurements are calculated based on the Virial theorem, dynamical mass estimates are sensitive to the choice of Virial constant (see \S \ref{sec:sigma}). \citet{cappellari:06} derived an analytic form for the Virial constant based on spherical jeans models of mass-follows-light S\'ersic models (Equation \ref{eq:kv_n}), but found that a single-valued constant $k=5$ is sufficient to describe the dynamical and structural properties within an effective radius of the elliptical galaxies in the local ATLAS3D sample. Furthermore, \citet{taylor:09} demonstrated the importance of accounting for structural differences, as quantified by S\'ersic index, when determining the dynamics of massive galaxies in the SDSS.  This latter work found that the dynamical mass of a galaxy is well-correlated with its stellar mass as long as one adopts a Virial constant that depends on S\'ersic index, despite the structural non-homology of the SDSS galaxy sample. Figures \ref{fig:mass_mass_sdss} and \ref{fig:mass_mass} explicitly show the relationship between stellar mass and dynamical masses in the SDSS and at $z\sim0.7$. The left panel of each figure shows the dynamical mass calculated with a single Virial constant relative to stellar mass, the right panel includes $k_{\mathrm{V}(n)}$.

In the left panel of Figure \ref{fig:mass_mass_sdss} we reproduce a main result from \citet{taylor:09}: the scatter is decreased and overall correlation between dynamical and stellar masses for SDSS galaxies is improved when structural non-homology is included in dynamical mass calculations (in the right panel relative to the left panel).  This relation is less apparent for galaxies in the $z\sim0.7$ sample (Figure \ref{fig:mass_mass}).  In this case, the scatter is similar within the errors for both relations and if anything the scatter increases in the right panel (with $k_{\mathrm{V}}(n)$).  We note that this difference suggests that perhaps the analytic form that we adopt for the S\'ersic-dependence of the Virial constant may no longer be optimal at high-z. However the modeling of these galaxies including differences in extraction apertures, seeing, velocity dispersion profiles used for aperture corrections, and possibly structural evolution is beyond the scope of this paper.


\begin{thebibliography}{106}
\expandafter\ifx\csname natexlab\endcsname\relax\def\natexlab#1{#1}\fi

\bibitem[{{Abazajian} {et~al.}(2009){Abazajian}, {Adelman-McCarthy},
  {Ag{\"u}eros}, {Allam}, {Allende Prieto}, {An}, {Anderson}, {Anderson},
  {Annis}, {Bahcall}, \& et~al.}]{dr7}
{Abazajian}, K.~N., {et~al.} 2009, \apjs, 182, 543

\bibitem[{{Barro} {et~al.}(2013{\natexlab{a}}){Barro}, {Faber},
  {P{\'e}rez-Gonz{\'a}lez}, {Koo}, {Williams}, {Kocevski}, {Trump}, {Mozena},
  {McGrath}, {van der Wel}, {Wuyts}, {Bell}, {Croton}, {Ceverino}, {Dekel},
  {Ashby}, {Cheung}, {Ferguson}, {Fontana}, {Fang}, {Giavalisco}, {Grogin},
  {Guo}, {Hathi}, {Hopkins}, {Huang}, {Koekemoer}, {Kartaltepe}, {Lee},
  {Newman}, {Porter}, {Primack}, {Ryan}, {Rosario}, {Somerville}, {Salvato}, \&
  {Hsu}}]{barro:13a}
{Barro}, G., {et~al.} 2013{\natexlab{a}}, \apj, 765, 104

\bibitem[{{Barro} {et~al.}(2013{\natexlab{b}}){Barro}, {Faber},
  {Perez-Gonzalez}, {Pacifici}, {Trump}, {Koo}, {Wuyts}, {Guo}, {Bell},
  {Dekel}, {Porter}, {Primack}, {Ferguson}, {Ashby}, {Caputi}, {Ceverino},
  {Croton}, {Fazio}, {Giavalisco}, {Hsu}, {Kocevski}, {Koekemoer},
  {Kurczynski}, {Kollipara}, {Lee}, {McIntosh}, {McGrath}, {Moody},
  {Somerville}, {Papovich}, {Salvato}, {Santini}, {Williams}, {Willner}, \&
  {Zolotov}}]{barro:13b}
---. 2013{\natexlab{b}}, arXiv:1311.5559

\bibitem[{{Barro} {et~al.}(2014){Barro}, {Trump}, {Koo}, {Dekel}, {Kassin},
  {Kocevski}, {Faber}, {van der Wel}, {Guo}, {Perez-Gonzalez}, {Toloba},
  {Fang}, {Pacifici}, {Simons}, {Campbell}, {Ceverino}, {Finkelstein},
  {Goodrich}, {Kassis}, {Koekemoer}, {Konidaris}, {Livermore}, {Lyke},
  {Mobasher}, {Nayyeri}, {Peth}, {Primack}, {Rizzi}, {Somerville}, {Wirth}, \&
  {Zolotov}}]{barro:14}
---. 2014, arXiv:1405.7042

\bibitem[{{Bell} \& {de Jong}(2001)}]{bell:01}
{Bell}, E.~F., \& {de Jong}, R.~S. 2001, \apj, 550, 212

\bibitem[{{Bell} {et~al.}(2004){Bell}, {Wolf}, {Meisenheimer}, {Rix}, {Borch},
  {Dye}, {Kleinheinrich}, {Wisotzki}, \& {McIntosh}}]{bell:04}
{Bell}, E.~F., {et~al.} 2004, \apj, 608, 752

\bibitem[{{Belli} {et~al.}(2014{\natexlab{a}}){Belli}, {Newman}, \&
  {Ellis}}]{belli:14a}
{Belli}, S., {Newman}, A.~B., \& {Ellis}, R.~S. 2014{\natexlab{a}}, \apj, 783,
  117

\bibitem[{{Belli} {et~al.}(2014{\natexlab{b}}){Belli}, {Newman}, {Ellis}, \&
  {Konidaris}}]{belli:14b}
{Belli}, S., {Newman}, A.~B., {Ellis}, R.~S., \& {Konidaris}, N.~P.
  2014{\natexlab{b}}, arXiv:1404.4872

\bibitem[{{Bertin} {et~al.}(2002){Bertin}, {Ciotti}, \& {Del
  Principe}}]{bertin:02}
{Bertin}, G., {Ciotti}, L., \& {Del Principe}, M. 2002, \aap, 386, 149

\bibitem[{{Bezanson} {et~al.}(2012){Bezanson}, {van Dokkum}, \&
  {Franx}}]{bezanson:12}
{Bezanson}, R., {van Dokkum}, P., \& {Franx}, M. 2012, \apj, 760, 62

\bibitem[{{Bezanson} {et~al.}(2009){Bezanson}, {van Dokkum}, {Tal},
  {Marchesini}, {Kriek}, {Franx}, \& {Coppi}}]{bezanson:09}
{Bezanson}, R., {van Dokkum}, P.~G., {Tal}, T., {Marchesini}, D., {Kriek}, M.,
  {Franx}, M., \& {Coppi}, P. 2009, \apj, 697, 1290

\bibitem[{{Bezanson} {et~al.}(2013){Bezanson}, {van Dokkum}, {van de Sande},
  {Franx}, {Leja}, \& {Kriek}}]{bezanson:13b}
{Bezanson}, R., {van Dokkum}, P.~G., {van de Sande}, J., {Franx}, M., {Leja},
  J., \& {Kriek}, M. 2013, \apjl, 779, L21

\bibitem[{{Bezanson} {et~al.}(2011){Bezanson}, {van Dokkum}, {Franx},
  {Brammer}, {Brinchmann}, {Kriek}, {Labb{\'e}}, {Quadri}, {Rix}, {van de
  Sande}, {Whitaker}, \& {Williams}}]{bezanson:11}
{Bezanson}, R., {et~al.} 2011, \apjl, 737, L31+

\bibitem[{{Bielby} {et~al.}(2012){Bielby}, {Hudelot}, {McCracken}, {Ilbert},
  {Daddi}, {Le F{\`e}vre}, {Gonzalez-Perez}, {Kneib}, {Marmo}, {Mellier},
  {Salvato}, {Sanders}, \& {Willott}}]{bielby:12}
{Bielby}, R., {et~al.} 2012, \aap, 545, A23

\bibitem[{{Binney}(1978)}]{binney:78}
{Binney}, J. 1978, \mnras, 183, 501

\bibitem[{{Blanton} \& {Roweis}(2007)}]{blanton:07}
{Blanton}, M.~R., \& {Roweis}, S. 2007, \aj, 133, 734

\bibitem[{{Blanton} {et~al.}(2003){Blanton}, {Hogg}, {Bahcall}, {Baldry},
  {Brinkmann}, {Csabai}, {Eisenstein}, {Fukugita}, {Gunn}, {Ivezi{\'c}},
  {Lamb}, {Lupton}, {Loveday}, {Munn}, {Nichol}, {Okamura}, {Schlegel},
  {Shimasaku}, {Strauss}, {Vogeley}, \& {Weinberg}}]{blanton:03}
{Blanton}, M.~R., {et~al.} 2003, \apj, 594, 186

\bibitem[{{Blanton} {et~al.}(2005){Blanton}, {Schlegel}, {Strauss},
  {Brinkmann}, {Finkbeiner}, {Fukugita}, {Gunn}, {Hogg}, {Ivezi{\'c}}, {Knapp},
  {Lupton}, {Munn}, {Schneider}, {Tegmark}, \& {Zehavi}}]{blanton:05}
---. 2005, \aj, 129, 2562

\bibitem[{{Blumenthal} {et~al.}(1986){Blumenthal}, {Faber}, {Flores}, \&
  {Primack}}]{blumenthal:86}
{Blumenthal}, G.~R., {Faber}, S.~M., {Flores}, R., \& {Primack}, J.~R. 1986,
  \apj, 301, 27

\bibitem[{{Brammer} {et~al.}(2011){Brammer}, {Whitaker}, {van Dokkum},
  {Marchesini}, {Franx}, {Kriek}, {Labb{\'e}}, {Lee}, {Muzzin}, {Quadri},
  {Rudnick}, \& {Williams}}]{brammer:11}
{Brammer}, G.~B., {et~al.} 2011, \apj, 739, 24

\bibitem[{{Brammer} {et~al.}(2012){Brammer}, {van Dokkum}, {Franx},
  {Fumagalli}, {Patel}, {Rix}, {Skelton}, {Kriek}, {Nelson}, {Schmidt},
  {Bezanson}, {da Cunha}, {Erb}, {Fan}, {F{\"o}rster Schreiber}, {Illingworth},
  {Labb{\'e}}, {Leja}, {Lundgren}, {Magee}, {Marchesini}, {McCarthy},
  {Momcheva}, {Muzzin}, {Quadri}, {Steidel}, {Tal}, {Wake}, {Whitaker}, \&
  {Williams}}]{3dhst}
---. 2012, \apjs, 200, 13

\bibitem[{{Brinchmann} {et~al.}(2004){Brinchmann}, {Charlot}, {White},
  {Tremonti}, {Kauffmann}, {Heckman}, \& {Brinkmann}}]{brinchmann:04}
{Brinchmann}, J., {Charlot}, S., {White}, S.~D.~M., {Tremonti}, C.,
  {Kauffmann}, G., {Heckman}, T., \& {Brinkmann}, J. 2004, \mnras, 351, 1151

\bibitem[{{Bruce} {et~al.}(2012){Bruce}, {Dunlop}, {Cirasuolo}, {McLure},
  {Targett}, {Bell}, {Croton}, {Dekel}, {Faber}, {Ferguson}, {Grogin},
  {Kocevski}, {Koekemoer}, {Koo}, {Lai}, {Lotz}, {McGrath}, {Newman}, \& {van
  der Wel}}]{bruce:12}
{Bruce}, V.~A., {et~al.} 2012, \mnras, 427, 1666

\bibitem[{{Bruzual} \& {Charlot}(2003)}]{bc:03}
{Bruzual}, G., \& {Charlot}, S. 2003, \mnras, 344, 1000

\bibitem[{{Cappellari} \& {Emsellem}(2004)}]{cappellari:04}
{Cappellari}, M., \& {Emsellem}, E. 2004, \pasp, 116, 138

\bibitem[{{Cappellari} {et~al.}(2006){Cappellari}, {Bacon}, {Bureau}, {Damen},
  {Davies}, {de Zeeuw}, {Emsellem}, {Falc{\'o}n-Barroso}, {Krajnovi{\'c}},
  {Kuntschner}, {McDermid}, {Peletier}, {Sarzi}, {van den Bosch}, \& {van de
  Ven}}]{cappellari:06}
{Cappellari}, M., {et~al.} 2006, \mnras, 366, 1126

\bibitem[{{Cappellari} {et~al.}(2013{\natexlab{a}}){Cappellari}, {Scott},
  {Alatalo}, {Blitz}, {Bois}, {Bournaud}, {Bureau}, {Crocker}, {Davies},
  {Davis}, {de Zeeuw}, {Duc}, {Emsellem}, {Khochfar}, {Krajnovi{\'c}},
  {Kuntschner}, {McDermid}, {Morganti}, {Naab}, {Oosterloo}, {Sarzi}, {Serra},
  {Weijmans}, \& {Young}}]{cappellari:13planes}
---. 2013{\natexlab{a}}, \mnras, 432, 1709

\bibitem[{{Cappellari} {et~al.}(2013{\natexlab{b}}){Cappellari}, {McDermid},
  {Alatalo}, {Blitz}, {Bois}, {Bournaud}, {Bureau}, {Crocker}, {Davies},
  {Davis}, {de Zeeuw}, {Duc}, {Emsellem}, {Khochfar}, {Krajnovi{\'c}},
  {Kuntschner}, {Morganti}, {Naab}, {Oosterloo}, {Sarzi}, {Scott}, {Serra},
  {Weijmans}, \& {Young}}]{cappellari:13}
---. 2013{\natexlab{b}}, \mnras, 432, 1862

\bibitem[{{Chabrier}(2003)}]{chabrier:03}
{Chabrier}, G. 2003, \pasp, 115, 763

\bibitem[{{Chang} {et~al.}(2013){Chang}, {van der Wel}, {Rix}, {Holden},
  {Bell}, {McGrath}, {Wuyts}, {H{\"a}ussler}, {Barden}, {Faber}, {Mozena},
  {Ferguson}, {Guo}, {Galametz}, {Grogin}, {Kocevski}, {Koekemoer}, {Dekel},
  {Huang}, {Hathi}, \& {Donley}}]{chang:13}
{Chang}, Y.-Y., {et~al.} 2013, \apj, 773, 149

\bibitem[{{Chevance} {et~al.}(2012){Chevance}, {Weijmans}, {Damjanov},
  {Abraham}, {Simard}, {van den Bergh}, {Caris}, \& {Glazebrook}}]{chevance:12}
{Chevance}, M., {Weijmans}, A.-M., {Damjanov}, I., {Abraham}, R.~G., {Simard},
  L., {van den Bergh}, S., {Caris}, E., \& {Glazebrook}, K. 2012, \apjl, 754,
  L24

\bibitem[{{Ciotti}(1991)}]{ciotti:91}
{Ciotti}, L. 1991, \aap, 249, 99

\bibitem[{{Cooper} {et~al.}(2012){Cooper}, {Newman}, {Davis}, {Finkbeiner}, \&
  {Gerke}}]{cooper:12}
{Cooper}, M.~C., {Newman}, J.~A., {Davis}, M., {Finkbeiner}, D.~P., \& {Gerke},
  B.~F. 2012, {spec2d: DEEP2 DEIMOS Spectral Pipeline}, astrophysics Source
  Code Library

\bibitem[{{Daddi} {et~al.}(2005){Daddi}, {Renzini}, {Pirzkal}, {Cimatti},
  {Malhotra}, {Stiavelli}, {Xu}, {Pasquali}, {Rhoads}, {Brusa}, {di Serego
  Alighieri}, {Ferguson}, {Koekemoer}, {Moustakas}, {Panagia}, \&
  {Windhorst}}]{daddi:05}
{Daddi}, E., {et~al.} 2005, \apj, 626, 680

\bibitem[{{Djorgovski} \& {Davis}(1987)}]{djorgovski:87}
{Djorgovski}, S., \& {Davis}, M. 1987, \apj, 313, 59

\bibitem[{{Dressler} {et~al.}(1987){Dressler}, {Lynden-Bell}, {Burstein},
  {Davies}, {Faber}, {Terlevich}, \& {Wegner}}]{dressler:87}
{Dressler}, A., {Lynden-Bell}, D., {Burstein}, D., {Davies}, R.~L., {Faber},
  S.~M., {Terlevich}, R., \& {Wegner}, G. 1987, \apj, 313, 42

\bibitem[{{Erben} {et~al.}(2009){Erben}, {Hildebrandt}, {Lerchster}, {Hudelot},
  {Benjamin}, {van Waerbeke}, {Schrabback}, {Brimioulle}, {Cordes}, {Dietrich},
  {Holhjem}, {Schirmer}, \& {Schneider}}]{erben:09}
{Erben}, T., {et~al.} 2009, \aap, 493, 1197

\bibitem[{{Faber}(1987)}]{faber:87}
{Faber}, S.~M., ed. 1987, {Nearly normal galaxies: From the Planck time to the
  present; Proceedings of the Eighth Santa Cruz Summer Workshop in Astronomy
  and Astrophysics, Santa Cruz, CA, July 21-Aug. 1, 1986}

\bibitem[{{Faber} \& {Jackson}(1976)}]{faberjackson}
{Faber}, S.~M., \& {Jackson}, R.~E. 1976, \apj, 204, 668

\bibitem[{{Faber} {et~al.}(2003){Faber}, {Phillips}, {Kibrick}, {Alcott},
  {Allen}, {Burrous}, {Cantrall}, {Clarke}, {Coil}, {Cowley}, {Davis}, {Deich},
  {Dietsch}, {Gilmore}, {Harper}, {Hilyard}, {Lewis}, {McVeigh}, {Newman},
  {Osborne}, {Schiavon}, {Stover}, {Tucker}, {Wallace}, {Wei}, {Wirth}, \&
  {Wright}}]{faber:deimos}
{Faber}, S.~M., {et~al.} 2003, in Society of Photo-Optical Instrumentation
  Engineers (SPIE) Conference Series, Vol. 4841, Instrument Design and
  Performance for Optical/Infrared Ground-based Telescopes, ed. M.~{Iye} \&
  A.~F.~M. {Moorwood}, 1657--1669

\bibitem[{{Fall} \& {Efstathiou}(1980)}]{fall:80}
{Fall}, S.~M., \& {Efstathiou}, G. 1980, \mnras, 193, 189

\bibitem[{{Franx} {et~al.}(2008){Franx}, {van Dokkum}, {Schreiber}, {Wuyts},
  {Labb{\'e}}, \& {Toft}}]{franx:08}
{Franx}, M., {van Dokkum}, P.~G., {Schreiber}, N.~M.~F., {Wuyts}, S.,
  {Labb{\'e}}, I., \& {Toft}, S. 2008, \apj, 688, 770

\bibitem[{{Furusawa} {et~al.}(2008){Furusawa}, {Kosugi}, {Akiyama}, {Takata},
  {Sekiguchi}, {Tanaka}, {Iwata}, {Kajisawa}, {Yasuda}, {Doi}, {Ouchi},
  {Simpson}, {Shimasaku}, {Yamada}, {Furusawa}, {Morokuma}, {Ishida}, {Aoki},
  {Fuse}, {Imanishi}, {Iye}, {Karoji}, {Kobayashi}, {Kodama}, {Komiyama},
  {Maeda}, {Miyazaki}, {Mizumoto}, {Nakata}, {Noumaru}, {Ogasawara}, {Okamura},
  {Saito}, {Sasaki}, {Ueda}, \& {Yoshida}}]{furusawa:08}
{Furusawa}, H., {et~al.} 2008, \apjs, 176, 1

\bibitem[{{Grogin} {et~al.}(2011){Grogin}, {Kocevski}, {Faber}, {Ferguson},
  {Koekemoer}, {Riess}, {Acquaviva}, {Alexander}, {Almaini}, {Ashby}, {Barden},
  {Bell}, {Bournaud}, {Brown}, {Caputi}, {Casertano}, {Cassata}, {Castellano},
  {Challis}, {Chary}, {Cheung}, {Cirasuolo}, {Conselice}, {Roshan Cooray},
  {Croton}, {Daddi}, {Dahlen}, {Dav{\'e}}, {de Mello}, {Dekel}, {Dickinson},
  {Dolch}, {Donley}, {Dunlop}, {Dutton}, {Elbaz}, {Fazio}, {Filippenko},
  {Finkelstein}, {Fontana}, {Gardner}, {Garnavich}, {Gawiser}, {Giavalisco},
  {Grazian}, {Guo}, {Hathi}, {H{\"a}ussler}, {Hopkins}, {Huang}, {Huang},
  {Jha}, {Kartaltepe}, {Kirshner}, {Koo}, {Lai}, {Lee}, {Li}, {Lotz}, {Lucas},
  {Madau}, {McCarthy}, {McGrath}, {McIntosh}, {McLure}, {Mobasher},
  {Moustakas}, {Mozena}, {Nandra}, {Newman}, {Niemi}, {Noeske}, {Papovich},
  {Pentericci}, {Pope}, {Primack}, {Rajan}, {Ravindranath}, {Reddy}, {Renzini},
  {Rix}, {Robaina}, {Rodney}, {Rosario}, {Rosati}, {Salimbeni}, {Scarlata},
  {Siana}, {Simard}, {Smidt}, {Somerville}, {Spinrad}, {Straughn}, {Strolger},
  {Telford}, {Teplitz}, {Trump}, {van der Wel}, {Villforth}, {Wechsler},
  {Weiner}, {Wiklind}, {Wild}, {Wilson}, {Wuyts}, {Yan}, \& {Yun}}]{candels}
{Grogin}, N.~A., {et~al.} 2011, \apjs, 197, 35

\bibitem[{{Guo} {et~al.}(2009){Guo}, {McIntosh}, {Mo}, {Katz}, {van den Bosch},
  {Weinberg}, {Weinmann}, {Pasquali}, \& {Yang}}]{guo:09}
{Guo}, Y., {et~al.} 2009, \mnras, 398, 1129

\bibitem[{{Hildebrandt} {et~al.}(2009){Hildebrandt}, {Pielorz}, {Erben}, {van
  Waerbeke}, {Simon}, \& {Capak}}]{hildebrandt:09}
{Hildebrandt}, H., {Pielorz}, J., {Erben}, T., {van Waerbeke}, L., {Simon}, P.,
  \& {Capak}, P. 2009, \aap, 498, 725

\bibitem[{{Hilz} {et~al.}(2012){Hilz}, {Naab}, \& {Ostriker}}]{hilz:12a}
{Hilz}, M., {Naab}, T., \& {Ostriker}, J.~P. 2012, ArXiv e-prints

\bibitem[{{Holden} {et~al.}(2010){Holden}, {van der Wel}, {Kelson}, {Franx}, \&
  {Illingworth}}]{holden:10}
{Holden}, B.~P., {van der Wel}, A., {Kelson}, D.~D., {Franx}, M., \&
  {Illingworth}, G.~D. 2010, \apj, 724, 714

\bibitem[{{Holden} {et~al.}(2012){Holden}, {van der Wel}, {Rix}, \&
  {Franx}}]{holden:12}
{Holden}, B.~P., {van der Wel}, A., {Rix}, H.-W., \& {Franx}, M. 2012, \apj,
  749, 96

\bibitem[{{Hopkins} {et~al.}(2009{\natexlab{a}}){Hopkins}, {Bundy}, {Murray},
  {Quataert}, {Lauer}, \& {Ma}}]{hopkins:09cores}
{Hopkins}, P.~F., {Bundy}, K., {Murray}, N., {Quataert}, E., {Lauer}, T.~R., \&
  {Ma}, C. 2009{\natexlab{a}}, \mnras, 398, 898

\bibitem[{{Hopkins} {et~al.}(2009{\natexlab{b}}){Hopkins}, {Hernquist}, {Cox},
  {Keres}, \& {Wuyts}}]{hopkins:09scaling}
{Hopkins}, P.~F., {Hernquist}, L., {Cox}, T.~J., {Keres}, D., \& {Wuyts}, S.
  2009{\natexlab{b}}, \apj, 691, 1424

\bibitem[{{Hyde} \& {Bernardi}(2009)}]{hyde:09}
{Hyde}, J.~B., \& {Bernardi}, M. 2009, \mnras, 396, 1171

\bibitem[{{Isobe} {et~al.}(1990){Isobe}, {Feigelson}, {Akritas}, \&
  {Babu}}]{isobe:90}
{Isobe}, T., {Feigelson}, E.~D., {Akritas}, M.~G., \& {Babu}, G.~J. 1990, \apj,
  364, 104

\bibitem[{{J{\o}rgensen} \& {Chiboucas}(2013)}]{jorgensen:13}
{J{\o}rgensen}, I., \& {Chiboucas}, K. 2013, \aj, 145, 77

\bibitem[{{J{\o}rgensen} {et~al.}(1996){Jorgensen}, {Franx}, \&
  {Kjaergaard}}]{jorgensen:96}
{Jorgensen}, I., {Franx}, M., \& {Kjaergaard}, P. 1996, \mnras, 280, 167

\bibitem[{{Kassin} {et~al.}(2007){Kassin}, {Weiner}, {Faber}, {Koo}, {Lotz},
  {Diemand}, {Harker}, {Bundy}, {Metevier}, {Phillips}, {Cooper}, {Croton},
  {Konidaris}, {Noeske}, \& {Willmer}}]{kassin:07}
{Kassin}, S.~A., {et~al.} 2007, \apjl, 660, L35

\bibitem[{{Koekemoer} {et~al.}(2007){Koekemoer}, {Aussel}, {Calzetti}, {Capak},
  {Giavalisco}, {Kneib}, {Leauthaud}, {Le F{\`e}vre}, {McCracken}, {Massey},
  {Mobasher}, {Rhodes}, {Scoville}, \& {Shopbell}}]{cosmosacs}
{Koekemoer}, A.~M., {et~al.} 2007, \apjs, 172, 196

\bibitem[{{Koekemoer} {et~al.}(2011){Koekemoer}, {Faber}, {Ferguson}, {Grogin},
  {Kocevski}, {Koo}, {Lai}, {Lotz}, {Lucas}, {McGrath}, {Ogaz}, {Rajan},
  {Riess}, {Rodney}, {Strolger}, {Casertano}, {Castellano}, {Dahlen},
  {Dickinson}, {Dolch}, {Fontana}, {Giavalisco}, {Grazian}, {Guo}, {Hathi},
  {Huang}, {van der Wel}, {Yan}, {Acquaviva}, {Alexander}, {Almaini}, {Ashby},
  {Barden}, {Bell}, {Bournaud}, {Brown}, {Caputi}, {Cassata}, {Challis},
  {Chary}, {Cheung}, {Cirasuolo}, {Conselice}, {Roshan Cooray}, {Croton},
  {Daddi}, {Dav{\'e}}, {de Mello}, {de Ravel}, {Dekel}, {Donley}, {Dunlop},
  {Dutton}, {Elbaz}, {Fazio}, {Filippenko}, {Finkelstein}, {Frazer}, {Gardner},
  {Garnavich}, {Gawiser}, {Gruetzbauch}, {Hartley}, {H{\"a}ussler},
  {Herrington}, {Hopkins}, {Huang}, {Jha}, {Johnson}, {Kartaltepe},
  {Khostovan}, {Kirshner}, {Lani}, {Lee}, {Li}, {Madau}, {McCarthy},
  {McIntosh}, {McLure}, {McPartland}, {Mobasher}, {Moreira}, {Mortlock},
  {Moustakas}, {Mozena}, {Nandra}, {Newman}, {Nielsen}, {Niemi}, {Noeske},
  {Papovich}, {Pentericci}, {Pope}, {Primack}, {Ravindranath}, {Reddy},
  {Renzini}, {Rix}, {Robaina}, {Rosario}, {Rosati}, {Salimbeni}, {Scarlata},
  {Siana}, {Simard}, {Smidt}, {Snyder}, {Somerville}, {Spinrad}, {Straughn},
  {Telford}, {Teplitz}, {Trump}, {Vargas}, {Villforth}, {Wagner}, {Wandro},
  {Wechsler}, {Weiner}, {Wiklind}, {Wild}, {Wilson}, {Wuyts}, \&
  {Yun}}]{candelsb}
---. 2011, \apjs, 197, 36

\bibitem[{{Kriek} {et~al.}(2009){Kriek}, {van Dokkum}, {Labb{\'e}}, {Franx},
  {Illingworth}, {Marchesini}, \& {Quadri}}]{kriek:09}
{Kriek}, M., {van Dokkum}, P.~G., {Labb{\'e}}, I., {Franx}, M., {Illingworth},
  G.~D., {Marchesini}, D., \& {Quadri}, R.~F. 2009, \apj, 700, 221

\bibitem[{{Martin} {et~al.}(2005){Martin}, {Fanson}, {Schiminovich},
  {Morrissey}, {Friedman}, {Barlow}, {Conrow}, {Grange}, {Jelinsky},
  {Milliard}, {Siegmund}, {Bianchi}, {Byun}, {Donas}, {Forster}, {Heckman},
  {Lee}, {Madore}, {Malina}, {Neff}, {Rich}, {Small}, {Surber}, {Szalay},
  {Welsh}, \& {Wyder}}]{martin:05}
{Martin}, D.~C., {et~al.} 2005, \apjl, 619, L1

\bibitem[{{Massey} {et~al.}(2010){Massey}, {Stoughton}, {Leauthaud}, {Rhodes},
  {Koekemoer}, {Ellis}, \& {Shaghoulian}}]{massey:10}
{Massey}, R., {Stoughton}, C., {Leauthaud}, A., {Rhodes}, J., {Koekemoer}, A.,
  {Ellis}, R., \& {Shaghoulian}, E. 2010, \mnras, 401, 371

\bibitem[{{Miller} {et~al.}(2012){Miller}, {Ellis}, {Sullivan}, {Bundy},
  {Newman}, \& {Treu}}]{miller:12}
{Miller}, S.~H., {Ellis}, R.~S., {Sullivan}, M., {Bundy}, K., {Newman}, A.~B.,
  \& {Treu}, T. 2012, \apj, 753, 74

\bibitem[{{Mo} {et~al.}(1998){Mo}, {Mao}, \& {White}}]{mmw:98}
{Mo}, H.~J., {Mao}, S., \& {White}, S.~D.~M. 1998, \mnras, 295, 319

\bibitem[{{Muzzin} {et~al.}(2009){Muzzin}, {Marchesini}, {van Dokkum},
  {Labb{\'e}}, {Kriek}, \& {Franx}}]{muzzin:09}
{Muzzin}, A., {Marchesini}, D., {van Dokkum}, P.~G., {Labb{\'e}}, I., {Kriek},
  M., \& {Franx}, M. 2009, \apj, 701, 1839

\bibitem[{{Muzzin} {et~al.}(2013){Muzzin}, {Marchesini}, {Stefanon}, {Franx},
  {McCracken}, {Milvang-Jensen}, {Dunlop}, {Fynbo}, {Brammer}, {Labb{\'e}}, \&
  {van Dokkum}}]{muzzin:13}
{Muzzin}, A., {et~al.} 2013, \apj, 777, 18

\bibitem[{{Nelson} {et~al.}(2014){Nelson}, {van Dokkum}, {Franx}, {Brammer},
  {Momcheva}, {F{\"o}rster Schreiber}, {da Cunha}, {Tacconi}, {Bezanson},
  {Kirkpatrick}, {Leja}, {Rix}, {Skelton}, {van der Wel}, {Whitaker}, \&
  {Wuyts}}]{nelson:14}
{Nelson}, E., {et~al.} 2014, arXiv:1406.3350

\bibitem[{{Newman} {et~al.}(2012){Newman}, {Cooper}, {Davis}, {Faber}, {Coil},
  {Guhathakurta}, {Koo}, {Phillips}, {Conroy}, {Dutton}, {Finkbeiner}, {Gerke},
  {Rosario}, {Weiner}, {Willmer}, {Yan}, {Harker}, {Kassin}, {Konidaris},
  {Lai}, {Madgwick}, {Noeske}, {Wirth}, {Connolly}, {Kaiser}, {Kirby},
  {Lemaux}, {Lin}, {Lotz}, {Luppino}, {Marinoni}, {Matthews}, {Metevier}, \&
  {Schiavon}}]{newmandeep:12}
{Newman}, J.~A., {et~al.} 2012, arXiv:1203.3192

\bibitem[{{Onodera} {et~al.}(2012){Onodera}, {Renzini}, {Carollo},
  {Cappellari}, {Mancini}, {Strazzullo}, {Daddi}, {Arimoto}, {Gobat}, {Yamada},
  {McCracken}, {Ilbert}, {Capak}, {Cimatti}, {Giavalisco}, {Koekemoer}, {Kong},
  {Lilly}, {Motohara}, {Ohta}, {Sanders}, {Scoville}, {Tamura}, \&
  {Taniguchi}}]{onodera:12}
{Onodera}, M., {et~al.} 2012, \apj, 755, 26

\bibitem[{{Patel} {et~al.}(2013){Patel}, {van Dokkum}, {Franx}, {Quadri},
  {Muzzin}, {Marchesini}, {Williams}, {Holden}, \& {Stefanon}}]{patel:13a}
{Patel}, S.~G., {et~al.} 2013, \apj, 766, 15

\bibitem[{{Peng} {et~al.}(2002){Peng}, {Ho}, {Impey}, \& {Rix}}]{galfit}
{Peng}, C.~Y., {Ho}, L.~C., {Impey}, C.~D., \& {Rix}, H.-W. 2002, \aj, 124, 266

\bibitem[{{Rousseeuw} \& {van Driessen}(2006)}]{rousseeuw:06}
{Rousseeuw}, P.~J., \& {van Driessen}, K. 2006, {Data Mining and Knowledge
  Discovery}, 12, 29

\bibitem[{{Shen} {et~al.}(2003){Shen}, {Mo}, {White}, {Blanton}, {Kauffmann},
  {Voges}, {Brinkmann}, \& {Csabai}}]{shen:03}
{Shen}, S., {Mo}, H.~J., {White}, S.~D.~M., {Blanton}, M.~R., {Kauffmann}, G.,
  {Voges}, W., {Brinkmann}, J., \& {Csabai}, I. 2003, \mnras, 343, 978

\bibitem[{{Simard} {et~al.}(2011){Simard}, {Mendel}, {Patton}, {Ellison}, \&
  {McConnachie}}]{simard:11}
{Simard}, L., {Mendel}, J.~T., {Patton}, D.~R., {Ellison}, S.~L., \&
  {McConnachie}, A.~W. 2011, \apjs, 196, 11

\bibitem[{{Skelton} {et~al.}(2014){Skelton}, {Whitaker}, {Momcheva}, {Brammer},
  {van Dokkum}, {Labbe}, {Franx}, {van der Wel}, {Bezanson}, {Da Cunha},
  {Fumagalli}, {Foerster Schreiber}, {Kriek}, {Leja}, {Lundgren}, {Magee},
  {Marchesini}, {Maseda}, {Nelson}, {Oesch}, {Pacifici}, {Patel}, {Price},
  {Rix}, {Tal}, {Wake}, \& {Wuyts}}]{skelton3dhst}
{Skelton}, R.~E., {et~al.} 2014, ArXiv:1403.3689

\bibitem[{{Straatman} {et~al.}(2014){Straatman}, {Labb{\'e}}, {Spitler},
  {Allen}, {Altieri}, {Brammer}, {Dickinson}, {van Dokkum}, {Inami},
  {Glazebrook}, {Kacprzak}, {Kawinwanichakij}, {Kelson}, {McCarthy},
  {Mehrtens}, {Monson}, {Murphy}, {Papovich}, {Persson}, {Quadri}, {Rees},
  {Tomczak}, {Tran}, \& {Tilvi}}]{straatman:14}
{Straatman}, C.~M.~S., {et~al.} 2014, \apjl, 783, L14

\bibitem[{{Szomoru} {et~al.}(2010){Szomoru}, {Franx}, {van Dokkum}, {Trenti},
  {Illingworth}, {Labb{\'e}}, {Bouwens}, {Oesch}, \& {Carollo}}]{szomoru:10}
{Szomoru}, D., {et~al.} 2010, \apjl, 714, L244

\bibitem[{{Tacconi} {et~al.}(2008){Tacconi}, {Genzel}, {Smail}, {Neri},
  {Chapman}, {Ivison}, {Blain}, {Cox}, {Omont}, {Bertoldi}, {Greve},
  {F{\"o}rster Schreiber}, {Genel}, {Lutz}, {Swinbank}, {Shapley}, {Erb},
  {Cimatti}, {Daddi}, \& {Baker}}]{tacconi:08}
{Tacconi}, L.~J., {et~al.} 2008, \apj, 680, 246

\bibitem[{{Taniguchi} {et~al.}(2007){Taniguchi}, {Scoville}, {Murayama},
  {Sanders}, {Mobasher}, {Aussel}, {Capak}, {Ajiki}, {Miyazaki}, {Komiyama},
  {Shioya}, {Nagao}, {Sasaki}, {Koda}, {Carilli}, {Giavalisco}, {Guzzo},
  {Hasinger}, {Impey}, {LeFevre}, {Lilly}, {Renzini}, {Rich}, {Schinnerer},
  {Shopbell}, {Kaifu}, {Karoji}, {Arimoto}, {Okamura}, \&
  {Ohta}}]{taniguchi:07}
{Taniguchi}, Y., {et~al.} 2007, \apjs, 172, 9

\bibitem[{{Taylor} {et~al.}(2010){Taylor}, {Franx}, {Brinchmann}, {van der
  Wel}, \& {van Dokkum}}]{taylor:10}
{Taylor}, E.~N., {Franx}, M., {Brinchmann}, J., {van der Wel}, A., \& {van
  Dokkum}, P.~G. 2010, \apj, 722, 1

\bibitem[{{Taylor} {et~al.}(2009){Taylor}, {Franx}, {van Dokkum}, {Quadri},
  {Gawiser}, {Bell}, {Barrientos}, {Blanc}, {Castander}, {Damen},
  {Gonzalez-Perez}, {Hall}, {Herrera}, {Hildebrandt}, {Kriek}, {Labb{\'e}},
  {Lira}, {Maza}, {Rudnick}, {Treister}, {Urry}, {Willis}, \&
  {Wuyts}}]{taylor:09}
{Taylor}, E.~N., {et~al.} 2009, \apjs, 183, 295

\bibitem[{{Toft} {et~al.}(2012){Toft}, {Gallazzi}, {Zirm}, {Wold}, {Zibetti},
  {Grillo}, \& {Man}}]{toft:12}
{Toft}, S., {Gallazzi}, A., {Zirm}, A., {Wold}, M., {Zibetti}, S., {Grillo},
  C., \& {Man}, A. 2012, \apj, 754, 3

\bibitem[{{Toft} {et~al.}(2014){Toft}, {Smol{\v c}i{\'c}}, {Magnelli}, {Karim},
  {Zirm}, {Michalowski}, {Capak}, {Sheth}, {Schawinski}, {Krogager}, {Wuyts},
  {Sanders}, {Man}, {Lutz}, {Staguhn}, {Berta}, {Mccracken}, {Krpan}, \&
  {Riechers}}]{toft:14}
{Toft}, S., {et~al.} 2014, \apj, 782, 68

\bibitem[{{Tomczak} {et~al.}(2013){Tomczak}, {Quadri}, {Tran}, {Labbe},
  {Straatman}, {Papovich}, {Glazebrook}, {Allen}, {Kacprzak},
  {Kawinwanichakij}, {Kelson}, {McCarthy}, {Mehrtens}, {Monson}, {Persson},
  {Spitler}, {Tilvi}, \& {van Dokkum}}]{tomczak:13}
{Tomczak}, A.~R., {et~al.} 2013, arXiv:1309.5972

\bibitem[{{Trujillo} {et~al.}(2006){Trujillo}, {F{\"o}rster Schreiber},
  {Rudnick}, {Barden}, {Franx}, {Rix}, {Caldwell}, {McIntosh}, {Toft},
  {H{\"a}ussler}, {Zirm}, {van Dokkum}, {Labb{\'e}}, {Moorwood},
  {R{\"o}ttgering}, {van der Wel}, {van der Werf}, \& {van
  Starkenburg}}]{trujillo:06}
{Trujillo}, I., {et~al.} 2006, \apj, 650, 18

\bibitem[{{Tully} \& {Fisher}(1977)}]{tullyfisher}
{Tully}, R.~B., \& {Fisher}, J.~R. 1977, \aap, 54, 661

\bibitem[{{van de Sande} {et~al.}(2011){van de Sande}, {Kriek}, {Franx}, {van
  Dokkum}, {Bezanson}, {Whitaker}, {Brammer}, {Labb{\'e}}, {Groot}, \&
  {Kaper}}]{sande:11}
{van de Sande}, J., {et~al.} 2011, \apjl, 736, L9

\bibitem[{{van de Sande} {et~al.}(2013){van de Sande}, {Kriek}, {Franx}, {van
  Dokkum}, {Bezanson}, {Bouwens}, {Quadri}, {Rix}, \& {Skelton}}]{sande:13}
---. 2013, \apj, 771, 85

\bibitem[{{van der Wel} {et~al.}(2004){van der Wel}, {Franx}, {van Dokkum}, \&
  {Rix}}]{wel:04}
{van der Wel}, A., {Franx}, M., {van Dokkum}, P.~G., \& {Rix}, H.-W. 2004,
  \apjl, 601, L5

\bibitem[{{van der Wel} {et~al.}(2011){van der Wel}, {Rix}, {Wuyts}, {McGrath},
  {Koekemoer}, {Bell}, {Holden}, {Robaina}, \& {McIntosh}}]{wel:11}
{van der Wel}, A., {et~al.} 2011, \apj, 730, 38

\bibitem[{{van der Wel} {et~al.}(2012){van der Wel}, {Bell}, {H{\"a}ussler},
  {McGrath}, {Chang}, {Guo}, {McIntosh}, {Rix}, {Barden}, {Cheung}, {Faber},
  {Ferguson}, {Galametz}, {Grogin}, {Hartley}, {Kartaltepe}, {Kocevski},
  {Koekemoer}, {Lotz}, {Mozena}, {Peth}, \& {Peng}}]{wel:12}
---. 2012, \apjs, 203, 24

\bibitem[{{van der Wel} {et~al.}(2014){van der Wel}, {Franx}, {van Dokkum},
  {Skelton}, {Momcheva}, {Whitaker}, {Brammer}, {Bell}, {Rix}, {Wuyts},
  {Ferguson}, {Holden}, {Barro}, {Koekemoer}, {Chang}, {McGrath}, {Haussler},
  {Dekel}, {Behroozi}, {Fumagalli}, {Leja}, {Lundgren}, {Maseda}, {Nelson},
  {Wake}, {Patel}, {Labbe}, {Faber}, {Grogin}, \& {Kocevski}}]{wel:14}
---. 2014, ArXiv:1404.2844

\bibitem[{{van Dokkum} \& {Franx}(1996)}]{dokkum:96}
{van Dokkum}, P.~G., \& {Franx}, M. 1996, \mnras, 281, 985

\bibitem[{{van Dokkum} {et~al.}(2009){van Dokkum}, {Kriek}, \&
  {Franx}}]{dokkumnature:09}
{van Dokkum}, P.~G., {Kriek}, M., \& {Franx}, M. 2009, \nat, 460, 717

\bibitem[{{van Dokkum} \& {van der Marel}(2007)}]{dokkummarel:07}
{van Dokkum}, P.~G., \& {van der Marel}, R.~P. 2007, \apj, 655, 30

\bibitem[{{van Dokkum} {et~al.}(2008){van Dokkum}, {Franx}, {Kriek}, {Holden},
  {Illingworth}, {Magee}, {Bouwens}, {Marchesini}, {Quadri}, {Rudnick},
  {Taylor}, \& {Toft}}]{dokkumnic:08}
{van Dokkum}, P.~G., {et~al.} 2008, \apjl, 677, L5

\bibitem[{{van Dokkum} {et~al.}(2010){van Dokkum}, {Whitaker}, {Brammer},
  {Franx}, {Kriek}, {Labb{\'e}}, {Marchesini}, {Quadri}, {Bezanson},
  {Illingworth}, {Muzzin}, {Rudnick}, {Tal}, \& {Wake}}]{dokkum:10}
---. 2010, \apj, 709, 1018

\bibitem[{{Vogt} {et~al.}(1996){Vogt}, {Forbes}, {Phillips}, {Gronwall},
  {Faber}, {Illingworth}, \& {Koo}}]{vogt:96}
{Vogt}, N.~P., {Forbes}, D.~A., {Phillips}, A.~C., {Gronwall}, C., {Faber},
  S.~M., {Illingworth}, G.~D., \& {Koo}, D.~C. 1996, \apjl, 465, L15

\bibitem[{{Warren} {et~al.}(2007){Warren}, {Hambly}, {Dye}, {Almaini}, {Cross},
  {Edge}, {Foucaud}, {Hewett}, {Hodgkin}, {Irwin}, {Jameson}, {Lawrence},
  {Lucas}, {Adamson}, {Bandyopadhyay}, {Bryant}, {Collins}, {Davis}, {Dunlop},
  {Emerson}, {Evans}, {Gonzales-Solares}, {Hirst}, {Jarvis}, {Kendall}, {Kerr},
  {Leggett}, {Lewis}, {Mann}, {McLure}, {McMahon}, {Mortlock}, {Rawlings},
  {Read}, {Riello}, {Simpson}, {Smith}, {Sutorius}, {Targett}, \&
  {Varricatt}}]{warren:07}
{Warren}, S.~J., {et~al.} 2007, \mnras, 375, 213

\bibitem[{{Weiner} {et~al.}(2006){Weiner}, {Willmer}, {Faber}, {Harker},
  {Kassin}, {Phillips}, {Melbourne}, {Metevier}, {Vogt}, \& {Koo}}]{weiner:06}
{Weiner}, B.~J., {et~al.} 2006, \apj, 653, 1049

\bibitem[{{Weinzirl} {et~al.}(2011){Weinzirl}, {Jogee}, {Conselice},
  {Papovich}, {Chary}, {Bluck}, {Gr{\"u}tzbauch}, {Buitrago}, {Mobasher},
  {Lucas}, {Dickinson}, \& {Bauer}}]{weinzirl:11}
{Weinzirl}, T., {et~al.} 2011, \apj, 743, 87

\bibitem[{{Whitaker} {et~al.}(2012){Whitaker}, {Kriek}, {van Dokkum},
  {Bezanson}, {Brammer}, {Franx}, \& {Labb{\'e}}}]{whitaker:12a}
{Whitaker}, K.~E., {Kriek}, M., {van Dokkum}, P.~G., {Bezanson}, R., {Brammer},
  G., {Franx}, M., \& {Labb{\'e}}, I. 2012, \apj, 745, 179

\bibitem[{Whitaker {et~al.}(2011)Whitaker, Labbe, van Dokkum, Brammer, Kriek,
  Marchesini, Quadri, Franx, Muzzin, Williams, Bezanson, Illingworth, Lee,
  Lundgren, Nelson, Rudnick, Tal, \& Wake}]{whitaker:11}
Whitaker, K.~E., {et~al.} 2011, \apj, 735, 86

\bibitem[{{Williams} {et~al.}(2009){Williams}, {Quadri}, {Franx}, {van Dokkum},
  \& {Labb{\'e}}}]{williams:09}
{Williams}, R.~J., {Quadri}, R.~F., {Franx}, M., {van Dokkum}, P., \&
  {Labb{\'e}}, I. 2009, \apj, 691, 1879

\bibitem[{{Zaritsky} {et~al.}(2006){Zaritsky}, {Gonzalez}, \&
  {Zabludoff}}]{zaritsky:06}
{Zaritsky}, D., {Gonzalez}, A.~H., \& {Zabludoff}, A.~I. 2006, \apj, 638, 725

\bibitem[{{Zaritsky} {et~al.}(2008){Zaritsky}, {Zabludoff}, \&
  {Gonzalez}}]{zaritsky:08}
{Zaritsky}, D., {Zabludoff}, A.~I., \& {Gonzalez}, A.~H. 2008, \apj, 682, 68

\bibitem[{{Zaritsky} {et~al.}(2011){Zaritsky}, {Zabludoff}, \&
  {Gonzalez}}]{zaritsky:07}
---. 2011, \apj, 727, 116

\end{thebibliography}
\end{document}